\documentclass[pra,a4paper,twocolumn,superscriptaddress,longbibliography,nofootinbib]{revtex4-2}
\usepackage[utf8]{inputenc}
\usepackage{amssymb}
\usepackage{amsmath}
\usepackage{amsfonts}
\usepackage{amsthm}
\usepackage{amsbsy}
\usepackage{bm}
\usepackage{bbold}

\usepackage{graphicx}
\usepackage[dvipsnames]{xcolor}
\usepackage{multirow}
\usepackage{float}
\usepackage{comment}
\usepackage{ulem}
\usepackage{relsize}
\usepackage{cmap}
\usepackage{physics}
\usepackage{cancel}
\usepackage[colorlinks]{hyperref}

\DeclareMathOperator{\TT}{\texttt{T}}

\newcommand{\DOS}{{LDoS}}
\newcommand{\NLSM}{{NL$\sigma$M}}
\newcommand{\iqHe}{{iqHe}}
\newcommand{\sqHe}{{sqHe}}
\begin{document}
	
	\title{Instanton analysis for the spin quantum Hall symmetry class: Non-perturbative corrections to physical observables and generalized multifractal spectrum}
	
       \author{M. V. Parfenov}

\affiliation{\mbox{L. D. Landau Institute for Theoretical Physics, Semenova 1-a, 142432, Chernogolovka, Russia}}
       
        \affiliation{Department of Physics, HSE University, 101000 Moscow, Russia}

\affiliation{Laboratory for Condensed Matter Physics, HSE University, 101000 Moscow, Russia}

\author{I. S. Burmistrov}

\affiliation{\mbox{L. D. Landau Institute for Theoretical Physics, Semenova 1-a, 142432, Chernogolovka, Russia}}

\affiliation{Laboratory for Condensed Matter Physics, HSE University, 101000 Moscow, Russia}

	\date{\today} 
	
\begin{abstract}
Recently, there has been renewed interest in studies of criticality in the spin quantum Hall effect, realized in the Altland-Zirnbauer symmetry class C of disordered, noninteracting fermions in two spatial dimensions. In our study, we develop a nonperturbative analysis of the replica two-dimensional nonlinear sigma model in class C. We explicitly construct the instanton solution with a unit topological charge. By treating fluctuations around the instanton at the Gaussian level, we calculate the instanton correction to the {\color{black} disorder-averaged logarithm of the partition function.}
We compute non-perturbative corrections to the anomalous dimensions of pure power-law scaling local operators, which determine the spectrum of generalized multifractality. We also calculate instanton corrections to the renormalized longitudinal and Hall spin conductivities and determine the topology of the phase diagram for class C. Our results demonstrate that the spin quantum Hall effect is indeed a close cousin of the integer quantum Hall effect.

\end{abstract}

	\maketitle
	%%%%%%%%%%%%%%%%%%%%%%%%%%%%%%%%%%%%

\section{Introduction}

The  prototypical example of a quantum phase transition in disordered noninteracting fermionic systems is the Anderson localization/delocalization transition \cite{Anderson58}. After more than sixty years of active research on Anderson localization, its physics and related phenomena are well understood (see Ref. \cite{Evers2008,Ryu2016} for a review). Perhaps the most intriguing discovery in the realm of Anderson localization was the discovery of the integer quantum Hall effect ({\iqHe}) \cite{Klitzing1980,Tsui1981} that realizes Anderson transitions between topologically nontrivial localized  phases. The understanding of the synergy between topology and quantum interference in the {\iqHe} has stimulated search of other topological Anderson transitions. It was found \cite{Wigner1951,Dyson1962a,Dyson1962b,Zirnbauer1996,Zirnbauer1997,Zirnbauer2005} that there are exist ten different (Altland-Zirnbauer) symmetry classes of disordered noninteracting Hamiltonians. In addition, each spatial dimension admits five out of ten symmetry classes with nontrivial topology \cite{Schnyder2008,Schnyder2009,Kitaev2009}. Physics in each of ten symmetry classes is described by the corresponding effective long-wave field theory -- nonlinear sigma model ({\NLSM}) (see Ref. \cite{Evers2008} for a review). The nontrivial topology is reflected by the presence of the topological term (either theta-term or Wess-Zumino-Witten-Novikov term) in the {\NLSM} action.

In two spatial dimensions the corresponding Anderson transitions between different topological phases occur  at strong coupling typically. Thus, criticality is not accessible within effective long-wave description in terms of {\NLSM}. Nevertheless, the non-perturbative (instanton) analysis of {\NLSM} in a weak coupling regime allows one to understand the structure of the phase diagram and explain quantization of the proper physical observable. A well-known example of 
such a situation is the integer quantum Hall effect (iqHe). The iqHe phase diagram and the quantization of the Hall conductance has been understood on the basis of existence of {\it instantons} -- topologically nontrivial solutions of classical equations of motion for the {\NLSM} action \cite{Levine1983,Khmelnitskii1983,pruisken1984localization,Pruisken1985,pruisken1987quasiparticles,pruisken1987quasiparticlesB,pruisken1995cracking,Pruisken2007}.

Recently, the interest to the {\iqHe} has been renewed in the context of a critical theory for the Anderson transition between different topological phases (plateau-plateau transition). For a long time it has been suggested \cite{Zirnbauer1999,Kettemann1999,Bhaseen2000,tsvelik2001wavefunctionsstatisticsquantum,Tsvelik2007,Zirnbauer2019} that 
the Wess--Zumino--Novikov--Witten models are an ultimate conformal critical theory for the Anderson transition in the  {\iqHe}. However, the critical theory predicts not only the localization length  exponent but also anomalous dimensions of various local operators. Examples of such operators are the disorder averaged moments of the local density of states ({\DOS}), $\nu(\bm{x})$, which demonstrate {\it pure} power-law scaling with the system size, $\langle \nu^q(\bm{x})\rangle {\sim} L^{{-}x_{\textsf{(q)}}}$, at criticality \cite{Wegner1980,Castellani1986,Lerner1988}. In addition to the moments of {\DOS} there are much more pure scaling observables~\cite{Wegner1986}. The corresponding operators can be expressed in terms of the disorder averages of specific combinations of wave functions \cite{Gruzberg2013,Karcher2022,Karcher2022b,Karcher2023}. The corresponding set of generalized multifractal exponents $x_{\bm{\lambda}}$ is unique for each symmetry class and dimensionality (see Refs. \cite{Mirlin2000,EversMirlin} for a review). Provided the local conformal invariance and Abelian fusion rules for the pure scaling operators hold, the generalized multifractal exponents $x_{\lambda}$ were proven to have a parabolic form with a single free parameter only \cite{Bondesan2017,Karcher2021,Padayasi2023}. 
The validity of this prediction has been debated in numerical simulations \cite{Obuse2008,Evers2008,Karcher2021}. 
One more interesting aspect of recent discussions of the {\iqHe} criticality is the numerical evidence that the magnitude of the localization length exponent in the {\iqHe} transition varies with a change of geometry of a random potential \cite{Sedrakyan2017,Sedrakyan2019,Sedrakyan2021,Dresselhaus2022,Topchyan2024}.

The iqHe has a close cousin -- the spin quantum Hall effect ({\sqHe}) that occurs in superconducing class C \cite{Volovik1997,Kagolovsky1999,Senthil1999}.  The analog of the integer quantized Hall conductivity is the spin Hall conductivity that describes the response of the spin current to the gradient of the external magnetic field.~\footnote{We note that similar relation between the spin current and the gradient of the magnetic field is realized in thin films of the superfluid ${}^3$He-A \cite{GEVolovik_1989,GEVolovik_1989_2}.}  The sqHe has very similar phenomenology as the iqHe, in particular, it has even integer quantized spin Hall conductivity, Anderson transitions between different topological phases, description in terms of the {\NLSM} (see Ref. \cite{Evers2008} for a review). Additionally, the sqHe has an advantage: an infinite set of anomalous dimensions of local operators has been computed analytically at criticality by mapping to the classical percolation problem \cite{Gruzberg1999,Beamond2002,Mirlin2003,Evers2003,Subramaniam2008,Karcher2022}. Recently, the {\sqHe} criticality has been intensively tested against description by the conformal field theory.   
It was shown \cite{Mirlin2003,Puschmann2021,Karcher2021,Karcher2022,Karcher2022b,Karcher2023a,Karcher2023} that although the numerical data for the generalized multifractal exponents $x_{\bm{\lambda}}$ reproduce exact analytical results obtained from mapping to percolation, there is a clear evidence for a violation of parabolicity. These results prove a lack of the local conformal invariance at the {\sqHe} transition in $d{=}2$. {\color{black} There is an alternative point of view on breaking of the generalized parabolicity {\color{black} and the presence of local conformal invariance}. As it 
{\color{black} has recently been proposed}
\cite{zirnbauer2024infraredlimito3nonlinear}, it can be  
{\color{black} explained by a} non-perturbative reconstruction of non-linear sigma-model manifold, when 
renormalization group (RG) flow reaches strong coupling limit.}

The absence of plausible candidates for the critical theory of the {\sqHe} transition makes non-perturbative weak coupling analysis of the corresponding {\NLSM} to be of interest.  
Surprisingly, the non-perturbative analysis of the {\NLSM} for class C is absent so far in the literature. In our paper we fill this gap by developing the instanton analysis of the replica {\NLSM} in the class C. In particular, 
\color{black}
\begin{itemize}
\item[(i)] we construct the instanton solution for the replica {\NLSM} in the class C, cf. Eqs. \eqref{eq:inst:gen} and \eqref{eq:Qinst:unrot};
\item[(ii)] employing the Pauli-Villars regularization scheme~\cite{PV1949}, we calculate the instanton correction to the partition function, cf. Eq. \eqref{eq:ZZ:inst:final};. 
\item[(iii)] we apply the developed methodology to computation of non-perturbative corrections to 
the average {\DOS}, cf. Eq. \eqref{eq:RG:LDOS:non-pert}, and to the anomalous dimensions of the local derivativeless eigen operators (with respect to the renormalization group), cf. Eq. \eqref{eq:anom:dim:inst:full:0}, that determine the spectrum of generalized multifractality;
\item[(iv)] we calculate instanton corrections to the renormalized longitudinal and Hall spin conductivities and extract the two-parameter non-perturbative renormalization group equations, cf. Eq. \eqref{eq: RGeq}. 
\end{itemize}
\color{black}
Our results support a general idea that the sqHe is in many ways very similar to the iqHe. 

The outline 
of the paper is  as follows. We start from the formalism of the replica Pruisken's {\NLSM} for class C (Sec. \ref{Sec:Formalism}). In Sec. \ref{Sec:Instantons} we present construction of instantons with topological charge $\pm 1$ for class C 
and analysis of Gaussian fluctuations around it. The instanton contribution to the partition function is computed in Sec. \ref{Sec:PartitionFunction}. 
Next  we present computation and analysis of instanton corrections to the anomalous dimensions of pure scaling eigen operators 
(Sec. \ref{sec: RGop}). In Sec. \ref{Sec:SpinCond} we compute instanton corrections to the longitudinal 
and Hall spin conductivities 
{\color{black} 
and rewrite them in terms of two-parameter renormalization group equations
}. We end the paper with discussions and conclusions (Sec. \ref{Sec:Final}). Details of lengthy calculations are given in Appendices.

\section{Pruisken's {\NLSM} for class C \label{Sec:Formalism}}
\subsection{{\NLSM} action \label{SubSec: NLSM}}

 We use formalism of Finkel'stein's {\NLSM} for the symmetry of class C (see Refs. \cite{Jeng2001a,Jeng2001,DellAnna2006,Liao2017,Babkin2022} for details) adapted for noninteracting electrons. We exclude interaction term, reduce the space of positive and negative Matsubara frequencies to the retarded-advanced (RA) space, leaving two frequencies only. In this way we obtain the following form of the Pruisken's {\NLSM}:
\begin{equation}\label{eq: act1}
    S_{\sigma} = -\frac{g}{16}\int_{\boldsymbol{x}} \text{Tr} \left(\nabla \mathcal{Q}\right)^2 + \frac{g_{H}}{16}\int_{\boldsymbol{x}} \text{Tr} \left[\varepsilon_{jk}\mathcal{Q} \nabla_j \mathcal{Q} \nabla_k \mathcal{Q}\right] ,
\end{equation}
where $\int_{\bm{x}} = \int d^2 \bm{x}$ and $\varepsilon_{jk}$ denotes Levi-Civita symbol with $\varepsilon_{xy}=-\varepsilon_{yx}=1$. The field $\mathcal{Q}$ is a traceless Hermitian matrix, defined on $N_r{\times}N_r$ replica, $2{\times}2$ retarded-advanced  and $2{\times}2$ spin spaces. It satisfies a nonlinear constraint and Bogoliubov -- de Gennes (BdG) symmetry constraint: 
\begin{equation}\label{eq: constr1}
    \begin{split}
        \mathcal{Q}^2 =1, \quad & \mathcal{Q} = -\overline{\mathcal{Q}}, \quad \overline{\mathcal{Q}} = \textsf{s}_2L_0 \mathcal{Q}^T L_0 \textsf{s}_2, \\
        L_0 &= \begin{pmatrix}
        0 & \textsf{s}_0 \otimes \hat{1}_{\rm r} &\\
        \textsf{s}_0 \otimes \hat{1}_{\rm r} & 0 &\\
        \end{pmatrix}_{\rm RA} .
    \end{split}
\end{equation}
Here subscript ${\rm RA}$ implies matrix acting in the RA space. The parameter $g$ ($g_H$) denotes bare dimensionless longitudinal (transverse) spin conductance, $\hat{1}_{\rm r}$ stands for the unit marix in replica space, and $\textsf{s}_j$ are standard Pauli matrices. We note that the  last (topological) term in the r.h.s of Eq. \eqref{eq: act1} has exactly the same form as the one in class A \cite{pruisken1984localization}. 

Non-linear constraint on $\mathcal{Q}$-matrix and BdG symmetry relation define $\sigma$-model target manifold: {\color{black}$\mathcal{Q} \in \text{Sp}(4 N_r)/\text{U}(2 N_r)$}
.  In the end of all calculation we should take the replica limit: $N_r \rightarrow 0$. 
In order to resolve non-linear constraint we can rewrite $\mathcal{Q}$-matrix in terms of non-uniform matrix rotations:
\begin{equation}
    \mathcal{Q} = \tilde{\mathcal{T}}^{-1} \Lambda \tilde{\mathcal{T}}, \quad \Lambda=   
    \sigma_3 \otimes \textsf{s}_0 \otimes \hat{1}_{\rm r} ,
\end{equation}
where {\color{black} $\tilde{\mathcal{T}} \in$ Sp(4$N_r$), } 
 $\sigma_{3}$ is the corresponding Pauli matrix in the RA space. The matrix $\Lambda$ is so-called metalic saddle-point. {\color{black} It is convenient to realize the rotation matrices $\tilde{\mathcal{T}}$ belonging to Sp(4$N_r$) as the $4N_r{\times}4N_r$ matrices satisfying the following conditions:
\begin{equation}
    \tilde{\mathcal{T}}^{-1} =  \tilde{\mathcal{T}}^{\dagger}, \quad  \left(\tilde{\mathcal{T}}^{-1}\right)^{\textsf{T}} L_0 \textsf{s}_2 = \textsf{s}_2 L_0  \tilde{\mathcal{T}} .
    \label{eq:T:cond:i}
 \end{equation}
 }\color{black}
The first relation in Eq. \eqref{eq:T:cond:i} restricts the number of independent real variables of the matrix $\tilde{\mathcal{T}}$ to be equal to $(4N_r)^2$ as given for U($4N_r$) group. The second condition in Eq. \eqref{eq:T:cond:i} reduces the number of independent real variables down to $2(2N_r)^2+(2N_r)$ as it should be for Sp(4$N_r$).
 \color{black}

\subsection{Non-unitary matrix rotation}

The model defined in Sec. \ref{SubSec: NLSM} reduces to the $4{\cross}4$ matrix theory in single-replica limit $N_r=1$. Even with such size of matrix $\mathcal{Q}$ calculations  
might become too tedious. In order to avoid this difficulty, 
we perform non-unitary rotation of the matrix basis, introducing new matrix $Q$ as
\begin{equation}\label{eq: utran}
\begin{split}
     \mathcal{Q} & = U^{-1}Q U, \qquad 
    U =  \begin{pmatrix}
        1 & 1 & 0 & 0 \\
        0 & 0 & -1 & 1 \\
        1 & -1 & 0 & 0 \\
        0 & 0 & 1 & 1
        \end{pmatrix}_{\rm RA,S} \otimes \frac{\hat{1}_{\rm r}}{\sqrt{2}} .
    \end{split}
\end{equation}
Here the subscript ${\rm RA,S}$ implies matrix acting in the combined {\rm RA} and spin spaces.
We use peculiarities of symplectic group   
to change the anti-symmetric matrix which defines BdG symmetry relation in such a way that it acts in the spin space only:
\begin{equation}\label{eq: constr2}
    Q = -\Bar{Q}, \quad \Bar{Q} = \textsf{s}_2 Q^{T}\textsf{s}_2 .
\end{equation}

Also we note that the rotation \eqref{eq: utran} changes the definition of the metallic saddle-point:
\begin{equation}
 \Lambda  
 \rightarrow \underline{\Lambda}= \hat{1}_{\rm RA}  \otimes \textsf{s}_{3} \otimes   \hat{1}_{\rm r}   .
\end{equation}
It is easy to check, that the transformation \eqref{eq: utran} does not change the form of the action \eqref{eq: act1}. After the transformation \eqref{eq: utran} we cannot distinguish RA and replica spaces, therefore, it is convenient to introduce new notation: $n = 2N_r$ for dimension of the combined RA/replica space. 

After the above transformation, the {\NLSM} reduces to the $2 \cross 2$ matrix theory in single-replica limit,  $n=1$.

\section{Instantons with topological charges $\mathcal{C}=\pm$ 1\label{Sec:Instantons}}

\subsection{Constuction of the instanton solution}
In order to obtain the saddle-point solutions with non-trivial topology we %should 
use  
Bogomolny inequality: 
\begin{equation}\label{}    \operatorname{Tr}\left(\nabla_x \mathcal{Q} \pm i \mathcal{Q}\nabla_y \mathcal{Q}\right)^2\geq 0 . 
\end{equation}
It can be rewritten equivalently as
\begin{equation}
    \frac{1}{16} \int_{\boldsymbol{x}} \operatorname{Tr}\left(\nabla \mathcal{Q}\right)^2 \geq \pi \left| \mathcal{C}[\mathcal{Q}]\right| .
\end{equation}
Here we  
introduce the topological charge 
\begin{equation}\label{eq: topch}
    \mathcal{C}[\mathcal{Q}] = \frac{1}{16 \pi i} \int_{\boldsymbol{x}}\Tr \varepsilon_{jk} \mathcal{Q} \nabla_j  \mathcal{Q} \nabla_k  \mathcal{Q} .
\end{equation}
Therefore, stable matrix field configurations which minimize the action should satisfy the self-duality equation:
\begin{equation}\label{eq: selfd}
    \nabla_x \mathcal{Q} \pm i \mathcal{Q}\nabla_y \mathcal{Q} =0 .
\end{equation}

We note that Eq. \eqref{eq: selfd} is invariant under transformation~\eqref{eq: utran}. Therefore, we 
construct solution of this equation in the rotated basis as follows.  
At first, we set the number of replicas in the rotated basis equal to unity, $n=1$.  
Then,  
similar to class A \cite{pruisken1987quasiparticles}, we use Belavin-Polyakov instanton to solve 
Eq. \eqref{eq: selfd} by the matrix 
\begin{equation}\label{eq: inst1:0}
    \Lambda_{\mathrm{inst}}^{(n=1)} = \begin{pmatrix}
        |e_1|^2 -e_0^2 & 2 e_0 e_1 \\
       2 e_0 e^{*}_1  & -\left(|e_1|^2 -e_0^2\right)
    \end{pmatrix} ,
\end{equation}  
where
\begin{equation}
    e_0=\frac{\lambda}{\sqrt{\left|z-z_0\right|^2+\lambda^2}}, \quad e_1=\frac{z-z_0}{\sqrt{\left|z-z_0\right|^2+\lambda^2}} .
\end{equation}
Here \color{black} $z=x+i y$ stands for the complex coordinate, the complex \color{black} $z_0$ denotes a position of the instanton's center and $\lambda$ stands for it's scale size. 
Generalization of  solution \eqref{eq: inst1:0}  
to the case of $n>1$  
is constructed in the form  
that explicitly violates the replica symmetry:
\begin{equation}
      \Lambda_{\mathrm{inst}}^{(n>1)} = \begin{pmatrix}
        \Lambda_{\mathrm{inst}}^{(n=1)} & 0  & \dots & 0\\ 
        0 & \textsf{s}_{3}   & \dots & 0  \\ 
     \dots & \dots & \dots & \dots \\
       0 & 0 & \dots & \textsf{s}_{3}
    \end{pmatrix}_{\rm r} .
    \label{eq: inst1}
\end{equation}
Here the lower index ${\rm r}$ denotes that the matrix is written in the replica space. Topological charge \eqref{eq: topch} for this soluton is equal to one: $\mathcal{C}[\Lambda_{\mathrm{inst}}^{(n>1)}]=1$. Solution with negative topological charge can be obtained by complex conjugation of solution \eqref{eq: inst1}.  
Therefore, full instanton manifold can be written in terms of the unitary rotations $\mathcal{T}_0$ and $\tilde{R}$ about the metalic saddle point $\underline{\Lambda}$:
\begin{equation}
\begin{split}
    Q_{\mathrm{inst}} &= \mathcal{T}_0^{-1}\Lambda_{\mathrm{inst}}^{(n>1)} \mathcal{T}_0= \mathcal{T}_0^{-1}\tilde{R}^{-1} \underline{\Lambda} \tilde{R}\mathcal{T}_0, \\ \tilde{R} &= \begin{pmatrix}
        R_0 & 0 \\ 
        0 & \hat{1} \\ 
    \end{pmatrix}_{\rm r} = \begin{pmatrix}
       e^{*}_1 & 0 &  e_0 & 0 \\ 
       0& \hat{1}_{\textrm{r}} & 0 & 0 \\ 
       -e_0 & 0 &  e_1 & 0 \\
         0 & 0 &  0& \hat{1}_{\rm r} \\
    \end{pmatrix}_{\text{S}, {\rm r}} .
    \end{split}
    \label{eq:inst:gen}
\end{equation}
Here $\mathcal{T}_0\in \text{Sp}(2n)$ stands for an arbitrary global unitary rotation, which describes the orientation of the instanton in the coset space $\text{Sp}(2n)/\text{U}(n)$. 
One can check that  
$Q_{\mathrm{inst}}$ satisfies BdG symmetry relation  
\eqref{eq: constr2}. The classical action for the instanton solution \eqref{eq:inst:gen} 
is finite:
\begin{equation}
    S_{\rm cl} = -\pi g  
    + i \pi g_{H} .
\end{equation}
We note that the classical action $S_{\rm cl}$ is independent of $z_0$, $\lambda$, and $\mathcal{T}_0$, i.e. they can be identified as zero modes.
In contrast 
 to the case of class A, it is convenient to split $g_H$ on even integer part and fractional part:
\begin{equation}\label{eq: thedef}
g_H = 2k + \vartheta/\pi,  \qquad k\in \mathbb{Z}, \qquad -\pi < \vartheta \leq \pi .
\end{equation}
We note that a change of $\vartheta$ from $-\pi$ to $\pi$ corresponds to change of $g_H$ on $2$ rather than on $1$ as in the class A.

\subsection{Fluctuations near the instanton solution}

In order to construct perturbation theory around the instanton solution we use exponential parametrization of the $Q$-matrix:
\begin{equation}
    Q = \tilde{R}^{-1} \mathcal{V}\tilde{R}, \quad \mathcal{V} = e^{-W/2} \underline{\Lambda} e^{W/2}
    \label{eq: fluctpar}
\end{equation}
The matrix $W$ has to satisfy the following constraints: $ \{W,\Lambda\}=0$, $W^\dag=-W$, and $W = - \overline{W}$. Consequently, the matrix $W$ can be parametrized by $n{\times} n$ complex symmetric matrix $\hat{w}$,
    \begin{equation}  
  W = \begin{pmatrix}
            0 & \hat{w} \\
            -\hat{w}^{*} & 0 \\
        \end{pmatrix}_{\text{sp}} , \qquad \hat{w}^T = \hat w .
    \end{equation}

The presence of instanton can be interpreted as appearance of the non-Abelian vector potential, $A_j = \tilde{R} \nabla_j \tilde{R}^{-1}$, in the {\NLSM} action, e.g.
\begin{equation}
\label{eq: acpot}
    -\frac{g}{16}\int_{\boldsymbol{x}} \text{Tr} \left(\nabla \mathcal{Q}\right)^2 =   -\frac{g}{16}\int_{\boldsymbol{x}} \text{Tr} \left(\nabla_j \mathcal{V}-\left[\mathcal{V}, A_j\right]\right)^2 .
\end{equation}
Using symmetries of the fields $\hat w$ and $\hat w^{*}$  we can expand action \eqref{eq: acpot}  to the second order in $\hat w,\; \hat w^{*}$:
\begin{gather}
  \delta S_\sigma=
  -
  \frac{g}{8} \int d \boldsymbol{x} \mu^2(r)\Biggl[w^{11} O^{(2)} w^{ *11}+\sum_{\alpha=2}^n w^{\alpha \alpha} O^{(0)} w^{*\alpha \alpha}  \notag \\
 + 2 \sum_{1<\alpha<\beta\leqslant n} w^{\alpha \beta} O^{(0)} w^{*\alpha\beta }+2\sum_{\alpha=2}^n w^{1\alpha} O^{(1)} w^{*1\alpha}
  \Biggr]  ,
  \label{eq: fluctact}
\end{gather}
where $r^2=x^2+y^2$ and the Greek indices denote matrix structure in the replica space. Here we define the  
operators
\begin{equation}
    O^{(a)} =-\frac{\left(r^2+\lambda^2\right)^2}{4 \lambda^2}\Biggl [\nabla_j +\frac{i a}{r^2+\lambda^2} \varepsilon_{j k} x_k\Biggr]^2
    -\frac{a}{2},   
\end{equation}
{\color{black} and the measure 
\begin{equation}\label{eq:mudef}
    \mu(r) = \frac{2\lambda}{r^2+\lambda^2} .
\end{equation}}
We note that the set of operators $ O^{(a)}$ and the measure $\mu(r)$ are exactly the same as the ones which arise in analysis of fluctuations around the instanton in class A~\cite{pruisken1987quasiparticles}.

The natural appearance of the measure $\mu^2(r)$ indicates that it is convenient to employ inverse stereographic projection from the flat space onto the sphere with a radius $\lambda$. Therefore, we should introduce new  
coordinates -- the spherical angles:
\begin{equation}
     \cos{\phi}= \frac{r^2-\lambda^2}{r^2+\lambda^2} =\eta , \quad \theta = \arctan\left(\frac{y}{x}\right) .
\end{equation}
In terms of
the spherical coordinates, the
quantities $e_1$ and $e_0$ can be written as
\begin{equation}
    e_0 = \sqrt{\frac{1-\eta}{2}}, \quad e_1 = \sqrt{\frac{1+\eta}{2}}e^{i\theta} ,
\end{equation}
while the operators $O^{(a)}$ become
\begin{gather}
     O^{(a)} = -\frac{\partial}{\partial \eta}\left[\left(1-\eta^2\right) \frac{\partial}{\partial \eta}\right] %+ 
     - \frac{1}{1-\eta^2}\frac{\partial^2}{\partial \theta^2}
     %- 
     +\frac{ia}{1-\eta}\frac{\partial}{\partial \theta} \notag\\ %- 
     +\frac{a^2}{4}\frac{1+\eta}{1-\eta}
     %+
     -\frac{a}{2} .
     \label{eq: kinop}
\end{gather}

Eigensystem for operators \eqref{eq: kinop} can be found in a standard way as the solution of Schr\"odinger-type equation
\begin{equation}
     O^{(a)}\Phi^{(a)}\left(\eta,\theta\right) = E^{(a)}\Phi^{(a)}\left(\eta,\theta\right) .
\end{equation}
Here the eigenfunctions are normalized with respect to the measure $d\eta d\theta$. 
The eigen functions are expressed 
in terms of Jacobi polynomials:
\begin{equation}
P_n^{\alpha, \beta}(\eta)=\frac{(-1)^n}{2^n n !} \frac{(1-\eta)^{-\alpha}}{(1+\eta)^\beta} \frac{d^n}{d \eta^n} \frac{(1-\eta)^{n+\alpha}}{(1+\eta)^{-n-\beta}} .
\end{equation}
as \cite{pruisken1987quasiparticles}
\begin{equation}
    \Phi^{(a)}_{J,M}{=} C_{J,M}^{(a)}e^{{-}iM\theta}(1{-}\eta)^{\frac{a}{2}}\left(1{-}\eta^2\right)^{\frac{M}{2}}P_{J{-}M{-}s_a}^{M{+}a,M}(\eta) ,
\end{equation}
where $s_a=0,1,1$ for $a=0,1,2$, respectively. 
The eigen states are enumerated by the angular momentum $J=0,1,2,\dots$ for $a=0$ and $J=1,2,3, \dots$ for $a=1,2$. The corresponding momentum projections satisfy $-J-a(a-1)/2\leqslant M \leqslant J - s_a$. The normalization constants read~\cite{Pruisken2005}
\begin{align}      C_{J,M}^{(a)}=& \frac{\sqrt{\Gamma\left(J{+}M{+}1{+}a(a{-}1)/2\right)\Gamma\left(J{-}M{+}1{-}s_a\right)}}{2^{M{+}1{+}a(a{-}1)/2}\sqrt{\pi}\Gamma\left(J\right)}
\notag \\
& \times
\begin{cases}
 \frac{\sqrt{2J+1}}{J+1}, & \quad a=0 ,\\
 1, & \quad a=1 ,\\
 \frac{\sqrt{2J+1}}{\sqrt{J(J+1)}},  & \quad a=2 .
\end{cases}
\end{align}
The eigen energies are given as
\begin{equation}
    E^{(a)}_J = 
    (J-s_a)(J+1-s_a+a) .
\end{equation}
We emphasize that the eigen energy vanishes for the smallest allowed angular momentum, $E^{(a)}_{J{=}s_a}=0$.

\begin{table}[b]
\caption{Number of fields for operators $O^{(a)}$}
\begin{tabular}{||c|c|c||}
\hline \hline \text { Operator } & \text { Number of fields } $w^{\alpha \beta}$ & \text { Degeneracy } \\
\hline $O^{(0)}$ & $(n^2-n)/2$ & $1$ \\
$O^{(1)}$ & $(n-1)$ & $2$ \\
$O^{(2)}$ & $1$ & $3$ \\
\hline \hline
\end{tabular}
\label{tab:modes}
\end{table}

\subsection{Analysis of the zero modes}

There are several zero-energy modes for the operators $O^{(a)}$. From Table \ref{tab:modes} we can compute the number of the zero modes to be equal to $n^2+3n+2$. 

We also note, that one can rewrite eigen functions corresponding to the modes with zero eigen energies in terms of $e_0$ and $e_1$:
\begin{multline}
    \Phi^{(0)}_{0,0}=\frac{1}{2\sqrt{\pi}}, \quad \Phi^{(1)}_{1,-1}= \frac{1}{\sqrt{2\pi}}e_1,\quad \Phi^{(1)}_{1,0}= \frac{1}{\sqrt{2\pi}}e_0,\\ \Phi^{(2)}_{1,-2}{=}\sqrt{\frac{3}{4\pi}}e_1^2,\quad \Phi^{(2)}_{1,-1}{=}\sqrt{\frac{3}{2\pi}}e_0 e_1,\quad \Phi^{(2)}_{1,0}{=}\sqrt{\frac{3}{4\pi}}e_0^2 .
\end{multline}

%It is important 
Now we show that each zero mode is related with corresponding instanton degree of freedom  (collective coordinate). For this purpose we introduce small deviations of instanton degrees of freedom: $\xi_j = \{\lambda, x_0,y_0 \}$ and generators $t$, which can be defined as an expansion of global rotation matrices $\mathcal{T}_0$ near the identity martix:
\begin{equation}\label{eq: gen}
     \mathcal{T}_0 = 1+it .
\end{equation}
In this way we find
\begin{equation}
\begin{split}
    & Q(\xi_j +  \delta \xi_j) =  \tilde{R}^{-1}(\xi_j)\left (\Lambda + \left[\Lambda, B\right]\right)\tilde{R}(\xi_j) , \\
    & B =   i \tilde{R}(\xi_j) t \tilde{R}^{-1}(\xi_j) - \delta \xi_j \tilde{R}(\xi_j)\left(\partial_{\xi_j} \tilde{R}^{-1}(\xi_j)\right) .
    \end{split}
\end{equation}
Comparing the above equations with Eq. \eqref{eq: fluctpar}, 
we relate fluctuation matrix $\hat w$ and the zero modes as
\begin{gather}
    w = 2 \left(i \tilde{R} t \tilde{R}^{-1} - \delta \xi_i \tilde{R}\left(\partial_{\xi_i} \tilde{R}^{-1}\right)\right)_{12} .
\end{gather}
Here the subscripts $12$
%lower indices 
correspond to the spin space structure in the rotated basis. Explicit expressions for matrices and symmetry relations for $\hat w$ allows us to obtain the following results  
\begin{gather}
    w^{11} {=} {-}4 e_0 e^{*}_1 \left(i t_{11}^{11} {-} \frac{\delta \lambda}{2\lambda}\right) {+} 2e_0^2\left(\frac{\delta z^{*}}{\lambda} {-} i t_{12}^{* 11}\right)  {+} 2 i  e^{*2}_1 t_{12}^{11}, \notag \\
    w^{1\alpha} = w^{\alpha1}=2i (e^{*}_1 t^{1 \alpha}_{12} - e_0 t^{* 1\alpha}_{11}), \notag \\
w^{\alpha \beta} = 2 i t_{12}^{\alpha \beta}, \quad \beta\geqslant\alpha>1 .
\label{eq: zmtofl1}
\end{gather}

\begin{figure}[b]
    \centering
\includegraphics[width=0.85\linewidth]{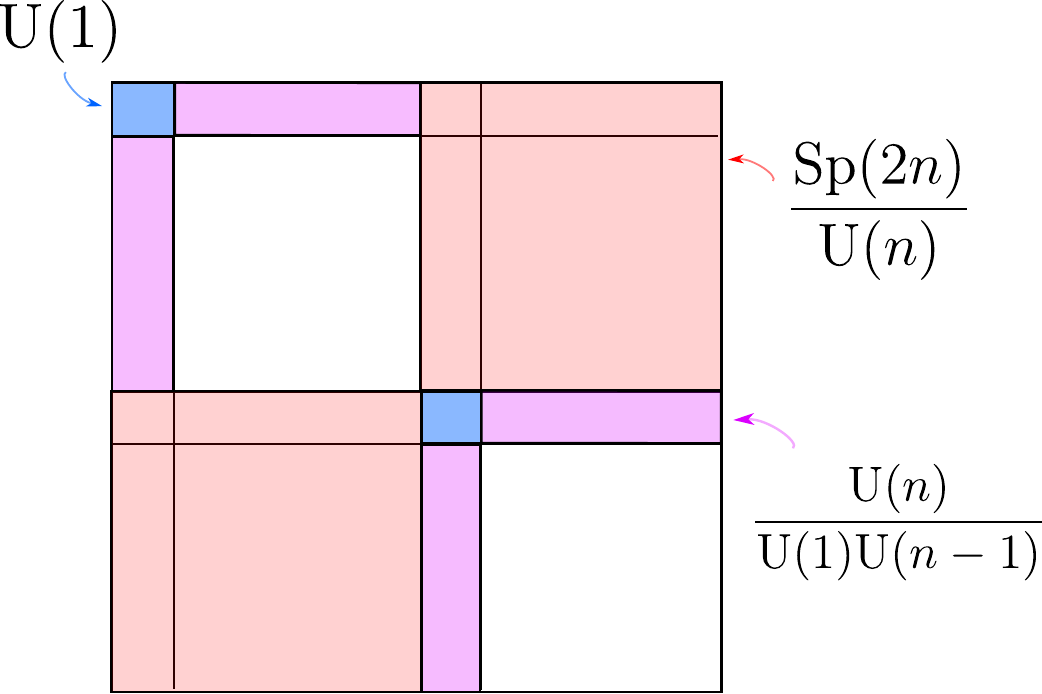}
    \caption{The symmetry-breaking pattern in terms of generators \eqref{eq: gen} for an arbitrary matrix %from 
    \color{black} $\mathcal{T}_0\in\text{Sp}(2n)$. The off-diagonal elements corresponding to $\mathcal{T}^\prime\in \text{Sp}(2n)/\text{U}(n)$ are shown by pink color. The diagonal elements corresponding to $\TT \in \text{U}(1)\cup \text{U}(n)/[\text{U}(1)\times\text{U}(n-1)]$ are shown in blue and violet colors.  \color{black}}
    \label{fig:Symbreak}
\end{figure}

If one considers a similar task for the trivial topological sector ($\mathcal{C}[\mathcal{Q}]=0$), 
i.e. one makes an expansion %near 
around the metalic saddle-point $\underline{\Lambda}$, one 
%can 
obtains
\begin{equation}\label{eq: zmtofl0}
    w^{\alpha \beta} = 2 i t_{12}^{\alpha \beta} .
\end{equation}
These are trivial zero modes corresponding to $\mathcal{N}_0=n^2+n$ real parameters. 
The number of zero modes for the instanton solution can be splitted as follows
\begin{multline}
     n^2+3n+2 = {\color{red}\underbrace{\color{black}{n^2 + n}}_{\text{\color{red} trivial 
     }}} + {\color{blue}\underbrace{\color{black}{2n + 2}}_{\text{\color{blue} instanton}}}=  \mathcal{N}_0  \\ +{\color{blue}\underbrace{\color{black}{2n - 2 +1}}_{\text{\color{blue} instanton rotations}}} +{\color{blue}\underbrace{\color{black}{3}}_{\text{\color{blue} $\xi_i$}}} .
\end{multline}
{\color{black}In the last line of the 
above equation we separate the number of zero modes corresponding to the instaton parameters $\xi_i=\{z_0, \lambda\}$ from the number of zero modes corresponding to generators of instaton rotations $t$}. The above zero mode structure corresponds to the symmetry-breaking 
pattern shown in Fig. 
\ref{fig:Symbreak}. The instanton breaks $U(n)$ down to $U(n{-}1)$ explicitly. \color{black} Therefore, we can present the zero modes corresponding to rotations $\mathcal{T}_0$ around the instanton as a product
$\mathcal{T}_0 = \TT \mathcal{T}^{\prime}$, where $\mathcal{T}^{\prime}\in \text{Sp}(2n)/\text{U}(n)$ and 
$\TT \in \text{U}(1) \cup \text{U}(n)/[\text{U}(1)\times\text{U}(n-1)]$ describes the additional rotational zero modes.
\color{black}

%
%
%%%%%%%%
%%%%%%%%%
%
%
%

\section{Calculation of the partition function\label{Sec:PartitionFunction}}

{\color{black} In order to calculate the partition function we use the fact, that total partition function can be written as sum over contributions from all topological sectors: 
\begin{equation}
    \mathcal{Z} = \sum_{\mathcal{C}=-\infty}^{\infty}  \mathcal{Z}_{\mathcal{C}} .
\end{equation}
where $\mathcal{C}$ is the integer-valued %quantized 
topological charge, see Eq.~\eqref{eq: topch}. For a reason to be explained shortly, it is more convenient to work with the quantity $\mathcal{F}=(1/(2n)) \ln \mathcal{Z}$. Analogously to the statistical mechanics, $\mathcal{F}$ can be referred as `the free energy'. In the replica limit, $n\to 0$, $\mathcal{F}$ determines the disorder-averaged logarithm of  the partition function. Due to exponential smallness of configurations with $|\mathcal{C}| >1$, we can expand the logarithm of the partition function in a series:
\begin{equation}
    \ln \mathcal{Z} \approx \ln \mathcal{Z}_{0} + \frac{\mathcal{Z}_{+1}}{\mathcal{Z}_{0}} + \frac{\mathcal{Z}_{-1}}{\mathcal{Z}_{0}} + ... 
\end{equation}
Therefore, our main goal for this section is computation of the instanton corrections to 
$\ln \mathcal{Z}$, which we call the normalized partition function in all calculations below.

}

Calculation of the partition function can be separated into two parts. The first one is calculation of the contribution of massive modes in Gaussian approximation. The second part is  integration over the zero modes, that should be carried out exactly due to divergence of the corresponding determinants in the Gaussian approximation.

Also, we notice that we are interested in calculation of the normalized partition function which is defined as
\begin{equation}\label{eq: part1}
    \frac{\mathcal{Z}_{\mathrm{inst}}}{\mathcal{Z}_0} = \frac{\int \mathcal{D}[w,w^{*}] \exp\left(S_{\rm cl}+\delta S_{\sigma}\right)}{\int\mathcal{D}[w,w^{*}] \exp\left(\delta S_{0}\right)} ,
\end{equation}
where {\color{blue}$\mathcal{Z}_{\mathrm{inst}} = \mathcal{Z}_{+1} = \mathcal{Z}^{*}_{-1} $} and $\delta S_0$ has the same form as $\delta S_{\sigma}$ \eqref{eq: fluctact} with all operators $O^{(a)}$ replaced by $O^{(0)}$.

\subsection{Determinants for the massive modes}

In this section we exclude all calculations related to the zero modes, concentrating on obtaining the determinant due to integration over the massive modes. Therefore, schematically we rewrite Eq. \eqref{eq: part1} as:
\begin{equation} \label{eq: part2}
    \frac{\mathcal{Z}_{\mathrm{inst}}}{\mathcal{Z}_0} = \color{black} \int_{\rm zm} \color{black} A_{\rm zm}[\lambda, z_0,g,\vartheta]e^{S_{\rm cl} + \mathcal{D}} .
\end{equation}
Here \color{black} $\int_{\rm zm}$ denotes the integral over the manifold of the zero modes, \color{black} $A_{zm}$ is a functional associated with the contribution of the zero modes, which will be defined below and $\mathcal{D}$ is the contribution of the massive modes. In this section our aim is to compute $\mathcal{D}$ in the Gaussian approximation.

It is convenient to introduce Green's functions \color{black} for the operators $O^{(a)}$ \color{black} in a standard way:
\begin{equation}
\mathcal{G}_a(\eta,\theta;\eta',\theta';\omega)=\sum_{J M} \frac{|J M\rangle_{(a)(a)}\langle J M|}{E_J^{(a)}+\omega} .
\end{equation}

Then we can calculate integral over $w$ fields explicitly, using expansion in terms of the eigenfunctions of operators $O^{(a)}$:
\begin{equation}
    w^{\alpha \beta}(\eta, \theta) = \frac{2\sqrt{2}}{\sqrt{g}}\sum_{J,M} u_{J M}^{\alpha \beta} \Phi^{(a)*}_{J,M} .
\end{equation}
Performing Gaussian integration over complex coefficients $u_{J,M}^{\alpha\beta}$
we obtain
\begin{multline}\label{eq: Ddef}
    \mathcal{D}= (n-1) \left(\operatorname{tr}\ln \frac{1}{2}\mathcal{G}_1(0)-\operatorname{tr}\ln \frac{1}{2}\mathcal{G}_0(0)\right) +\operatorname{tr}\ln\mathcal{G}_2(0) \\ -\operatorname{tr}\ln\mathcal{G}_0(0)= -(n-1)D^{(1)}-D^{(2)} .
\end{multline}
Appearance of factors 1/2 under $\Tr\ln$  after integration over $u_{J M}^{\alpha \beta}$ is due to last two terms in Eq. \eqref{eq: fluctact}. It reflects symmetry relations between matrix elements of fluctuation matrices $W$, cf. Eq. \eqref{eq: fluctpar}.

\color{black}
The traces in Eq. \eqref{eq: Ddef} can be readily written in terms of the eigen values of the operators $O^{(a)}$. However, these sums are divergent.
This fact reflects the infrared divergencies which are well-known to appear in the course of the perturbative background field renormalization of the {\NLSM} action. Usually, such divergences are treated by means of the dimensional regularization scheme. However, since the instanton solution exists strictly in two spatial dimensions we cannot use dimensional regularization scheme. Fortunately, there exists other regularization scheme -- Pauli-Villars method \cite{PV1949} -- that can be employed in two dimensions. We note that it was used for computation of non-perturbative corrections in Yang-Mills theory due to Belavin-Polyakov-Schwarz-Tyupkin instanton \cite{tHooft1976}. Later this methodology has been adapted to the studies of instanton effects in the {\iqHe} \cite{pruisken1987quasiparticles,Pruisken2005,pruisken2007theta}. Below we will sketch the derivation of the results for the regularized determinants. More details can be found in Ref. \cite{Pruisken2005}. 

The idea of the Pauli-Villars method is to introduce $K+1$ copies of the quantum theory $\delta S_\sigma$. In each copy the operators $O^{(a)}$ are supplemented by the mass term $\mathcal{M}_f^2$, $f=0,\dots,K$, i.e.  
$O^{(a)}\to O^{(a)}+\mathcal{M}_f^2$. It is assumed that $\mathcal{M}_0 =0$ while $\mathcal{M}_f \gg 1$ for $f=1,\dots, K$. The additional $K$ copies of the theory are served to cancel all the divergencies except the logarithmic one. In order to extract the logarithmically divergent contriburion some of the additional copies are assumed to contribute to the logarithm of the determinant, $\mathcal{D}$, as if they result from the integration over Grassmanian variables, i.e. contributing with the opposite sign in front of $\tr\ln$ in Eq. \eqref{eq: Ddef}. In other words, one has to use the following substitution in Eq. \eqref{eq: Ddef},
\begin{equation}
     \operatorname{tr}\ln r_b^{{-}1} \mathcal{G}_a(0) \rightarrow \sum_{f=0}^{K}\varepsilon_f \operatorname{tr}\ln r_b^{-1} \mathcal{G}_a(\mathcal{M}_f^2) .
\end{equation}
Here $r_b=1,2$ (see Eq. \eqref{eq: Ddef}), $\varepsilon_0=1$, and $\varepsilon_f =\pm 1$ for $f=1,\dots, K$. The particular sign of each $\varepsilon_f$ is chosen to be able to cancel all divergencies except the logarithmic one. Then the regularized versions of the functions $D^{(1,2)}$ become
\begin{equation}
    D_{\mathrm{reg}}^{(a)}= \lim\limits_{\Lambda\to \infty} \Bigl [ \Phi_\Lambda\left (\frac{1+a}{2},r_a\right )-
    \Phi_\Lambda\left (\frac{1}{2},r_a\right )\Bigr ] ,
\end{equation}    
where $r_1=2$ and $r_2=1$, and
\begin{gather}
\Phi_\Lambda (p,r) = \sum_{J=p+1}^{\Lambda} 2J \ln[r (J^2-p^2)] +\sum_{f=1}^{K}\varepsilon_f \notag \\ \times \sum_{J=p}^{\Lambda} 2J \ln[r (J^2-p^2+\mathcal{M}_f^2)] .
\end{gather}
Evaluating the function $\Phi_\Lambda (p,r)$ under assumptions that $\Lambda \gg \mathcal{M}_f\gg 1$, we find (see Ref. \cite{Pruisken2005} for details)
\begin{gather}
\Phi_\Lambda (p,r) {=} {-}2 p \ln r {+}  4 \sum_{J{=}1}^{\Lambda} J\ln J {+} 2 p^{2} {-} 2 \sum
\limits_{J{=}1}^{2 p} (J{-}p) \ln J \notag  \\
{-} 2p^2 \ln \Lambda 
{+} 2 \ln r \sum_{J=p{+}1}^\Lambda J{+}
\sum_{f=1}^K \varepsilon_f \Bigl [ 2 \Lambda ( \Lambda {+}1) \ln
\Lambda {-} \Lambda^{2} \notag \\
 {+} \frac{\ln e \Lambda }{3} {+} 2 \ln r \sum_{J=p{+}1}^\Lambda J 
{-} \frac{1{-}6 p}{3} \ln {\mathcal M}_f {-} 2 {\mathcal M}_f^2 \ln {\mathcal M}_f
\notag \\ {+}2 {\mathcal M}_f^2 \ln \Lambda {-}2p^2 \ln \Lambda \Bigr ].
\label{eq:res:Phi:L:0}
\end{gather}
Now we choose $K=5$ and set $\varepsilon_1=\varepsilon_2=1$ and $\varepsilon_3=\varepsilon_4=\varepsilon_5=-1$. The masses satisfy the following two equations
\begin{equation}\label{eq: PVdef:0}
    \sum_{f=1}^{K} \varepsilon_f \mathcal{M}_f^2=0, \quad
    \sum_{f=1}^{K}\varepsilon_f \mathcal{M}_f^2 \ln \mathcal{M}_f = 0 .
\end{equation}
Then, the above expression for the function $\Phi_\Lambda (p,r)$ simplifies drastically,
\begin{gather}
\Phi_\Lambda (p,r) {=} {-}2 p \ln r {+} \frac{1{-}6 p}{3} \ln {\mathcal M}{+} 2 p^{2} {-} 2 \sum
\limits_{J{=}1}^{2 p} (J{-}p) \ln J \notag  \\
 {+} 4 \sum_{J{=}1}^{\Lambda} J\ln J  {-}
 2 \Lambda ( \Lambda {+}1) \ln
\Lambda {+} \Lambda^{2} 
 {-} \frac{\ln e \Lambda }{3} 
, \label{eq:res:Phi:L}
\end{gather}
where we introduced the so-called Pauli-Villars mass
\begin{equation}\label{eq: PVdef}
\ln \mathcal{M} =-
    \sum_{f=1}^{K}\varepsilon_f \ln \mathcal{M}_f .
\end{equation}
\color{black}
Next, using the result \eqref{eq:res:Phi:L}, we can compute the regularized determinants $D_{\mathrm{reg}}^{(a)}$. Interestingly, they can be expressed in terms of  regularized determinants for the similar problem in class A, see Ref.~\cite{pruisken2005instanton},  
 \begin{equation}
\begin{aligned}
& D^{(1)}_{\text{reg}} = D^{(1)}_{\text{reg},\,A} -\ln 2, \quad \color{black} D^{(2)}_{\text{reg}} = D^{(2)}_{\text{reg},\,A},\color{black}\\ 
& D^{(1)}_{\text{reg},\,A} =-\ln \mathcal{M} +\frac{3}{2}-2 \ln 2, \\
& D_{\text {reg }}^{(2)}=-2\ln \mathcal{M} +4- \ln 2 -3 \ln 3 .
\end{aligned}
\end{equation}
Hence we obtain the final result for the regularized determinant, coming from the integration over the massive modes
\begin{multline}
     \mathcal{D}=  (n+1) \ln \mathcal{M}e^{\gamma-1/2}-\frac{3 (n-1)}{2}+(3 n -2)\ln 2-4 \\ +3 \ln 3 -(n+1)(\gamma-1/2) .
\end{multline}
Here for the purposes that will be clear further, we added and subtracted a constant $(n+1)(\gamma-1/2)$, where $\gamma \approx 0.577$ stands for Euler constant. \color{black} We emphasize that the Pauli-Villars mass $\mathcal{M}$ controls divergencies in the fluctuation determinant $\mathcal{D}$. As we discussed above such divergences arise similar to divergencies in the course of the perturbative renormalization of the action due to elimination of the fast fluctuations on the top of slow background field configuration.
\color{black}

\subsection{Jacobian for the zero modes}

Detailed calculation of the Jacobian for collective modes can be found in Ref.~\cite{pruisken1987quasiparticles}. In this section we present resulting expression and explain pecularities arising in the case of class C. In order to derive the zero-mode Jacobian we use  explicit expressions for coefficients $u_{J,M}^{\alpha\beta}$  in terms of the instanton degrees of freedom \eqref{eq: zmtofl1}:
\begin{multline}
    u^{11}_{1,-1}=-\sqrt{\frac{4\pi g}{3}}\left(i t^{11}_{11}-\frac{\delta \lambda}{2\lambda}\right), \quad  u^{11}_{1,-2}= \sqrt{\frac{2 \pi g}{3}} it_{12}^{11}, \\ 
    u^{11}_{1,0}= \sqrt{\frac{2 \pi g}{3}} \left(\frac{\delta z^*}{\lambda}-it_{12}^{*11}\right), \quad u^{1 \alpha}_{1,-1} = \sqrt{ \pi g} it^{1 \alpha}_{12}, \\ 
    u^{1 \alpha}_{1,0} = -\sqrt{\pi g}i t_{11}^{*1\alpha},\quad u^{\alpha \beta}_{0,0}=\sqrt{2 \pi g} i t^{\alpha \beta}_{12}, \quad \beta\geqslant\alpha>1 .
\end{multline}
The above expressions can be rewritten in terms of Jacobi matrices with block structure in replica space:
\begin{equation}
    \left( \text{Re}\,u_{1,-1}^{11}, \; \text{Im}\,u_{1,-1}^{11}\right) = \begin{pmatrix}
        \sqrt{\frac{ \pi g}{3 \lambda^2}} & 0   \\ 
        0   &  -\sqrt{\frac{4 \pi g}{3}}
    \end{pmatrix} \begin{pmatrix}
        \delta \lambda \\
        t_{11}^{11}
    \end{pmatrix} ,
    \label{eq:JJ:1}
\end{equation}
\begin{multline}
    \left( \text{Re}\,u_{1,-2}^{11}, \; \text{Im}\,u_{1,-2}^{11}, \;\text{Re}\,u_{1,1}^{10}, \; \text{Im}\,u_{1,1}^{10}\right) = \\ \begin{pmatrix}
        0& -\sqrt{\frac{2\pi g}{3}} & 0 & 0  \\ 
        \sqrt{\frac{2\pi g}{3}}  & 0 & 0 &0 \\ 
        0 & -\sqrt{\frac{2\pi g}{3}}& \sqrt{\frac{2\pi g}{3 \lambda^2}}&0\\
        -\sqrt{\frac{2\pi g}{3}} & 0 & 0 & -\sqrt{\frac{2\pi g}{3 \lambda^2}}
    \end{pmatrix} \begin{pmatrix}
        \text{Re}\,t_{12}^{11} \\
        \text{Im}\,t_{12}^{11} \\ 
        x_0 \\ 
        y_0
    \end{pmatrix} ,
     \label{eq:JJ:2}
\end{multline}
\begin{multline}
    \left( \text{Re}\,u_{1,-1}^{1\alpha}, \; \text{Im}\,u_{1,-1}^{1\alpha}, \;\text{Re}\,u_{1,0}^{1\alpha}, \; \text{Im}\,u_{1,0}^{1\alpha}\right) = \\ \begin{pmatrix}
        0& -\sqrt{\pi g} & 0 & 0  \\ 
        \sqrt{\pi g}  & 0 & 0 &0 \\ 
        0 & 0 & 0 & -\sqrt{\pi g}\\
        0 & 0 & -\sqrt{\pi g} & 0
    \end{pmatrix} \begin{pmatrix}
        \text{Re}\,t_{12}^{1\alpha} \\
        \text{Im}\,t_{12}^{1\alpha} \\ 
        \text{Re}\,t_{11}^{1\alpha} \\
        \text{Im}\,t_{11}^{1\alpha}
    \end{pmatrix} ,
     \label{eq:JJ:3}
\end{multline}
\begin{equation}
    \left( \text{Re}\,u_{0,0}^{\alpha \beta}, \; \text{Im}\,u_{0,0}^{\alpha \beta}\right) = \begin{pmatrix}
        0 & -\sqrt{2\pi g}   \\ 
        \sqrt{2\pi g}   &  0
    \end{pmatrix} \begin{pmatrix}
        \text{Re}\,t_{12}^{\alpha \beta} \\
        \text{Im}\,t_{12}^{\alpha \beta}
    \end{pmatrix} ,
     \label{eq:JJ:4}
\end{equation}
with $\beta\geqslant\alpha>1$. Using these explicit expressions for the blocks of the full Jacobi matrix, we  derive the following Jacobian for collective modes:
\begin{equation}
    \left|J_{\mathrm{inst}}\right| = \frac{2 g}{3 \lambda} \cdot \frac{4 g}{9 \lambda^2}\cdot g^{2n-2}\cdot \left(2 g\right)^{n(n-1)/2}  .
\end{equation}
Here we present the result as a product of four factors which correspond to block matrices \eqref{eq:JJ:1} -- \eqref{eq:JJ:4}. Jacobian for the trivial topological sector can be derived in similar way, using Eq. \eqref{eq: zmtofl0}:
\begin{equation}
   |J_0| = \left(2 g\right)^{n(n+1)/2} .
\end{equation}
Therefore, resulting answer for the contribution from the zero modes acquires the following form:
\begin{equation}
\int_{\rm zm}  A_{\mathrm{zm}}= \frac{8 g^{n+1}}{27 \cdot2^{n}} \int \frac{d\boldsymbol{r_0} d\lambda}{\lambda^3}  \frac{\int \mathcal{D}[\mathcal{T}^{\prime}] \int \mathcal{D}[\TT]}{\int \mathcal{D}[\mathcal{T}^{\prime}]} ,
\end{equation}
where {\color{black} global rotations $\mathcal{T}^{\prime}\in \text{Sp}(2n)/\text{U}(n)$ corresponding to the zero modes in the absence of the instanton,
while $\TT \in \text{U}(1)\cup \text{U}(n)/[\text{U}(1){\times}\text{U}(n{-}1)]$ are additional rotational zero mode %matrices 
induced by the presence of the instanton (see Fig. \ref{fig:Symbreak}).}

\subsection{The partition function}

Using Eq. \eqref{eq: part2}, we calculate the instanton correction to the partition function. 
Since $\mathcal{T}^\prime$ and $\TT$ describe the zero modes, integration over them is reduced to the volume of the corresponding manifolds, which were computed, for example, in Refs.~\cite{marinov1980invariant,boya2003volumes}:
\color{black}
\begin{equation}
\int \mathcal{D}[\TT]   =
\text{vol}\left[\text{U}(1)\right]  \text{vol}\left[\frac{\text{U}(n)}{\text{U}(1){\times} \text{U}(n{-}1)}\right]
= \frac{2\pi^{n}}{\Gamma(n)}  .
\end{equation}
\color{black}
After all substitutions we find the following one-instanton contribution to the partition function
\begin{gather}
\frac{\mathcal{Z}_{\rm inst}}{\mathcal{Z}_0}  = \frac{n (n+1)}{2}G_n\int \frac{d\boldsymbol{r}_0 d\lambda}{\lambda^3} (\pi g)^{n+1} e^{-\pi g(\mathcal{M}) + i \vartheta },\notag \\
    G_n  = \frac{2^{2n+3}}{\pi e^{2}}\frac{e^{-n-(n+1)\gamma}}{\Gamma(n+2)} .
    \label{eq:ZZ:inst:final}
\end{gather}
Here we introduce one-loop renormalized spin conductance in Pauli-Villars regularization scheme (see Appendix \ref{App:2}),
\begin{gather}
g(\mathcal{M})= g - \beta_0 \ln \mathcal{M}e^{\gamma-1/2}, \qquad \beta_0 = \frac{n+1}{\pi} .\label{eq:g:pert:PV}   
\end{gather}
In the replica limit, $n\to 0$, the correction $\mathcal{Z}_{\rm inst}/\mathcal{Z}_0$ is proportional to $n$.

The result \eqref{eq:ZZ:inst:final} is derived within the Gaussian theory for the fluctuations around the instanton solutions. Therefore, $g$ in the prefactor of the exponent under the intergral sign in Eq. \eqref{eq:ZZ:inst:final} remains unrenormalized. \color{black} It is natural to expect \cite{Morris1985a,Morris1985b,Morris1986a,Morris1986b} \color{black} that treatment of fluctuations beyond the Gaussian approximation will result in substitution of $g$ by $g(\mathcal{M})$ in the preexponential factor also. However, such a calculation is beyond the scope of the present paper.

\color{black} In order the result \eqref{eq:ZZ:inst:final} for $\mathcal{Z}_{\rm inst}/\mathcal{Z}_0$ becomes operative, one has to relate the Pauli-Villars mass $\mathcal{M}$ with the instanton size $\lambda$. Fortunately, the form of Eq. \eqref{eq:g:pert:PV} suggests the form of such a relation. A key observation is that the correction to the conductance in the flat space due to elimination of fast fluctuations with spatial scales between the ultra-violet length scale $\ell$ and the running scale $\lambda$ has exactly the same form as Eq. \eqref{eq:g:pert:PV} but with $\lambda/\ell$ under the logarithm instead of $\mathcal{M}$. Therefore, one concludes that $\mathcal{M}= \zeta\lambda/\ell$ where $\zeta$ is some constant. Within a particular scheme, this constant can be fixed to the magnitude $\zeta=e/2$ (see Appendix \ref{AppF}). Therefore, the correction $\mathcal{Z}_{\rm inst}/\mathcal{Z}_0$ depends on a particular value of $\zeta$, i.e. on a method of transformation from the theory of fluctuations on the sphere and in the flat space. This known to occur in the class A also \cite{Pruisken2005}.      
\color{black}

\section{Instanton corrections for pure scaling eigenoperators\label{sec: RGop}}

\color{black}
As was outlined in Introduction, there is an infinite set of local operators demonstrating pure scaling behavior at the {\sqHe} criticality. In this section, we will compute instanton corrections to the anomalous dimensions of  gradientless local operators $\mathcal{K}_{\bm{\lambda}}[\mathcal{Q}]$. Each such operator is an eigenoperator with respect to the renormalization group and, consequently, the corresponding physical observable demonstrates a pure scaling behavior at criticality characterized by a critical exponent $x_{\bm{\lambda}}$, 
$\mathcal{K}_{\bm{\lambda}}[\mathcal{Q}]\sim L^{-x_{\bm{\lambda}}}$ where $L$ is the system size.
The operators $\mathcal{K}_{\bm{\lambda}}[\mathcal{Q}]$ can be enumerated by a tuple $\bm{\lambda}=(\lambda_1,\dots,\lambda_s)$ of integer numbers, $\lambda_1\geqslant\lambda_2\geqslant\dots\lambda_s>0$, which are the highest weight of corresponding irreducible representation of $\text{Sp}(2n)/\text{U}(n)$ \cite{Wegner1986}. Each operator $\mathcal{K}_{\bm{\lambda}}[\mathcal{Q}]$ involves $|\bm{\lambda}|=\lambda_1+\dots+\lambda_s$ matrices $\mathcal{Q}$.
The simplest example of such pure scaling observables is the disorder-averaged moments of the {\DOS}, $\langle \nu^q\rangle$. They correspond to the operators $\mathcal{K}_{\bm{\lambda}}[\mathcal{Q}]$ with $\bm{\lambda}=\textsf{(q)}$ (see Ref. \cite{Gruzberg2013} for details).

\color{black}

An operator $\mathcal{O}$ averaged over the {\NLSM} can be written as a sum over topological sectors,
\begin{equation}\label{eq: instav1}
     \left\langle \mathcal{O}\right\rangle \approx  \left\langle \mathcal{O}\right\rangle_0 \left(1-\frac{\mathcal{Z}_{\rm inst}+\mathcal{Z}_{\rm inst}^*}{\mathcal{Z}_0}  \right) + \left\langle \mathcal{O}\right\rangle_{+1}+ \left\langle \mathcal{O}\right\rangle_{-1}+\dots
\end{equation}
For computation of averages at non-trivial topological sectors, we employ the saddle-point approximation near the instanton solution, taking into account the Gaussian fluctuations in the action only. \color{black} As above, we restrict our considerations by the contribution from the topological sector with $\mathcal{C}=\pm 1$. One has to take into account that in fact there are many instanton solutions with a given topological charge parametrized by the zero-mode manifold of $\lambda$, $z_0$, and $\TT$. Therefore, one has to sum up contributions to $\left\langle \mathcal{O}\right\rangle_{\pm1}$ from all such instanton solutions (for more detailed discussion see Ref. \cite{Pruisken2005}). The weight of each contribution is fixed by the expression \eqref{eq:ZZ:inst:final} for $\mathcal{Z}_{\rm inst}/\mathcal{Z}_0$. 
\color{black}
Then, we find
\begin{gather}
\left\langle \mathcal{O}\right\rangle_{\pm 1} \simeq
\frac{n (n{+}1)}{2}G_n\int \frac{d\boldsymbol{r}_0 d\lambda}{\lambda^3}
\left\langle
\mathcal{O}\right\rangle_{\TT}
\bigl (\pi g\bigr )^{n+1} 
\notag \\
\times e^{-\pi g(\mathcal{M}) \pm i \vartheta } ,
\end{gather}
where we introduce
{\color{black}\begin{equation}
  \left\langle \mathcal{O}\right\rangle_{\TT} = \frac{ \int \mathcal{D}[\TT]  \mathcal{O}[\mathcal{Q}_{\rm inst}]}{\text{vol}\TT}, 
\end{equation}
We remind that here $\TT \in \text{U}(1)\cup \text{U}(n)/[\text{U}(1){\times} \text{U}(n{-}1)]$.
 
As it was shown in Ref. 
\cite{pruisken2005instanton}, for computation of anomalous dimensions of pure-scaling operators without derivatives,  
it is enough to restrict instanton zero-modes manifold to only such rotational zero modes, 
which commute with $\underline{\Lambda}$, 
i.e. to the diagonal blocks 
in Fig. \ref{fig:Symbreak}. They are precisely $\TT$ rotations.} 
 We took explicitly into account that the operator $\mathcal{O}$ evaluated on the instanton solution depends, in general, on the unitary rotations $\TT$. We note that below we will work with operators expressed in terms of the original $\mathcal{Q}$ matrices.

\subsection{{\DOS}}

\color{black}
The disorder-averaged {\DOS} $\langle \nu \rangle$ corresponds to the pure scaling operator $\mathcal{K}_{\textsf{(1)}}[\mathcal{Q}]$ which involves a single $\mathcal{Q}$-matrix. It can be written explicitly as \cite{Babkin2022}
\color{black}
\begin{equation}
      \mathcal{K}_{\textsf{(1)}}\left[\mathcal{Q}\right]   =  \frac{1}{4} 
     \Bigl [  \text{tr}\mathcal{Q}_{RR}^{\alpha \alpha}-\text{tr}\mathcal{Q}_{AA}^{\alpha \alpha}\Bigr ]
      ,
\end{equation}
where $\text{tr}$ is trace over spin space and $\alpha$ is a fixed replica index. We note that the average {\DOS}  
\color{black} is determined as 
$\langle \nu\rangle = \nu_0 \langle \mathcal{K}_{\textsf{(1)}}\left[\mathcal{Q}\right]  \rangle$ where $\nu_0$ is the bare value of the {\DOS}. As well-known \cite{Evers2008} the disorder-averaged {\DOS} depends on the energy as a power-law at the {\sqHe} criticality as the system size tends to infinity, $L\to 0$. To simplify calculations we will study the scaling of  $\langle \nu\rangle$ at zero energy with the system size $L$. One can relate the scaling with the energy and the system size by standard means, comparing $L$ with divergent correlation length depending on the energy. 

The lowest order perturbative treatment of the {\DOS} results in the following expression: 
\begin{equation}
\langle \nu\rangle=\nu_{\rm pert}(L) = \nu_0 \left(1 +\frac{\gamma^{(0)}_{\textsf{(1)}}}{g}\ln \frac{L}{\ell} \right), \quad \gamma^{(0)}_{\textsf{(1)}} = -\frac{1+n}{\pi} .
\label{eq:gamma1:0}
\end{equation}
\color{black}

It is convenient to use parametrization of the
instanton solution in terms of deviation from a ``metalic'' saddle-point $\Lambda$: 
 \begin{multline}\label{eq: qparam}
    \mathcal{Q}_{\rm{inst}} = \Lambda + U^{-1} \TT^{ -1} \rho \TT U, \\ \TT \in \frac{\text{U}(n)}{\text{U}(1){\times} \text{U}(n{-}1)} \cup  \text{U}(1) .
\end{multline}
Here, we remind, the rotation matrix $U$ is defined in Eq.~\eqref{eq: utran}. 
The matrix $\rho$ has only four non-zero matrix elements:
\begin{equation}
\rho_{00}^{11}  =-\rho_{-1-1}^{11}=-2 e_0^2, \quad
\rho_{0-1}^{11} =({\rho}_{-10}^{11})^*=2 e_0 e_1 .
\end{equation}
We write down expansion for the operator $\mathcal{K}_{\textsf{(1)}}$ computed on the instanton solution \eqref{eq: qparam} and averaged over $\TT$-rotations:
\begin{align}
\left\langle
\mathcal{K}_{\textsf{(1)}}[\mathcal{Q}_{\rm inst}]
\right\rangle_{\TT}
& =  \frac{1}{4}
\text{tr}\Bigl [\Lambda_{RR}^{\alpha \alpha}-\Lambda_{A A}^{\alpha \alpha}\Bigr ]
\notag
\\+  \frac{1}{4}\sum_{p_1=\pm}p_1\!&\!\left\langle\text{tr}\left[U^{-1}\TT^{-1}\rho \TT U\right]^{\alpha \alpha}_{p_1 p_1}\right\rangle_{\TT} .
\label{eq:K1:TT:va}
\end{align}
\color{black} We note that the matrix $\TT$ acts as a $2n\times 2n$ block-diagonal matrix in the spin space, see Fig. \ref{fig:Symbreak}. Therefore, we can write $\TT=\TT_+ (1+\textsf{s}_3)/2+\TT_-(1-\textsf{s}_3)/2$, where $\TT_+$ is $n\times n$ matrix belonging to $\TT \in \text{U}(1)\cup \text{U}(n)/[\text{U}(1){\times} \text{U}(n{-}1)]$, while $\TT_-=(\TT^{ -1}_+)^T$.

\color{black}
{\color{black} For averaging of the second line in Eq. \eqref{eq:K1:TT:va} over  
$\TT$-rotations
we use the following relations (see Refs. \cite{mello1990averages} and \cite{pruisken2007theta}):
\begin{equation}
\left\langle\left(\TT^{-1}\right)^{\alpha 1}_+ \TT^{ 1 b}_+\right\rangle_{\TT} = \frac{\delta^{\alpha \beta}}{n}  ,
\end{equation}
\begin{equation}\label{unav}
\left\langle\left(\TT^{-1}\right)^{\alpha 1}_+ \TT^{ 1 \beta}_+\left(\TT^{ -1}\right)^{\gamma 1}_+ \TT^{ 1 \delta}_+\right\rangle_{\TT}  = \frac{\left[\delta^{\beta \gamma} \delta^{\alpha\delta} +\delta^{\alpha\beta}  \delta^{\gamma\delta} \right] }{n+n^2} .
\end{equation}
}After averaging, we should subtract the similar contribution from the trivial topological sector proportional to the normalized partition function. Then we obtain
\begin{multline}
\left\langle
\mathcal{K}_{\textsf{(1)}}[\mathcal{Q}_{\rm inst}]
\right\rangle_{\TT} =   -  \frac{(n+1)}{2} G_n\int \frac{d\lambda}{\lambda} \bigl (\pi g\bigr )^{n+1} \\ \times e^{-\pi g(\mathcal{M})+i\vartheta}   \int d\boldsymbol{r}_0\frac{\mu(r_0)}{\lambda} .
\label{eq: K1av}
\end{multline}
{\color{black} Here $\mu(r_0)$ is the measure, induced by instanton, see Eq.~\eqref{eq:mudef}}. The term in the second line of Eq. \eqref{eq: K1av} has an ultra-violet divergence \color{black} due to integration over the instanton position $\bm{r_0}$.  To treat this divergence we are forced to take into account the Gaussian fluctuations around the instanton in the pre-exponential factor. After performing straightforward calculations (see Appendix \ref{App:5}), we find the following expression for the instanton correction to the disorder-averaged {\DOS}, 
\begin{gather}
    \delta \nu_{\rm inst} {=} \nu_0 \left(\left\langle \mathcal{K}_{\textsf{(1)}}[\mathcal{Q}_{\rm inst}]\right\rangle_{\TT}{+} \left\langle \mathcal{K}_{\textsf{(1)}}[\mathcal{Q}^*_{\rm inst}]\right\rangle_{\TT}\right) {\simeq} \pi \gamma^{(0)}_{\textsf{(1)}}G_n  \notag \\
    \times \int \frac{d\lambda}{\lambda} \bigl (\pi g\bigr )^{n+1}  e^{-\pi g(\mathcal{M})}\cos\vartheta  \int d\boldsymbol{r}_0\frac{\mu(r_0)}{\lambda} 
    \notag \\
    \times 
    \nu_0 \left (1 + \frac{\gamma^{(0)}_{\textsf{(1)}}}{g} \ln \mathcal{M}\right ) .
    \label{eq:deltaNu:inst:00}
\end{gather}
We note that the last line of Eq. \eqref{eq:deltaNu:inst:00} coincides with the perturbative renormalization of the {\DOS} in the flat space, cf. Eq. \eqref{eq:gamma1:0}. The Pauli-Villars masses in Eq. \eqref{eq:deltaNu:inst:00} can be translated into the expression for the flat space by 
\color{black}
 a trick with a spatial varying mass method \cite{pruisken1995cracking}. 
  \color{black} Physical idea behind this method is that the instanton solution centered at the spatial point $r_0$ acts for the Gaussian fluctuations of $\mathcal{Q}(\bm{r}=0)$ as a slow-varying background field with a spatial scale $1/\mu(r_0)$. Taking this into account, we write
  \color{black}
\begin{gather}
    \delta \nu_{\rm inst} {=} \nu_0 \left(\left\langle \mathcal{K}_{\textsf{(1)}}[\mathcal{Q}_{\rm inst}]\right\rangle_{\TT}{+} \left\langle \mathcal{K}_{\textsf{(1)}}[\mathcal{Q}^*_{\rm inst}]\right\rangle_{\TT}\right) {\simeq} \pi \gamma^{(0)}_{\textsf{(1)}}G_n  \notag \\
    \times \int \frac{d\lambda}{\lambda}  \bigl (\pi g\bigr )^{n+1}  e^{-\pi g(\zeta \lambda)}\cos\vartheta  \int d\boldsymbol{r}_0\frac{\mu(r_0)}{\lambda} \nu_{\rm pert}(1/\mu(r_0)).
    \label{eq:deltaNu:inst:0}
\end{gather}
\color{black}
Here we also substituted the Pauli-Villars mass $\mathcal{M}$ by $\zeta \lambda/\ell$ in the argument of $g$ in the exponent. To be precise, we define (cf. Eq. \eqref{eq:g:pert:PV})
\begin{equation}
 g(\lambda)= g - \beta_0 \ln(e^{\gamma-1/2} \lambda/\ell) . 
 \label{eq:beta:00}
\end{equation}
\color{black}

\begin{figure}
    \centering
    \includegraphics[width=1.01\linewidth]{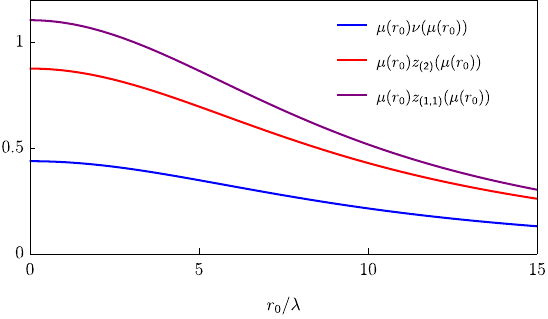}
    \caption{Schematic plot of $r_0$ integrands in Eqs. \eqref{eq:deltaNu:inst:0} and \eqref{eq: Kaver}}
    \label{fig:RGflow:r0}
\end{figure}

The decay of $\nu_{\rm pert}(1/\mu)$ at small $\mu$ (at long length scales) makes the integral over $\bm{r}_0$ \color{black} in the last line of Eq. \eqref{eq:deltaNu:inst:0} \color{black} convergent (see Fig. \ref{fig:RGflow:r0}). Performing integration over $\bm{r}_0$ in Eq. \eqref{eq:deltaNu:inst:0} (see Ref. \cite{Pruisken2005} and Appendix~\ref{AppF}), we find 
\begin{equation}
     \delta \nu_{\rm inst}
     {=}
     G_n \int\limits \frac{d \lambda}{\lambda}(\pi g)^{n{+}1}\nu(\zeta \lambda)\mathcal{H}_{\textsf{(1)}}(g(\zeta \lambda))e^{-\pi g(\zeta \lambda)}\cos \vartheta ,
\label{eq:deltaNu:inst:1}
\end{equation}
\color{black}
where $\zeta=e/2$, and we introduced 
\begin{equation}
     \nu(\lambda)  
= \nu_0 \Bigl [ 1+\frac{\gamma_{\textsf{(1)}}^{(0)}}{g}\ln(e^{\gamma-1/2} \lambda/\ell) \Bigr ] .
\label{eq:LDOS:pert:Lambda}
\end{equation}
Also we defined the function 
\begin{equation}
    \mathcal{H}_{\textsf{(1)}}(g) {=} \frac{2 \pi^2 g \gamma^{(0)}_{\textsf{(1)}} }{\beta_0{-}\gamma^{(0)}_{\textsf{(1)}}} \equiv \pi^2 g .
    \label{eq:Hlambda:def:00}
\end{equation}
\color{black}
\color{black}
It is worthwhile to mention that the function $\mathcal{H}_{\textsf{(1)}}$ is proportional to $g$. This fact can be understood in a following way. The integral over $r_0$ in Eq. \eqref{eq: K1av} diverges logarithmically  at large $r_0$. However, the one-loop RG Eq. \eqref{eq:beta:00} implies a dynamically generated localization length $\sim \exp(g/\beta_0)$. This length serves as a natural infra-red cut off for the overwise logarithmically divergent integral in Eq. \eqref{eq: K1av} thus resulting in a finite contribution proportional to $g$.

\color{black}
Combining together the perturbative contribution \eqref{eq:LDOS:pert:Lambda} and instanton correction \eqref{eq:deltaNu:inst:1}, we find
\begin{gather}
 \frac{\langle \nu\rangle}{\nu_0} = 1 + \int^{\zeta L}_{\zeta \ell} \frac{d\lambda}{\lambda}  \gamma_{\textsf{(1)}}(g(\lambda),\vartheta) .   
\end{gather}
Here  the function 
\begin{equation}
    \gamma_{\textsf{(1)}}(g,\vartheta) = \frac{\gamma_{\textsf{(1)}}^{(0)}}{g} + G_n (\pi g )^{n+1}\mathcal{H}_{\textsf{(1)}}(g)e^{-\pi g}\cos\vartheta 
\end{equation}
can be interpreted as 
\color{black} 
the anomalous dimension of the disorder-averaged {\DOS}, which determines its scaling with the system size $L$, 
\color{black}
\begin{equation}
   \frac{d\ln \langle\nu\rangle}{d\ln L} =  \gamma_{\textsf{(1)}}(g,\vartheta) .
\end{equation}
\color{black}
Taking into account explicit expressions for  $\gamma^{(0)}_{\textsf{(1)}}$ and $\beta_0$, see Eqs. \eqref{eq:gamma1:0} and \eqref{eq:beta:00}, we find in the replica limit $n\to 0$,
\begin{equation}
    \gamma_{\textsf{(1)}}(g,\vartheta) = -\frac{1}{\pi g} - \pi G_0 (\pi g )^{2}e^{-\pi g}\cos\vartheta  .
\label{eq:RG:LDOS:non-pert}
\end{equation}

\subsection{Operators with two $\mathcal{Q}$-matrices}

\color{black}
There exist only two eigenoperators for $|\bm{\lambda}|=2$. They involve two $\mathcal{Q}$-matrices and correspond to two tuples: $\bm{\lambda}=\textsf{(2)}$ and $\bm{\lambda}=\textsf{(1,1)}$. While the former corresponds to the second moment of the {\DOS}, the later describes more involved correlations of four wave functions of a random Hamiltonian of symmetry class C \cite{Karcher2021}. These two eigen operators can be constructed explicitly as \cite{Babkin2022}
\color{black}
\begin{equation}\label{eq: K2def}
    \mathcal{K}_{\bm{\lambda}}[\mathcal{Q}]=\frac{1}{16} \sum_{p_1, p_2=R/A} (\sigma_3)_{p_1p_1}(\sigma_3)_{p_2p_2} \mathcal{P}_{\bm{\lambda}}^{\alpha_1 \alpha_2; p_1 p_2} .
\end{equation}
Here the correlation function $\mathcal{P}_{\bm{\lambda}}^{\alpha_1 \alpha_2 ; p_1 p_2}$ is defined as
\begin{gather}
\mathcal{P}_{\bm{\lambda}}^{\alpha_1 \alpha_2; p_1 p_2}=  \operatorname{tr} \mathcal{Q}_{p_1 p_1}^{\alpha_1 \alpha_1}(\boldsymbol{x}) \operatorname{tr} \mathcal{Q}_{p_2 p_2}^{\alpha_2 \alpha_2}(\boldsymbol{x})
\notag  \\ + \mu_{\bm{\lambda}}\operatorname{tr}\left[\mathcal{Q}_{p_1 p_2}^{\alpha_1 \alpha_2}(\boldsymbol{x}) \mathcal{Q}_{p_2 p_1}^{\alpha_2 \alpha_1}(\boldsymbol{x})\right] .
%,\quad \mu_2 =\{2,-1\}
\label{eq: P2def}
\end{gather}
The coefficients 
$\mu_{\bm{\lambda}}$ \color{black} can be fixed by the condition that the operator $\mathcal{P}_{\bm{\lambda}}^{\alpha_1 \alpha_2; p_1 p_2}$ is the eigen operator with respect to the renormalization group transformation \cite{Babkin2022}. In particular, they \color{black}
are given as $\mu_{\textsf{(2)}}=-1$ and $\mu_{\textsf{(1,1)}}=2$. 

\color{black}
In order to study perturbative and non-perturbative renormalization of the pure scaling operators $\mathcal{K}_{\bm{\lambda}}[\mathcal{Q}]$ it is convenient to introduce the physical observable 
$z_{\bm{\lambda}} = z_{\bm{\lambda}}^{(0)} \left\langle K_{\bm{\lambda}}[\mathcal{Q}]\right\rangle$ with the $z_{\bm{\lambda}}^{(0)}=1$. The lowest-order perturbative renormalization of the operators yields \color{black} 
\begin{equation}
  z_{\bm{\lambda}}(L) = z_{\bm{\lambda}}^{(0)} \Bigl(1 + \frac{\gamma^{(0)}_{\bm{\lambda}}}{ g}\ln \frac{L}{\ell}\Bigr ) , 
  \label{eq:zL:LL}
\end{equation}
where the one-loop coefficients are given as \cite{Karcher2021}
\begin{equation}
    \gamma^{(0)}_{\textsf{(2)}} = -\frac{1+2n}{\pi},
    \quad 
    \gamma^{(0)}_{\textsf{(1,1)}} =
    -\frac{4+2n}{\pi} .
\end{equation}

\color{black} We repeat exactly the same steps as for computation of the instanton corrections to {\DOS} in the previous section.  \color{black} After substitution of parametrization \eqref{eq: qparam} into Eq. \eqref{eq: P2def} and with the help of Eq. \eqref{unav}, we obtain 
\begin{gather}
    \delta z_{\bm{\lambda}}^{\rm (inst)} {=} z_{\bm{\lambda}}^{(0)} \left(\left\langle K_{\bm{\lambda}}[\mathcal{Q}_{\rm inst}]\right\rangle_{\TT}{+}\left\langle K_{\bm{\lambda}}[\mathcal{Q}^*_{\rm inst}]\right\rangle_{\TT}\right)  {\simeq} \pi\gamma^{(0)}_{\bm{\lambda}} G_n
    \notag \\
    \times \int \frac{d\lambda}{\lambda} (\pi g)^{n+1}  e^{-\pi g(\zeta \lambda)}\cos\vartheta \int d\boldsymbol{r}_0\frac{\mu(r_0)}{\lambda} z_{\bm{\lambda}}\bigl (1/\mu(r_0)\bigr ).
    \label{eq: Kaver}
\end{gather}
We note %emphasize 
that under the integral over $\bm{r_0}$ in Eq. \eqref{eq: Kaver} we omitted terms 
which do not diverge in the ultra violet, i.e. at $r_0\to \infty$. 
\color{black} Surprisingly, the structure of Eq. \eqref{eq: Kaver} repeats exactly the structure of instanton correction to the {\DOS}. The specifics of the eigen operator is hidden into the one-loop coefficient $\gamma^{(0)}_{\bm{\lambda}}$ which appears in two places: as an overall factor and in the expression for $z_{\bm{\lambda}}\bigl (1/\mu(r_0)\bigr )$. 

After integration over $\bm{r_0}$, we find
\begin{equation}
\delta z_{\bm{\lambda}}^{\rm (inst)} 
 {=}
     G_n\! \int\! \frac{d \lambda}{\lambda}(\pi g)^{n{+}1}z_{\bm{\lambda}}(\zeta \lambda)\mathcal{H}_{\bm{\lambda}}(g(\zeta \lambda))e^{-\pi g(\zeta \lambda)}\cos \vartheta ,   
     \label{eq:zL:Inst} 
\end{equation}
where the function
\begin{equation}
  \mathcal{H}_{\bm{\lambda}}(g) = \frac{2 \pi^2 g \gamma^{(0)}_{\bm{\lambda}} }{\beta_0{+}|\gamma^{(0)}_{\bm{\lambda}}|} .
  \label{eq:Hlambda:def:0}
\end{equation}
We note that in the derivation of expression \eqref{eq:Hlambda:def:0}  (see Appendix~\ref{AppF}) the negative sign of $\gamma^{(0)}_{\bm{\lambda}}$ has been important. It is this negative sign results in decay of the integrand at $r_0\to\infty$, see Fig. \ref{fig:RGflow:r0}. However, one can extend the expression to the positive $\gamma^{(0)}_{\bm{\lambda}}$ as well \cite{Pruisken2005}. The expression \eqref{eq:Hlambda:def:0} holds for both cases.      
\color{black}

Repeating the same steps as in the previous section for the {\DOS}, using Eqs. \eqref{eq:zL:Inst} and \eqref{eq:zL:LL}, we find the instanton correction to the anomalous dimension of the operators $K_{\textsf{(2)}}$ and $K_{\textsf{(1,1)}}$ as 
\begin{gather}
    \gamma_{\bm{\lambda}} = 
    \frac{d\ln z_{\bm{\lambda}}}{d\ln L} = \frac{\gamma^{(0)}_{\bm{\lambda}}}{g} + \frac{2\pi \gamma^{(0)}_{\bm{\lambda}}}{\beta_0+|\gamma^{(0)}_{\bm{\lambda}}|} G_n (\pi g)^{n+2} e^{-\pi g} \cos\vartheta .
    \label{eq:anom:dim:inst:full}
\end{gather}

\subsection{Operators with an arbitrary number of $\mathcal{Q}$-matrices}\label{subsec: arbQand}

As one can check, Eq. \eqref{eq: Kaver} is valid for an arbitrary pure scaling operator corresponding to the tuple $\bm{\lambda}$ (see Appendix \ref{App:3}). Therefore, the result \eqref{eq:anom:dim:inst:full} holds also for an arbitrary eigen operator. In the replica limit, $n\to 0$, the perturbative coefficient $\gamma^{(0)}_{\bm{\lambda}}$ for $\bm{\lambda}=(\lambda_1,\dots, \lambda_s)$ is given as \cite{Karcher2021}
\begin{equation}
\gamma^{(0)}_{\bm{\lambda}} = \frac{1}{2\pi} \sum_{j=1}^{s} \lambda_j (\lambda_j + c_j), \quad c_j=1-4j .    
\end{equation}
Then, we find the following result for the anomalous dimension of the pure scaling operator in the replica limit, $n\to 0$,
\begin{equation}
\gamma_{\bm{\lambda}} = \frac{\bm{\lambda}(\bm{\lambda}+\bm{c})}{2\pi g} + \frac{\bm{\lambda}(\bm{\lambda}+\bm{c})}{2+|\bm{\lambda}(\bm{\lambda}+\bm{c})|} \mathcal{C}_C \left(\pi g\right)^2 e^{-\pi g} \cos\vartheta,    
\label{eq:anom:dim:inst:full:0}
\end{equation}
where $\mathcal{C}_{C} = 16 e^{-2-\gamma}$ and we introduce the vector $\bm{c}=(c_1,\dots,c_s)$. \color{black} We note that the result \eqref{eq:anom:dim:inst:full} holds also for the eigen operator $\mathcal{K}_{\textsf{(1)}}$ corresponding to the {\DOS}.
\color{black}
We emphasize that the instanton correction is expressed via the quadratic Casimir operator $\bm{\lambda}(\bm{\lambda}+\bm{c})$ similar to the one-loop perturbative correction. 
Therefore, the result \eqref{eq:anom:dim:inst:full:0} remains invariant under symmetry transformations which are consequence of Weyl-group invariance.

We note that the instanton contribution to the anomalous dimensions $\gamma_{\bm{\lambda}}$ at $\vartheta=\pi$ is of opposite sign with respect to the one-loop perturbative correction. It implies that the instanton effects reduce the multifractal behavior at $\vartheta=\pi$.

\section{Corrections to the spin conductivities\label{Sec:SpinCond}}

Our next aim is to compute instanton corrections to conductivities with the help of {\NLSM} formalism. 
The Kubo-type expressions for longitudinal spin conductivity can be written as follows (see Appendix \ref{App:4})
\begin{gather}
g^\prime = g + \frac{g}{8 n(n+1)}\left\langle \frac{1}{2} \operatorname{Tr} [\Lambda, \mathcal{Q}]^2 +\left(\operatorname{Tr} \Lambda \mathcal{Q}\right)^2 -(\operatorname{Tr}1)^2\right\rangle \notag \\
- \frac{g^2}{64 n(n+1)} \int d\bm{x}^\prime \Bigl\langle \Tr [\Lambda, \boldsymbol{\mathcal{J}}(\bm{x})] [\Lambda, \boldsymbol{\mathcal{J}}(\bm{x}^\prime)]\Bigr\rangle .
\label{eq:spin:g:K}
\end{gather}
Here we introduce the matrix current $\boldsymbol{\mathcal{J}}=\mathcal{Q}\nabla \mathcal{Q}$. We note that Eq. \eqref{eq:spin:g:K} produces correct perturbative renormalization for $g$.

Similar expression can be derived for the transverse spin conductivity (see Appendix \ref{App:4}),
\begin{gather}
   g_H^\prime {=} \color{black} g_H \color{black} {+}\frac{g^2}{8n(n{+}1)}\int\limits_{\boldsymbol{x}^{\prime}}\varepsilon_{\mu\nu}\Bigl\langle \operatorname{Tr}\left[\Lambda_{-}\mathcal{J}_{\mu}(\bm{x}) \Lambda_{+} \mathcal{J}_{\nu}(\bm{x}^\prime)  \right]\Bigr\rangle ,
   \label{eq:spin:gH:K}
\end{gather}
where $\Lambda_{\pm} = (1 \pm \Lambda)/2$ stands for  the projector on the retarded and advanced blocks.

\subsection{Longitudinal spin conductivity}

We start from calculation of the instanton corrections to $g^\prime$. Here we will proceed in a similar way as in Sec.~\ref{sec: RGop}. We use the approximation \eqref{eq: instav1} in our calculations below. An important remark is in order here. \color{black} The longitudinal spin conductivity involves two operators, cf. Eq.~\eqref{eq:spin:g:K}. These operators individually are not the eigenoperators of the renormalization group. This situation is similar to the eigen operator $\mathcal{K}_{\textsf{(2)}}$, cf. Eq.~\eqref{eq: P2def}, which is composed from two operators, each of which is not eigenoperator under the action of the renormalization group. 
\color{black}
Nevertheless, the full operator of the spin conductivity is the RG eigenoperator. In other words, within the background field renormalization method, the renormalized conductivity can be written as
\begin{multline}
  \delta g\left[\mathcal{Q}\right]  \rightarrow Z_g \delta g\left[\mathcal{Q}_0\right] = Z_g\left(g_{dm}\left[\mathcal{Q}_0\right]+ g_{j-j}\left[\mathcal{Q}_0\right]\right)  ,
\end{multline}
    where $\mathcal{Q}_0 = \mathcal{T}^{-1}_0 \Lambda  \mathcal{T}_0$ is a ``slow'' field and $Z_g$ is a renormalization factor for the conductivity. Here $g_{dm}$ corresponds to the operator in the first line of Eq~\eqref{eq:spin:g:K} (diamagnetic contribution) while $g_{j-j}$ is the operator in the second line of the same equation (current-current contribution).
    Therefore, our task is simplified and it is enough to compute the instanton contribution to $Z_g$ from the one of the operators. For reasons to be explained shortly, we compute the instanton contribution to the current-current part of the spin conductivity, $g_{j-j}$. It reads 
\begin{gather}
- \frac{g^2}{64 n(n+1)} \int d\bm{x}^\prime \left\langle \Tr [\Lambda, \boldsymbol{\mathcal{J}}(\bm{x})] [\Lambda, \boldsymbol{\mathcal{J}}(\bm{x}^\prime)]\right\rangle
\notag \\
\to     -G_n \int \frac{d\lambda}{\lambda}(\pi g)^{n+3}e^{-\pi g\left(\mathcal{M}\right)}\cos \vartheta .
\end{gather}
Hence we obtain the final expression to the single instanton correction to the disspative spin conductance
\begin{equation}
      \color{black}
     \delta g_{\rm inst}
    \color{black}
    = - G_n \int\frac{d\lambda}{\lambda}(\pi g)^{n+3}e^{-\pi g(\zeta\lambda)}\cos \vartheta ,
\end{equation}
where,
we remind, $g(\zeta \lambda)$ is the spin conductance renormalized %through 
within the perturbation theory, cf. Eq. \eqref{eq:beta:00}. 
\color{black} As in the previous section, \color{black} one can interpret the above correction in terms of the non-perturbative contribution to the beta-function (for the finite replica number $n$)
\begin{equation}\label{eq: betag}
   \beta_g(g,\vartheta) = -\frac{dg}{d\ln L} = \frac{1+n}{\pi} + G_n(\pi g)^{n+3}e^{-\pi g}\cos \vartheta .
\end{equation}

\subsection{Transverse spin conductivity}

Our next step is to compute the renormalization of the transverse spin conductivity. We note that there are no perturbative corrections to $g_{H}$. Using Eq.~\eqref{eq:spin:gH:K} and the approximation \eqref{eq: instav1}, we find
\begin{multline}
     \frac{g^2}{8n(n+1)}\left\langle\int_{\boldsymbol{x}^{\prime}}\varepsilon_{\mu\nu}\operatorname{Tr}\left[\Lambda_{-}\mathcal{J}_{\mu}(\bm{x}) \Lambda_{+} \mathcal{J}_{\nu}(\bm{x}^\prime)  \right]\right\rangle \\ \rightarrow -G_n \int \frac{d\lambda}{\lambda}(\pi g)^{n+3}e^{-\pi g\left(\mathcal{M}\right)}\sin \vartheta .
\end{multline}
With the help of Eq.~\eqref{eq: thedef}, we write the  
\color{black}
instanton correction to 
\color{black}
the theta-angle, cf. Eq. \eqref{eq: thedef},  as
\begin{equation} 
\color{black} 
\delta \vartheta_{\rm inst} = 
-\pi \color{black} G_n \int \frac{d\lambda}{\lambda}(\pi g)^{n+3}e^{-\pi g\left(\zeta \lambda\right)}\sin \vartheta .
\end{equation}
The corresponding beta-function takes the following form:
\begin{equation}\label{eq:beta:gH:n}
   \beta_{\vartheta}(g,\vartheta)=-\frac{d(\vartheta/2\pi)}{d\ln L} = \frac{G_n}{2}(\pi g)^{n+3}e^{-\pi g}\sin \vartheta .
\end{equation}

\subsection{The renormalization group equations for the spin conductivities}

\begin{figure}[t]
    \centering
    \includegraphics[width=0.8\linewidth]{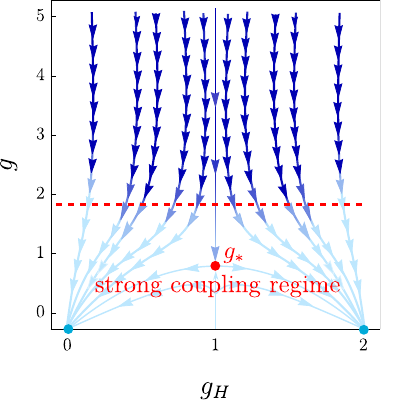}
    \caption{Sketch of the RG-flow diagram for spin conductivities in class C, see Eqs. \eqref{eq: RGeq}}
    \label{fig:RGflow}
\end{figure}

In the replica limit, $n\to 0$, using Eqs. \eqref{eq: betag} and \eqref{eq:beta:gH:n},  the RG equations for the spin conductivities acquire the following form:
\begin{equation}
\begin{split}
    \frac{dg}{d\ln L} & = -\frac{1}{\pi}-\frac{2}{\pi^2 g}-\mathcal{D}_{\text{C}}(\pi g)^3 e^{-\pi g}\cos \vartheta,  \\
    \frac{d\vartheta}{d\ln L} & = -\pi\mathcal{D}_{\text{C}}(\pi g)^3 e^{-\pi g}\sin \vartheta ,
    \end{split}
    \label{eq: RGeq}
\end{equation}
where $\mathcal{D}_{\text{C}}=8e^{-2-\gamma}/\pi\color{black}\approx 0.2$. Here we added the two-loop perturbative correction to the longitudinal spin conductivity~\cite{Hikami1981}. \color{black} Equations \eqref{eq: RGeq} describe the flow of the longitudinal conductivity $g$ and the fractional part of the transverse spin conductivity $\vartheta$, cf. Eq. \eqref{eq: thedef}, with increase of the system size $L$. 
\color{black}

Although the RG Eqs. \eqref{eq: RGeq} have been derived in the weak coupling regime, $g\gg1$, they help to understand the overall phase diagram of the class C and quantization of the transverse spin conductance (see Fig.~\ref{fig:RGflow}). Similar to the case of class A, the instantons provide the mechanism for the scale dependence of the theta-angle $\vartheta$, which is a fractional part of $g_H$. RG Eqs.~\eqref{eq: RGeq} predict that $\vartheta=0$ ($g_H=2k$) is the  stable fixed line of the RG flow, while $\vartheta=\pi$ ($g_H=2k+1$) is the unstable fixed line. It is the latter that corresponds to the transition between the spin quantum Hall phases with $g_H=2k$ where $k\in \mathbb{Z}$. Although the instanon correction to the renormalization of the longitudinal spin conductivity is of antilocalization character at $\vartheta=\pi$, its magnitude is not sufficient to compensate the perturbative localization corrections. This situation is not surprising since, as we have already mentioned above, RG Eqs. \eqref{eq: RGeq} are applicable at weak coupling, $g\gg 1$, only. \color{black} Based on numerical \cite{Evers1997} and analytical \cite{Cardy2000} studies of the critical spin conductance at the {\sqHe}   we know that there exists an unstable fixed point at $\vartheta=\pi$ and $g=\sqrt{3}/2$. \color{black} This fixed point describes the transition between different topological phases in class C (see Fig.~\ref{fig:RGflow}).    
\color{black}

\section{Discussions and conclusions\label{Sec:Final}}

Before closing the paper, we will discuss some important aspects of the obtained results.

\subsection{Instanton %configuration 
solution for unrotated $\mathcal{Q}-$matrix}

Although instanton configuration has been discussed for the $Q$-matrix, it is interesing to see how it looks like before the rotation. Performing the inverse transformation, cf. Eq. \eqref{eq: utran}, we obtain the instanton solution $\mathcal{Q}_{\text{inst}}$ in the following form (we set %$n=1$
\color{black} $N_r=1$ and $\bm{r_0}=0$)\color{black}
\begin{equation}\mathcal{Q}_{\text{inst}}=\left(\begin{array}{cc}\frac{r^2 \boldsymbol{s}_0-\lambda^2 \boldsymbol{s}_1}{r^2+\lambda^2} & \frac{r \lambda e^{i \theta}\left(i \boldsymbol{s}_2-\boldsymbol{s}_3\right)}{r^2+\lambda^2} \\-\frac{r \lambda e^{-i \theta}\left(i \boldsymbol{s}_2+\boldsymbol{s}_3\right)}{r^2+\lambda^2} & -\frac{r^2 \boldsymbol{s}_0+\lambda^2 \boldsymbol{s}_1}{r^2+\lambda^2}\end{array}\right)_{RA} . \label{eq:Qinst:unrot}
\end{equation}

It is crucial that the solution \eqref{eq:Qinst:unrot} has nontrivial structure in the spin space. In other words rotations in the spin and RA spaces are entangled. The naive construction of the instanton solution 
\color{black} for class C would be to place \color{black}
the class A instanton in the upper diagonal block of matrix $\mathcal{Q}_{\rm inst}$ and, then, construct the rest in a way consistent with the BdG symmetry. \color{black} However, such a procedure \color{black} results in the instanton solution with the topological charge 2. \color{black} Therefore, the structure \eqref{eq:Qinst:unrot} is not a trivial generalization of the class A instanton. 
\color{black}

\subsection{Manifestation of Weyl symmetry}

The anomalous dimensions $\gamma_{\bm{\lambda}}$, cf. Eq. \eqref{eq:anom:dim:inst:full:0}, determine the %scaling \color{blue} 
flow of the physical observables $z_{\bm{\lambda}}$ (corresponding to the 
\color{black}
eigen operators $\mathcal{K}_{\bm{\lambda}}$) with the system size $L$.  
At criticality, \color{black} the dependence becomes a power law,  \color{black}
\begin{equation}
\color{black} z_{\bm{\lambda}}
\color{black}
\sim L^{-x_{\bm{\lambda}}}, \quad x_{\bm{\lambda}}
= |\bm{\lambda}| x_{\textsf{(1)}}+\Delta_{\bm{\lambda}} ,
\label{eq:scaling:OP}
\end{equation}
{\color{black} where the scaling exponents $x_{\bm{\lambda}}$ are given by magnitudes of $-\gamma_{\bm{\lambda}}$ at the critical point}. \color{black} We note that anomalous dimension $\Delta_{\bm{\lambda}}$ describes the scaling of the pure scaling observables normalized to the proper power of the disorder-averaged {\DOS}, $z_{\bm{\lambda}}/\langle \nu\rangle^{|\bm{\lambda}|}$. 
\color{black}
\color{black}
Important outcome of our result \eqref{eq:anom:dim:inst:full:0} for the instanton conributions to the anomalous dimensions of the eigenoperators under the action of RG is that they preserve the Weyl-group symmetry relations. 
This symmery relates the anomalous dimensions of operators which can be obtained from each other by the following symmetry operations acting on the %vector 
tuple $\bm{\lambda}=(\lambda_1,\dots,\lambda_s)$: reflection $\lambda_j \rightarrow -c_j-\lambda_j$ and permutation of some pair: $\lambda_{j/i} \rightarrow \lambda_{i/j}+(c_{i/j}-c_{j/i})/2$. Since it is known \cite{Gruzberg2013} that the \color{black} existence of the \color{black} Weyl symmetry is not limited to the criticality, the Weyl symmetry of our nonperturbative result \eqref{eq:anom:dim:inst:full:0} provides a strong consistency check.   

We note also that the instanton correction, Eq. \eqref{eq:anom:dim:inst:full:0}, breaks the  generalized parabolicity (anomalous dimension is not  a linear function of the combination $\bm{\lambda}(\bm{\lambda}+\bm{c})$). Thus, instanton analysis signals breaking of the generalized parabolicity already in weak coupling regime. This fact is consistent with numerical data and analytical results from mapping to percolation at criticality that demonstrate clear evidence of a  
violation of generalized parabolicity for the multifractal exponents $x_{\bm{\lambda}}$ \cite{Mirlin2003,Puschmann2021,Karcher2021,Karcher2022,Karcher2022b,Karcher2023a,Karcher2023,Babkin2023}. 

\color{black}

\subsection{Comparison with the integer quantum Hall effect}

\color{black}
The instanton effect in the {\sqHe} discussed in this paper is a counterpart of the similar non-perturbative effect in the  {\iqHe} which has been studied with the help of the same technique -- NL$\sigma$M \cite{Pruisken2005}. Below, we will highlight the main distinctions between these two cases. 

Firstly, we note the different target manifolds of NL$\sigma$M. In the case of the {\iqHe}, the $Q$ matrices lie in the coset U(2$n$)/U($n$)$\times$U($n$), while, as mentioned above, the target manifold for class C has the form Sp(2$n$)/U($n$). The latter is a consequence of the presence of the additional BdG symmetry, cf.  Eq. \eqref{eq: constr1}. The difference in the target manifolds affects 
the mechanism of symmetry breaking due to the instanton (see Fig \ref{fig:Symbreak}). Due to the presence of an additional constraint on the diagonal blocks of the $Q$ matrix in class C, the volume of the zero mode manifold scales as $\sim n$ at $n\to 0$. Hence the instanton correction to the partition function turns out to be linear in the replica number, that guarantees the presence of a non-zero correction to the  
{\color{black} average logarithm of the partition function and to }
the disorder-averaged {\DOS}. In contrast, in class A, the volume of the zero mode manifold scales $\sim n^2$ in the replica limit, $n\to 0$ 
such that all corrections to {\color{black} the average logarithm of the partition function} and to   
the average {\DOS} (both perturbative and non-perturbative) vanish.

The second difference is the structure of the RG equations \eqref{eq: RGeq}. The non-perturbative corrections to the beta functions for the spin conductivities, \eqref{eq: betag} and \eqref{eq:beta:gH:n}, have a stronger power-law dependence on $g$ in the pre-exponential factor than it is in class A ($g^2$ versus $g^3$) \cite{pruisken1987quasiparticles,Pruisken2005}. 

This suggests that in a weak coupling regime, $g\gg 1$, the rate of change of the theta-angle in the class C is larger than that in the class A. Interestingly, the situation remains similar at criticallity. There the rate of RG flow of the theta-angle is controlled by inverse of the localization length exponent $\nu$. As known \cite{Kagolovsky1999,Gruzberg1999,Slevin2014}, the magnitude of $\nu$ for the {\iqHe} is almost two times large than in the case of the {\sqHe}.
Also we mention  that  the relative magnitude of the instanton correction with respect to the perturbative one is the same for both classes. The $g^3$ prefactor of the instanton correction in  the case of class C is compensated by the large perturbative contribution, $\sim g^0$ (in contrast with class A, where it is $\sim 1/g$ only).

\color{black}

\subsection{Future directions} 

Our results pave the way for a future research in the {\sqHe}. Firstly, it would be interesting to understand the instanton corrections in the {\NLSM} approach through the lens of the percolation mapping.

Secondly, it is known \cite{Subramaniam2006, mildenberger2007, Subramaniam2008, Evers2008,Obuse2008} that 
scaling with the system size of wave functions at the boundary of a system undergoing bulk Anderson transition is different from the corresponding scaling in the bulk. Recently, it was shown to be true for  generalized multifractality exponents \cite{Babkin2023}. Thus it would be interesting to understand how to treat instanton configurations in the presence of a boundary and to compute instanton corrections to the anomalous dimensions of pure scaling local operators at the boundary.

Thirdly, as known in the theory of {\iqHe} \cite{Pruisken2005,Pruisken2007}, the fluctuations of the topological term at the boundary corresponds to the edge theory of chiral spinless fermions. It would be interesting to study the edge theory that follows from the fluctuations of the topological term at the boundary for the {\sqHe}. Also, it would be challenging to relate the thus derived edge theory with microscopic theory of the {\sqHe}.  

Fourth, in the context of the {\iqHe} the instanon analysis has been extended to the Finkel'stein {\NLSM} that takes into account the electron-electron interaction \cite{Pruisken2007}. Recently, the generalized multifractality for the {\sqHe} has been extended to the interacting case \cite{Babkin2022}. It would be interesting to adapt the instanton analysis for the class C presented in this paper to include the electron-electron interaction.

Fifth, as known, by breaking the SU$(2)$ spin-rotational symmetry, the class C transforms into the class D which hosts the thermal quantum Hall effect in two dimensions \cite{Senthil2000}. The corresponding {\NLSM} looks similar to the {\NLSM} for class C and involves the topological theta-term. It would be interesting to develop the instanton analysis for the case of class D, in particular, since in that class there exists other topological objects -- domain walls -- describing jumps of the $\mathcal{Q}$-matrix between two disconnected pieces of the {\NLSM} target manifold~\cite{Bocquet2000,Chalker2001,Read2001,Gruzberg2005}.  

Sixth, class A can be obtained from the class C by breaking %BdG 
\color{black} SU$(2)$ symmetry down to U$(1)$. \color{black} It would be interesting to implement such a symmetry breaking into {\NLSM} and to study transformation of the instanton solution \eqref{eq:Qinst:unrot} into the instanton of class A.

\subsection{Summary}

In conclusion, we summarize all results which were discussed above. We developed the non-perturbative analysis of topological Anderson transition in the {\sqHe}. Using {\NLSM} for the class C we found instanton solution with non-trivial topological charge equal to $\pm 1$, \color{black} cf. Eqs. \eqref{eq:inst:gen} and \eqref{eq:Qinst:unrot}. \color{black} We identified all collective coordinates (zero modes) of the instanton and integrated over them exactly. We integrated over fluctuations around the instanton within the Gaussian approximation. Thus we derived the instanton correction to the {\color{black} logarithm of the partition function, cf. Eq. \eqref{eq:ZZ:inst:final}.} 
Remarkably, due to the structure of the {\NLSM} manifold in class C, Sp$(2N)$/U$(N)$, this correction survives in the replica limit (in contrast to the vanishing correction in class A). In addition, applying the same methodology, we computed the instanton corrections to the anomalous dimensions of all pure scaling local operators which determine the generalized multifractal spectrum in the {\sqHe}, \color{black} cf. Eq. \eqref{eq:anom:dim:inst:full:0}. \color{black} Interestingly, instanton corrections do not spoil the Weyl-group symmetry relations between anomalous dimensions \color{black} of different eigen operators\color{black}. We computed also instanton corrections to the longitudinal and Hall spin conductivities. Interpreting the derived results as corrections to the two parameter renormalization group equations, \color{black} cf. Eq. \eqref{eq: RGeq}, \color{black} we constructed the phase diagram for the {\sqHe} (Fig. \ref{fig:RGflow}). Finally, we listed several new directions which our work opens.

\begin{acknowledgements}
We thank A. Belavin, S. Bera, A. Gorsky, M. Yu. Lashkevich, A. Sedrakyan, and, especially, I. Gruzberg, for useful discussions. We are extremely grateful to Pavel Ostrovsky for the original idea of the non-unitary rotation \eqref{eq: utran} that allows us to construct the instanton solution \eqref{eq: inst1:0} in the form of $2\times 2$ matrix. I.S.B. is grateful to I. Gornyi, I. Gruzberg, and H. Obuse for collaboration on a related project.  The research was  supported by Russian Science Foundation (grant No. 22-42-04416). M.V.P. and I.S.B. acknowledge the hospitality during the 
``Nor-Amberd School in Theoretical Physics 2024'' where part of this work has been performed. M.V.P. and I.S.B. acknowledge personal support from the Foundation for the Advancement of Theoretical Physics and Mathematics ``BASIS''. 
\end{acknowledgements}

%%%%%%%%%%%%%%%%%%%%%%%%%%%%%%%%%%%%%%%%%%%%%%%%%%%%%%%%%%%%%%%%%%%%%%
%%%%%%%%%%%%%%%%%%%%%%%%%%%%%%%%%%%%%%%%%%%%%%%%%%%%%%%%%%%%%%%%%%%%%%
\appendix
    \section{Evaluation of {\DOS} and the spin conductivity in Pauli-Villars regularization \label{App:2}}
    
In this Appendix we present the calculation of {\DOS} and the spin conductivity in the trivial topological sector with the help of the  Pauli-Villars regularization scheme for class C. We benefit from similar calculations for class A \cite{Pruisken2005,Pruisken2007}. We find
\begin{align}
   \nu_{\rm pert}(\mathcal{M}) & = \frac{\nu}{\mathcal{K}_1(\Lambda)}\int \mathcal{D}Q \; \text{Tr}\underline{\Lambda} Q \;e^{-S[Q]} \notag \\
   & = \nu - \frac{\nu}{4n }\int \mathcal{D}W \text{Tr} W^2 e^{-S_0[W]}\notag \\ 
   & = \nu-\frac{\nu}{2n} \left\langle w^{\alpha \beta} w^{*\alpha\beta}\right\rangle_0 \notag \\
   & =
\nu \left(1- \frac{2(1+n)}{g}\mathcal{G}_0(\eta\theta;\eta\theta;0)\right).
\end{align}
We can rewrite the Green's function in coinciding points in terms of new function:
\begin{equation}
  \mathcal{G}_0(\eta\theta;\eta\theta;0) = \frac{1}{4 \pi}Y^{(0)}, \; Y^{(s)}=\sum_{J=J_{s}}^{\infty}\frac{2J+(1-s)^2}{E_{J}^{(s)}} ,
\end{equation}
where $J_{s}=2-(s-2)(s-1)/2$. Next 
we introduce the function
\begin{equation}
Y^{(\Lambda)}(p)=\sum_{J=p}^{\Lambda} \frac{2 J}{J^2-p^2} .
\end{equation}
Similar to \eqref{eq:res:Phi:L:0}, the regularized function $Y_{\text {reg }}^{(\Lambda)}(p)$ is given as follows
\begin{equation}
Y_{\text {reg }}^{(\Lambda)}(p)=\sum_{f=1}^K \varepsilon_f \sum_{J=p}^{\Lambda} \frac{2 J}{J^2-p^2+\mathcal{M}_f^2}+\sum_{J=p+1}^{\Lambda} \frac{2 J}{J^2-p^2} .
\end{equation}
We note that 
\begin{equation}\label{eq: Ydef}
Y_{\mathrm{reg}}^{(s)}=\lim _{\Lambda \rightarrow \infty} Y_{\mathrm{reg}}^{(\Lambda)}\left(\frac{1+s}{2}\right) .
\end{equation}
As in the main text, we assume existence of the cut-off $\Lambda \gg \mathcal{M}_f$. Applying the 
Euler-Maclaurin formula,
\begin{equation}\label{eq: EMform}
\sum_{J=p+1}^{\Lambda} g(J)=\int_{p+1}^{\Lambda} g(x) d x+\frac{g(\Lambda)+g(p+1)}{2}+\left.\frac{g^{\prime}(x)}{12} \right|_{p+1} ^{\Lambda},
\end{equation}
 and after some straightforward calculations we obtain
\begin{equation}
\lim _{\Lambda \rightarrow \infty} Y_{\text {reg }}^{(\Lambda)}(p)=2 \ln \mathcal{M}+\gamma -\psi(1+2p) ,
\end{equation}
where $\psi(z)$ stands for the digamma function. Therefore, the regularized expressions for $Y^{(s)}$ 
are given as
\begin{equation}
Y_{\text {reg }}^{(0)}=2 \ln \mathcal{M}+2 \gamma-1, \quad Y_{\text {reg }}^{(1)}=2 \ln \mathcal{M}+2 \gamma-\frac{3}{2} .
\end{equation}
Using the above results, we find the \color{black} perturbative correction to the disorder-averaged \color{black} {\DOS}:
\begin{equation}\label{eq: tdospv}
    \nu_{\rm pert}(\mathcal{M})  \approx    \nu \left(1+ \frac{\gamma^{(0)}_{1}}{ g}\ln \mathcal{M}e^{\gamma-1/2}\right) .
\end{equation}

Next, we consider the one-loop renormalization of the  spin conductivity in the Pauli-Villars regularization scheme. We start from considering the diamagnetic part of Eq. \eqref{eq:spin:g:K}. Using exponential parametrization for fluctuations, Eq. \eqref{eq: fluctpar}, we  write
\begin{equation}
    \delta g_{dm} = \frac{g}{8}(V_{1,1}+V_2)\left\langle \operatorname{Tr}W^2\left(2+\operatorname{Tr}1\right)\right \rangle_{S_0} .
\end{equation}
As one can see this expression is analogous to the expression obtained for the {\DOS}, cf. Eq.\eqref{eq: tdospv}. Therefore we obtain immediately,
\begin{equation}\label{eq: diamagg}
   \delta g_{dm} = - \frac{(1+n)}{\pi } \ln \mathcal{M}e^{\gamma-1/2} .
\end{equation}

At the next step, we take into account the current-current part ($j-j$) of the  correlation function for the spin conductivity. There are only three non-zero contributions, which can be written as
\begin{equation}
\begin{split}
& \left\langle \operatorname{Tr}\nabla_{\mu}W(\boldsymbol{x})\nabla_{\mu^{\prime}}W(\boldsymbol{x}^{\prime})\right \rangle, \\ - 2 & \left\langle \operatorname{Tr}\left(\nabla_{\mu}W(\boldsymbol{x}) W(\boldsymbol{x^{\prime}})\left[\nabla_{\mu^{\prime}}W(\boldsymbol{x^{\prime}})\right]W(\boldsymbol{x^{\prime}})\right)\right\rangle,  \\ \frac{2}{3} & \left\langle \operatorname{Tr}\left(\nabla_{\mu}W(\boldsymbol{x}) \nabla_{\mu^{\prime}}W^3(\boldsymbol{x^{\prime}})\right)\right\rangle
\end{split}
\end{equation}
All of these expressions can be obtained from expansion of the following term:
\begin{gather}
     \int d \boldsymbol{x} \operatorname{Tr} \left(\boldsymbol{E} \nabla Q(\boldsymbol{x}) \right) = \int\limits_{\partial \Omega} d s \; \boldsymbol{n} \div \operatorname{Tr} \left(\boldsymbol{E} Q(\boldsymbol{x}) \right), \notag \\ \boldsymbol{E} = \int d \boldsymbol{x}^\prime Q(\boldsymbol{x}^\prime)\nabla Q(\boldsymbol{x}^\prime)\Lambda .
\end{gather}
Using the fact that the $Q$ matrix at the boundary is the constant, $Q(\boldsymbol{x})|_{\partial \Omega}=Q_b = \text{const}$, to be consistent with the quantization of the topological charge, we obtain
\begin{equation}
    \int\limits_{\partial \Omega} d s \; \boldsymbol{n} \div \operatorname{Tr} \left(\boldsymbol{E} Q_b \right)= 0, \quad \div \operatorname{Tr} \left(\boldsymbol{E} Q_b \right) = 0 .
\end{equation}
Therefore, there are no full-derivative contributions to the longitudinal spin conductivity in Pauli-Villars regularization. Finally, we obtain that only one relevant one-loop correction to longitudinal spin conductivity $g$ comes from diamagnetic part, cf. Eq. \eqref{eq: diamagg}. It reads
\begin{equation}\label{eq: diamagg:1}
   g(\mathcal{M}) =g  - \frac{(1+n)}{\pi } \ln \mathcal{M}e^{\gamma-1/2} .
\end{equation}

\section{Renormalization of RG eigenoperators with three $\mathcal{Q}$-matrices \label{App:3}}

In this appendix we confirm the fulfillment of \color{black} Eq. \eqref{eq:anom:dim:inst:full} for the \color{black}
eigenoperators with three $\mathcal{Q}$ matrices. Following Ref.~\cite{Babkin2022}, we define RG eigenoperators in terms of correlation functions $\mathcal{P}_{\bm{\lambda}}^{\alpha_1 \alpha_2 \alpha_3 ; p_1 p_2 p_3}$:
\begin{gather}
    \mathcal{K}_{\bm{\lambda}}[\mathcal{Q}]=\frac{1}{64} \color{black} \sum_{p_j= R/A} (\sigma_3)_{p_1p_1} (\sigma_3)_{p_2p_2} (\sigma_3)_{p_3p_3} \color{black} 
    \notag \\
    \times \mathcal{P}_{\bm{\lambda}}^{\alpha_1 \alpha_2 \alpha_3 ; p_1 p_2 p_3},
\end{gather}
where $|\bm{\lambda}|=3$ and  
$\mathcal{P}_{\bm{\lambda}}^{\alpha_1 \alpha_2 \alpha_3 ; p_1 p_2 p_3}$ can be written in terms of the $\mathcal{Q}$-matrices as follows:
\begin{multline}
    \mathcal{P}_{\bm{\lambda}}^{\alpha \beta \mu ; p_1 p_2 p_3} =  \operatorname{tr} \mathcal{Q}_{p_1 p_1}^{\alpha \alpha} \operatorname{tr} \mathcal{Q}_{ p_2  p_2}^{\beta \beta} \operatorname{tr} \mathcal{Q}_{ p_3 p_3}^{\mu \mu} \\ +  \mu^{(\bm{\lambda})}_{2,1} \operatorname{tr} \mathcal{Q}_{p_1 p_1}^{\alpha \alpha}  \operatorname{tr} \mathcal{Q}_{ p_2 p_3}^{\beta \mu} \mathcal{Q}_{p_3  p_2}^{\mu \beta}\\ +\mu^{(\bm{\lambda})}_3 \operatorname{tr} \mathcal{Q}_{p_1 p_2}^{\alpha \beta} \mathcal{Q}_{ p_2 p_3}^{\beta \mu} \mathcal{Q}_{p_3 p_1}^{\mu \alpha} .
\end{multline}
Here $\alpha ,\beta, \mu$ denote fixed different replica indices.  In order to $\mathcal{K}_{\bm{\lambda}}[\mathcal{Q}]$ be the RG eigen operator, the constants $\mu^{(\bm{\lambda})}_{2,1}$ and $\mu^{(\bm{\lambda})}_3$ should take only some specific values:
\begin{equation}
    \begin{aligned}
\boldsymbol{\lambda} = \textsf{(3)}:   \quad \mu^{\textsf{(3)}}_{2,1}=-3,\quad \mu^{\textsf{(3)}}_3=2 , \\
\boldsymbol{\lambda} = \textsf{(2,1)}:   \quad \mu^{\textsf{(2,1)}}_{2,1}=1,\quad \mu^{\textsf{(2,1)}}_3=-2 , \\ 
\boldsymbol{\lambda} = \textsf{(1,1,1)}:   \quad \mu^{\textsf{(1,1,1)}}_{2,1}=6,\quad \mu^{\textsf{(1,1,1)}}_3=8 .
\end{aligned}
\end{equation}
Our aim is to compute non-perturbative renormalization of $\mathcal{K}_{\bm{\lambda}}[\mathcal{Q}]$. It can be performed with the help of the saddle-point approximation \eqref{eq: instav1}. Using the 
parametrization of $\mathcal{Q}$ in terms of a deviation from the trivial saddle-point $\Lambda$, Eq. \eqref{eq: qparam}, and taking into account fluctuations  in the action only, we obtain
\begin{multline}
\left\langle\mathcal{K}_{\bm{\lambda}}\right\rangle_{\pm 1} {=} \left \langle {-}\frac{\mu^{(\bm{\lambda})} _{2,1}|\rho _{12}|^2}{2 n (n+1)}{+}\frac{3 \rho _{11}}{n} {-} \frac{2 \mu^{(\bm{\lambda})}_{2,1}{+}3 \mu^{(\bm{\lambda})}_3|\rho _{12}|^2\rho _{11}}{8 (n+2)} \right.\\ \left.  +\frac{\rho _{11}^2 \left(\mu^{(\bm{\lambda})}_{2,1}+6\right)}{2 n (n+1)}  +\frac{\rho _{11}^3 \left(2 \mu^{(\bm{\lambda})}_{2,1}+\mu^{(\bm{\lambda})}_3+4\right)}{4 n (n+1) (n+2)} \right \rangle_{\pm 1}.
\label{eq:AppB:K3}
\end{multline}
\color{black} We note that we averaged over $\TT$ rotations in Eq. \eqref{eq:AppB:K3}. Technically, it \color{black} can be performed with the help of the generalization of the expression \eqref{unav} to the case with six 
%uniform rotation 
unitary matrices, which can be found, for example, in Ref. \cite{mello1990averages}. In Eq. \eqref{eq:AppB:K3} we use some specific notation for the functions averaged over the instanton manifold
\begin{equation}
    \langle f \rangle_{\pm 1} =  G_n\int \frac{d\lambda}{\lambda^3} (\pi g)^{n+1} e^{-\pi g(\mathcal{M})+i\vartheta} \int d\boldsymbol{r}_0f(\boldsymbol{r}_0).
\end{equation}
Next step is to omite contributions, which are finite in the ultra violet. Surprisingly, it leads to Eq. \eqref{eq: Kaver}, where we should replace only expressions for the one-loop coefficients $ \gamma^{(0)}_{\bm{\lambda}}$: 
\begin{gather}
    \gamma^{(0)}_{\textsf{(3)}}(g) = -\frac{3n}{\pi }, \quad \gamma^{(0)}_{\textsf{(2,1)}}(g) = -\frac{4+3n}{\pi }, \\  \gamma^{(0)}_{\textsf{(1,1,1)}}(g) = -\frac{9+3n}{\pi }.
\end{gather}
Repeteating the spatial-varying mass procedure we obtain instanton correction to the anomalous dimensions of the eigen oeprators $\mathcal{K}_{\bm{\lambda}}$  with $|\bm{\lambda}|=3$ \color{black} in the form of Eq.~\eqref{eq:anom:dim:inst:full}. \color{black}

%%%%%
\section{
Derivation of Kubo formula for the spin conductivity 
\label{App:4}}

Our aim is to derive Kubo-type formulas for spin conductivities, cf. Eqs. \eqref{eq:spin:g:K} and \eqref{eq:spin:gH:K}.
It can be done in several ways. Here we present derivation based on  
Matsubara Kubo formula, which was obtained in Ref. \cite{Babkin2022}.
Firstly, we introduce the generalization of {\NLSM} to the case of Matsubara frequency space (Finkel'stein {\NLSM}). The field $\hat{\mathcal{Q}}$ becomes a traceless Hermitian matrix, defined on $N_r{\times}N_r$ replica, $2N_m{\times}2N_m$ Matsubara  and $2{\times}2$ spin spaces. It satisfies the same nonlinear constraint, $\hat{\mathcal{Q}}^2=1$, and BdG symmetry relation: 
\begin{gather}\label{eq: constr1:app}
\hat{\mathcal{Q}} = -\overline{\hat{\mathcal{Q}}}, \,\, \overline{\hat{\mathcal{Q}}} = \textsf{s}_2 \hat{L}_0 \hat{\mathcal{Q}}^T  \hat{L}_0 \textsf{s}_2, \notag
\\
         \left(\hat{L}_0\right)_{n m}^{\alpha \beta} 
         =\delta_{\varepsilon_n,-\varepsilon_m} \delta^{\alpha \beta} \textsf{s}_0 .
\end{gather}
Here we define fermionic Matsubara frequencies in a standart way: $\varepsilon_n = \pi T\left(2n+1\right)$. Therefore, the trivial saddle-point of {\NLSM}, taking into account Matsubara frequency space, has the following form \cite{Babkin2022}: 
\begin{equation}
    \hat{\Lambda}_{n m}^{\alpha \beta}=\operatorname{sgn} \varepsilon_n \delta_{n m} \delta^{\alpha \beta} \mathrm{s}_0 .
\end{equation}

In such extended representation, Kubo formula for longitudinal spin conductivity can be written in terms of two operators:
\begin{widetext}
\begin{equation}\label{eq: matslc}
    \sigma_s\left(i \omega_k\right)=-\frac{g}{8 k L^2}\int_{\boldsymbol{x}}\left\langle\operatorname{Tr}\left[I_k^\alpha \mathbf{s}_3, \hat{\mathcal{Q}}\right]\left[I_{-k}^\alpha \mathbf{s}_3, \hat{\mathcal{Q}}\right]\right\rangle+\frac{g^2}{32 k L^2} \int_{\boldsymbol{x}^{\prime}}\int_{\boldsymbol{x}}\left\langle\left\langle\operatorname{Tr} I_k^\alpha \mathrm{s}_3 \hat{\mathcal{Q}}(\boldsymbol{x}) \nabla \hat{\mathcal{Q}}(\boldsymbol{x}) \operatorname{Tr} I_{-k}^\alpha \mathrm{s}_3 \hat{\mathcal{Q}}\left(\boldsymbol{x}^{\prime}\right) \nabla \hat{\mathcal{Q}}\left(\boldsymbol{x}^{\prime}\right)\right\rangle\right\rangle ,
\end{equation}
where $\left(I^{\gamma}_{k}\right)_{nm}^{\alpha \beta} = \delta_{n-m,k}\delta^{\alpha \beta}\delta^{\alpha\gamma}\mathbf{s}_0$. 
Next step is to average these two operators over unitary rotations of $\hat{\mathcal{Q}}$ which commute with $\hat{\Lambda}$. It can be performed with the help of the expressions similiar to Eq. \eqref{unav}, where we should replace all indices on vector index $a\rightarrow \{a_{\text{r}},a_{\text{S}},a_{\text{M}}\}$ in replica, spin, and Matsubara spaces. The averaged longitudinal spin conductivity operator consists of three different averages only:
$\left\langle \operatorname{Tr}I_k^\alpha \mathbf{s}_3 \mathcal{U}^{-1} A \mathcal{\mathcal{U}} \right\rangle_{\mathcal{U}} = 0$,
\begin{multline}\label{eq: trav}
    \left\langle \operatorname{Tr}I_k^\alpha \mathbf{s}_3 \mathcal{U}^{-1} A \mathcal{U} I_{-k}^\alpha \mathbf{s}_3 \mathcal{U}^{-1}B \mathcal{U}\right\rangle_{\mathcal{U}} = V_{1,1}\left[\left(N_m-k\right)\left(\operatorname{Tr}A\operatorname{Tr}B+\operatorname{Tr}\hat{\Lambda} A\operatorname{Tr}\hat{\Lambda} B + \operatorname{Tr}\left[A\bar{B}-\hat{\Lambda} A\hat{\Lambda}\bar{B}\right]\right)+ \right.\\ \left. 2k \left(\operatorname{Tr}\hat{\Lambda}_{-}A\operatorname{Tr}\hat{\Lambda}_{+}B + \operatorname{Tr}\hat{\Lambda}_{-}A\hat{\Lambda}_{-}\bar{B}\right)\right] + V_2 \left[2\left(N_m-k\right)\operatorname{Tr}\hat{\Lambda} A \hat{\Lambda} B + \right. \\ \left. 2k \left(\operatorname{Tr}\hat{\Lambda}_{-}A\operatorname{Tr}\hat{\Lambda}_{+}B + \operatorname{Tr}\hat{\Lambda}_{-}A\hat{\Lambda}_{-}\bar{B}\right) \right] ,
\end{multline}
and
\begin{multline}\label{eq: trtrav}
    \left\langle \operatorname{Tr}I_k^\alpha \mathbf{s}_3 \mathcal{U}^{-1} A \mathcal{U} \operatorname{Tr}I_{-k}^\alpha \mathbf{s}_3 \mathcal{U}^{-1}B \mathcal{U}\right\rangle_\mathcal{U} = V_{1,1}\left[\left(N_m-k\right) \operatorname{Tr}\left[A\left(B-\Bar{B} \right)+\hat{\Lambda} A \hat{\Lambda} \left(B-\Bar{B} \right)\right]+2k\operatorname{Tr}\hat{\Lambda}_{-}A\hat{\Lambda}_{+}\left(B-\Bar{B} \right)\right] + \\V_2 \left[2\left(N_m-k\right)\operatorname{Tr}\hat{\Lambda} A\operatorname{Tr}\hat{\Lambda} B +2k\operatorname{Tr}\hat{\Lambda}_{-}A\hat{\Lambda}_{+}\left(B-\Bar{B} \right)  \right] .
\end{multline}
Here the coefficients $V_{a,b}$ are given as follows
\begin{equation}
    V_1 = \frac{1}{n}, \quad V_{1,1} = \frac{1}{n^2-1}, \quad V_{2} = -\frac{1}{n(n^2-1)} .
\end{equation}
Here $n=2N_r N_m$ and $\hat{\Lambda}_{\pm} = \left(1 \pm \hat{\Lambda}\right)/2$ denote projectors on the positive and negative Matsubara frequences.  After that, choosing matrices $A$ and $B$ to be equal to $\hat{\mathcal{Q}}$ in Eq. \eqref{eq: trav} and $\hat{\mathcal{Q}}\nabla \hat{\mathcal{Q}}$ in Eq. \eqref{eq: trtrav}, we are able to write down the full expressions for the spin conductivity. For convenience, we divide the derived expression into two parts: diamagnetic (without spatial derivatives) and current-current correlation function (with derivatives):
\begin{multline}\label{eq: gdmM}
    g_{dm}\left(i \omega_k\right)= -\frac{g}{4k} \left\langle V_{1,1} \left[\left(N_m-k\right)\left( \operatorname{Tr} \left(\hat{\Lambda} \hat{\mathcal{Q}}\right)^2-\operatorname{Tr}1 +\left(\operatorname{Tr} \hat{\Lambda} \hat{\mathcal{Q}}\right)^2\right)-\frac{k}{2}\left(\operatorname{Tr}1+ \operatorname{Tr} \left(\hat{\Lambda} \hat{\mathcal{Q}}\right)^2 +\left(\operatorname{Tr} \hat{\Lambda} \hat{\mathcal{Q}}\right)^2\right)\right] + \right.\\\left. V_2\left[2\left(N_m-k\right)\operatorname{Tr}\left(\hat{\Lambda} \hat{\mathcal{Q}}\right)^2 -\frac{k}{2}\left(
\operatorname{Tr}1+ \operatorname{Tr} \left(\hat{\Lambda} \hat{\mathcal{Q}}\right)^2 +\left(\operatorname{Tr} \hat{\Lambda} \hat{\mathcal{Q}}\right)^2\right)\right] - 4(2N_m-k)\right\rangle
\end{multline}
and
\begin{multline}\label{eq: gjjM}
    g_{j-j}\left(i \omega_k\right) = \frac{g^2}{32k}\int_{\boldsymbol{x}^{\prime}}\left\langle V_{1,1}\left[2(N_m-k)\left(\operatorname{Tr}\boldsymbol{\mathcal{J}}(\bm{x})\cdot  \boldsymbol{\mathcal{J}}(\bm{x}^\prime) + \operatorname{Tr}\hat{\Lambda} \boldsymbol{\mathcal{J}}(\bm{x})\cdot  \hat{\Lambda}\boldsymbol{\mathcal{J}}(\bm{x}^\prime)\right) \right. \right.\\ \left. \left. + k\left(\operatorname{Tr}\boldsymbol{\mathcal{J}}(\bm{x})\cdot  \boldsymbol{\mathcal{J}}(\bm{x}^\prime) - \operatorname{Tr}\hat{\Lambda} \boldsymbol{\mathcal{J}}(\bm{x})\cdot  \hat{\Lambda}\boldsymbol{\mathcal{J}}(\bm{x}^\prime)\right) \right] +  V_2 \left[2\left(N_m-k\right)\left( \operatorname{Tr}\hat{\Lambda} \boldsymbol{\mathcal{J}}(\bm{x})\cdot  \operatorname{Tr}\hat{\Lambda}\boldsymbol{\mathcal{J}}(\bm{x}^\prime)\right) +k\left(\operatorname{Tr}\boldsymbol{\mathcal{J}}(\bm{x})\cdot \boldsymbol{\mathcal{J}}(\bm{x}^\prime)  \right. \right. \right.\\\left. \left. \left. -\operatorname{Tr}\hat{\Lambda} \boldsymbol{\mathcal{J}}(\bm{x})\cdot  \hat{\Lambda}\boldsymbol{\mathcal{J}}(\bm{x}^\prime)\right)\right] \right\rangle ,
\end{multline}
\end{widetext}
respectively. 
Here we introduce the matrix current $\boldsymbol{\mathcal{J}}(\bm{x}) = \hat{\mathcal{Q}}(\bm{x})\nabla\hat{\mathcal{Q}}(\bm{x})$. After performing the averaging, we reduce the Matsubara frequency space to a single frequency and, thus, set $N_m =1$. Our consideration restricts only RA space, therefore, we can set $k=1$, in other words we consider processes, which change energy only at the first bosonic Matsubara frequency. A strong consistency check of this reasoning is that the entire part of Eqs. \eqref{eq: gdmM} and \eqref{eq: gjjM} proportional to $N_m-k$ vanishes during the background field renormalzation procedure. Therefore, for $N_m=k =1$, Eqs. \eqref{eq: gdmM} and \eqref{eq: gjjM} reduce to Eq. \eqref{eq:spin:g:K} after some straightforward algebra. 

Kubo formula for the transverse spin conductivity has the form: 
\begin{multline}
    \sigma^{t}_s(i \omega_k) = \frac{g^2}{32 k L^2} \int_{\boldsymbol{x}^{\prime}}\int_{\boldsymbol{x}}\left\langle\left\langle\varepsilon_{\mu \nu}\operatorname{Tr} I_k^\alpha \mathrm{s}_3 \mathcal{Q}(\boldsymbol{x}) \nabla_{\mu} \mathcal{Q}(\boldsymbol{x}) \right.\right.\\ \left.\left. \times\operatorname{Tr} I_{-k}^\alpha \mathrm{s}_3 \mathcal{Q}\left(\boldsymbol{x}^{\prime}\right) \nabla_{\nu} \mathcal{Q}\left(\boldsymbol{x}^{\prime}\right)\right\rangle\right\rangle .
\end{multline}
In order to average the above expression, it is enough to use Eq. \eqref{eq: trtrav} only. After that we set $N_m=k =1$ and obtain Eq. \eqref{eq:spin:gH:K}.

\section{One-loop corrections to {\DOS} on the instanton background \label{App:5}}

In this Appendix we justify the apperance of the renormalized obvervables in pre-exponential factors for instanton corrections to the RG eigen operators. We will take into account quantum fluctuations near instanton saddle-point in the pre-exponential factor. Using Eq. \eqref{eq: instav1}, we write the following expression for the renormalized {\DOS}:
\begin{gather}
\nu^{\prime}\left(\mathcal{M}\right) = \frac{\nu}{2n}\left\langle \operatorname{Tr}\Lambda \mathcal{Q} \right\rangle_0 \left(1-\frac{\mathcal{Z}_{\rm inst}+\mathcal{Z}_{\rm inst}^*}{\mathcal{Z}_0}  \right) \notag \\ + \frac{\nu}{2n}\left\langle \operatorname{Tr}\Lambda \mathcal{Q} \right\rangle_{+1}  +  \frac{\nu}{2n}\left\langle \operatorname{Tr}\Lambda \mathcal{Q} \right\rangle_{-1} .
\label{eq:App:nuN:0}
\end{gather}
The first term in the right hand side of the above equation was calculated previously (see Appendix \ref{App:2}). Here we focus on calculation of the last two terms. Using exponential parametrization for $Q$, we find
\begin{gather}
    \left\langle \operatorname{Tr}\Lambda \mathcal{Q} \right\rangle_{+1} =\left\langle \operatorname{Tr}\underline{\Lambda} Q \right\rangle_{+1} \approx  \left\langle \operatorname{Tr} \tilde{R} \underline{\Lambda} \tilde{R}^{-1} \underline{\Lambda}\right\rangle_{+1} \notag \\ +\frac{1}{2}\left\langle \operatorname{Tr} \tilde{R} \underline{\Lambda} \tilde{R}^{-1} \underline{\Lambda} W^2\right\rangle_{+1} .
    \label{eq: appE1}
\end{gather}
We start from the term without fluctuation matrix field $W$:
\begin{gather}
    \left\langle \operatorname{Tr} \tilde{R} \underline{\Lambda} \tilde{R}^{-1} \underline{\Lambda}\right\rangle_{+1} = 2n\frac{\mathcal{Z}_{\rm inst}}{\mathcal{Z}_0}  -  n(n+1)G_n \notag \\ \times \int \frac{d \lambda}{\lambda}(\pi g)^{n+1}e^{-\pi g\left(\mathcal{M}\right)+ i\theta}\int d \boldsymbol{r}_0 \frac{\mu(r_0)}{\lambda} .
\end{gather}
We note that the the first term, proportional to $\mathcal{Z}_{\rm inst}$ can be rewritten as $\left\langle \operatorname{Tr} \underline{\Lambda} \underline{\Lambda}\right\rangle_{0}\mathcal{Z}_{\rm inst}/\mathcal{Z}_0$ and it cancels out with the second term in brackets in Eq. \eqref{eq:App:nuN:0}. The second contribution in Eq. \eqref{eq: appE1}, which depends on quantum fluctuations, can be calculated as follows:
\begin{widetext}
\begin{multline}
   \frac{1}{2} \left\langle \operatorname{Tr} \tilde{R} \underline{\Lambda} \tilde{R}^{-1} \underline{\Lambda} W^2\right\rangle_{+1} = -\left( \left\langle\left(|e_1|^2-e_0^2\right)w^{11}w^{*11} \right\rangle_{+1} + \sum_{\alpha=2}^{n}\left\langle\left(|e_1|^2-e_0^2 +1\right)w^{1\alpha}w^{*1\alpha}\right\rangle_{+1}  + \right. \\ \left.\sum_{\alpha=2}^{n} \left\langle w^{\alpha\alpha}w^{*\alpha\alpha}\right\rangle_{+1}   + 2 \sum_{1<\alpha<\beta\leqslant n} \left\langle w^{\alpha\beta}w^{*\alpha\beta}\right\rangle_{+1} \right) .
\end{multline}
The above averages can be rewritten in terms of the Green's functions at coinciding points:
\begin{multline}
\frac{1}{2} \left\langle \operatorname{Tr} \tilde{R} \underline{\Lambda} \tilde{R}^{-1} \underline{\Lambda} W^2\right\rangle_{+1}= -\left(\left\langle\left(\left|e_1\right|^2-e_0^2\right) \mathcal{G}_2\right\rangle_{+1}+\frac{n-1}{2}\left\langle\left(\left|e_1\right|^2-e_0^2+1\right) \mathcal{G}_1\right\rangle_{+1}+(n-1)\left\langle\mathcal{G}_0\right\rangle_{+1} \right.\\ \left. +\frac{(n-1)(n-2)}{2}\left\langle\mathcal{G}_0\right\rangle_{+1}\right) .
\end{multline}
We should subtract from this equation the contribution of the trivial topological sector $\frac{1}{2}\left\langle \operatorname{Tr} \underline{\Lambda}  \underline{\Lambda} W^2\right\rangle_{0} \mathcal{Z}_{\rm inst}/\mathcal{Z}_0$ (the second term in brackets in eq. \eqref{eq:App:nuN:0}), after that we obtain:
\begin{equation}
 -\left(  \left\langle\left(|e_1|^2-e_0^2\right)\mathcal{G}_2 \right\rangle_{+1} - \frac{\mathcal{Z}_{\rm inst}}{\mathcal{Z}_0} \mathcal{G}_0+ \frac{n-1}{2}\left\langle\left(|e_1|^2-e_0^2 +1\right)\mathcal{G}_1\right\rangle_{+1}  - (n-1)\frac{\mathcal{Z}_{\rm inst}}{\mathcal{Z}_0} \mathcal{G}_0\right) .
\end{equation}
Using the Pauli-Villars regularization scheme and representation of regularized Green's functions in terms of $Y$-function, cf. Eq. \eqref{eq: Ydef}, we  obtain with logarithmic accuracy, i.e. with neglect of all terms without large Pauli-Villars mass $\mathcal{M}$:
\begin{equation}
     \left(\delta\nu^{\prime}\left(\mathcal{M}\right)\right)_{+1}  = \frac{(1+n)^2\ln \mathcal{M}}{2 \pi g}G_n \int \frac{d \lambda}{\lambda}(\pi g)^{n+1} e^{-\pi g\left(\mathcal{M}\right)+ i\vartheta}\int d \boldsymbol{r_0} \frac{\mu(r_0)}{\lambda} .
\end{equation}
Taking into account a similar contribution with the negative topological charge, we obtain expression \eqref{eq:deltaNu:inst:0} for renormalized {\DOS}.
\end{widetext}

\section{Curing the ultra-violet divergences by means of the spatial variating mass method\label{AppF}}

In this Appendix, we discuss the transformation of Eq. \eqref{eq:deltaNu:inst:00} into Eq. \eqref{eq:deltaNu:inst:0}. 
In order to cure the ultra-violet divergences, we employ the scheme of spatially varying mass adapted from Ref. \cite{Pruisken2005}.  
Firstly, \color{black} we transfer from the curved space in which the quantum correction to the {\DOS} is controlled by the Pauli-Villars mass $\mathcal{M}$ to the flat space in which a relevant length scale limiting the quantum fluctuations is $1/\mu(r_0)$, \color{black}
\begin{gather}
     \nu(\mathcal{M})  \rightarrow \nu(\mathcal{M})\left(1 - \frac{\gamma_{\textsf{(1)}}^{(0)}}{g} \ln [\mu(r_0) \ell \mathcal{M}]\right) \notag \\  = \nu\left(1+\frac{\gamma_{\textsf{(1)}}^{(0)}}{g}\ln \frac{e^{\gamma-1/2}}{\mu(r_0) \ell}\right)\equiv \nu(1/\mu(r_0)) .
\end{gather}
This expression we substitute into integral over instanton position $r_0$ in Eq. \eqref{eq:deltaNu:inst:0}. Next we convert the perturbative correction with spatially dependent mass into the perturbative correction at the length scale $\zeta \lambda$ where $\zeta=e/2$,
\begin{gather}
     \nu(\zeta \lambda)  = \frac{1}{4\pi} \int d\boldsymbol{r}_0 \, \mu^2(r_0)\nu(\mu(r_0)) \notag \\  
      = \nu_0 \left(1+\frac{\gamma_{\textsf{(1)}}^{(0)}}{g}\ln\lambda \mu_0 \zeta e^{\gamma-1/2}\right), \quad \zeta = \frac{e}{2} .
\end{gather}
\color{black} The above expression suggests the following \color{black} 
correspondence between the Pauli-Villars mass and \color{black} the instanton size $\lambda$\color{black}
\begin{equation}
    \mathcal{M} \rightarrow \zeta \lambda \mu_0 .
\end{equation}
Next we rewrite Eq. \eqref{eq:deltaNu:inst:0} in a more transparent form
\begin{equation}
    \delta \nu_{\rm inst} = G_n \int \frac{d \lambda}{\lambda} (\pi g)^{n+1} e^{-\pi g\left(\zeta \lambda\right)} \mathcal{A}_{\textsf{(1)}}\cos \vartheta .
\end{equation}
 Here we introduce the amplitude $\mathcal{A}_{\bm{\lambda}}$ 
\begin{gather}
    \mathcal{A}_{\bm{\lambda}} {=} \pi \gamma^{(0)}_{\bm{\lambda}} \int\limits_0^L d\bm{r}_0 \frac{\mu(r_0)}{\lambda} z_{\bm{\lambda}}\left(1/\mu(r_0)\right)  {=} {-} 2 \pi^2 \gamma^{(0)}_{\bm{\lambda}} z_{\bm{\lambda}}(1/\mu(0)) \notag \\ \times \int \limits_{\mu(0)}^{\mu\left(L\right)} d\left[\ln \mu(r_0)\right] \frac{z_{\bm{\lambda}}\left(1/\mu(r_0)\right)}{z_{\bm{\lambda}}\left(1/\mu(0)\right)} ,
\end{gather}
where the integral over $r_0$ runs until the system size $L$.
We assume that $\mathcal{A}_{\bm{\lambda}}$ corresponds to the operator with negative anomalous dimension, $\gamma_{\bm{\lambda}}<0$. In particular, this is the case of {\DOS}. Using 
\color{black}
definition of the anomalous dimension $\gamma_{\bm{\lambda}}$,
\color{black} 
we can rewrite the integrand as follows:
\begin{equation}
    \frac{z_{\bm{\lambda}}\left(\mu(r_0)\right)}{z_{\bm{\lambda}}\left(\mu(0)\right)} = \exp \left(- \int\limits_{\ln \mu(0)}^{\ln \mu(r_0)}d\left[\ln \mu\right]\gamma_{\bm{\lambda}}\right) .
\end{equation}
We note that the condition $\gamma_{\bm{\lambda}}<0$ guarantees the convergence of integral over $\mu(r_0)$ as $\mu(L)\to 0$. 

Now it is convenient to change integral variable from $\ln \mu$ to $g$ with the help of perturbative RG equation for $g$, see Eq. \eqref{eq: betag}. Then we obtain $\mathcal{A}_{\bm{\lambda}} = z_{\bm{\lambda}}\left(1/\mu(0)\right) \mathcal{H}_{\bm{\lambda}}\left(g\left(1/\mu(0)\right)\right)$, 
where we introduce the  function
\begin{equation}\label{eq: Hdefapp}
    \mathcal{H}_{\bm{\lambda}} {=} {-} 2 \pi^2 \gamma^{(0)}_{\bm{\lambda}} \int \limits_{g\left(1/\mu(0)\right)}^{g\left(1/\mu\left(L\right)\right)} \!\!\frac{dg}{\beta_g(g)} \exp \left({-} \!\! \int\limits_{g(1/\mu(0))}^{g(1/\mu(r_0))}\frac{dg^{\prime}\,\gamma_{\bm{\lambda}}(g^{\prime})}{\beta_g(g^{\prime})}\right) .
\end{equation}
Using the one-loop results for $\gamma_{\bm{\lambda}}$ and $\beta_g$ and integrating over $g^\prime$
 and $g$, we obtain the expression \eqref{eq:Hlambda:def:0}. 
 The result in the case of positive anomalous dimensions $\gamma_{\bm{\lambda}}$ can be obtained with the help of a kind of analytic continuation (see Ref. \cite{Pruisken2005} for details).

\bibliography{literature_classC_top}	

\begin{thebibliography}{89}%
\makeatletter
\providecommand \@ifxundefined [1]{%
 \@ifx{#1\undefined}
}%
\providecommand \@ifnum [1]{%
 \ifnum #1\expandafter \@firstoftwo
 \else \expandafter \@secondoftwo
 \fi
}%
\providecommand \@ifx [1]{%
 \ifx #1\expandafter \@firstoftwo
 \else \expandafter \@secondoftwo
 \fi
}%
\providecommand \natexlab [1]{#1}%
\providecommand \enquote  [1]{``#1''}%
\providecommand \bibnamefont  [1]{#1}%
\providecommand \bibfnamefont [1]{#1}%
\providecommand \citenamefont [1]{#1}%
\providecommand \href@noop [0]{\@secondoftwo}%
\providecommand \href [0]{\begingroup \@sanitize@url \@href}%
\providecommand \@href[1]{\@@startlink{#1}\@@href}%
\providecommand \@@href[1]{\endgroup#1\@@endlink}%
\providecommand \@sanitize@url [0]{\catcode `\\12\catcode `\$12\catcode
  `\&12\catcode `\#12\catcode `\^12\catcode `\_12\catcode `\%12\relax}%
\providecommand \@@startlink[1]{}%
\providecommand \@@endlink[0]{}%
\providecommand \url  [0]{\begingroup\@sanitize@url \@url }%
\providecommand \@url [1]{\endgroup\@href {#1}{\urlprefix }}%
\providecommand \urlprefix  [0]{URL }%
\providecommand \Eprint [0]{\href }%
\providecommand \doibase [0]{https://doi.org/}%
\providecommand \selectlanguage [0]{\@gobble}%
\providecommand \bibinfo  [0]{\@secondoftwo}%
\providecommand \bibfield  [0]{\@secondoftwo}%
\providecommand \translation [1]{[#1]}%
\providecommand \BibitemOpen [0]{}%
\providecommand \bibitemStop [0]{}%
\providecommand \bibitemNoStop [0]{.\EOS\space}%
\providecommand \EOS [0]{\spacefactor3000\relax}%
\providecommand \BibitemShut  [1]{\csname bibitem#1\endcsname}%
\let\auto@bib@innerbib\@empty
%</preamble>
\bibitem [{\citenamefont {Anderson}(1958)}]{Anderson58}%
  \BibitemOpen
  \bibfield  {author} {\bibinfo {author} {\bibfnamefont {P.~W.}\ \bibnamefont
  {Anderson}},\ }\bibfield  {title} {\bibinfo {title} {Absence of diffusion in
  certain random lattices},\ }\href
  {http://link.aps.org/doi/10.1103/PhysRev.109.1492} {\bibfield  {journal}
  {\bibinfo  {journal} {Phys. Rev.}\ }\textbf {\bibinfo {volume} {109}},\
  \bibinfo {pages} {1492} (\bibinfo {year} {1958})}\BibitemShut {NoStop}%
\bibitem [{\citenamefont {Evers}\ \emph {et~al.}(2008)\citenamefont {Evers},
  \citenamefont {Mildenberger},\ and\ \citenamefont {Mirlin}}]{Evers2008}%
  \BibitemOpen
  \bibfield  {author} {\bibinfo {author} {\bibfnamefont {F.}~\bibnamefont
  {Evers}}, \bibinfo {author} {\bibfnamefont {A.}~\bibnamefont
  {Mildenberger}},\ and\ \bibinfo {author} {\bibfnamefont {A.~D.}\ \bibnamefont
  {Mirlin}},\ }\bibfield  {title} {\bibinfo {title} {Multifractality at the
  quantum {Hall} transition: {Beyond} the parabolic paradigm},\ }\href
  {https://doi.org/10.1103/PhysRevLett.101.116803} {\bibfield  {journal}
  {\bibinfo  {journal} {Phys. Rev. Lett.}\ }\textbf {\bibinfo {volume} {101}},\
  \bibinfo {pages} {116803} (\bibinfo {year} {2008})}\BibitemShut {NoStop}%
\bibitem [{\citenamefont {Chiu}\ \emph {et~al.}(2016)\citenamefont {Chiu},
  \citenamefont {Teo}, \citenamefont {Schnyder},\ and\ \citenamefont
  {Ryu}}]{Ryu2016}%
  \BibitemOpen
  \bibfield  {author} {\bibinfo {author} {\bibfnamefont {C.-K.}\ \bibnamefont
  {Chiu}}, \bibinfo {author} {\bibfnamefont {J.~C.~Y.}\ \bibnamefont {Teo}},
  \bibinfo {author} {\bibfnamefont {A.~P.}\ \bibnamefont {Schnyder}},\ and\
  \bibinfo {author} {\bibfnamefont {S.}~\bibnamefont {Ryu}},\ }\bibfield
  {title} {\bibinfo {title} {Classification of topological quantum matter with
  symmetries},\ }\href {https://doi.org/10.1103/RevModPhys.88.035005}
  {\bibfield  {journal} {\bibinfo  {journal} {Rev. Mod. Phys.}\ }\textbf
  {\bibinfo {volume} {88}},\ \bibinfo {pages} {035005} (\bibinfo {year}
  {2016})}\BibitemShut {NoStop}%
\bibitem [{\citenamefont {von Klitzing}\ \emph {et~al.}(1980)\citenamefont {von
  Klitzing}, \citenamefont {Dorda},\ and\ \citenamefont
  {Pepper}}]{Klitzing1980}%
  \BibitemOpen
  \bibfield  {author} {\bibinfo {author} {\bibfnamefont {K.}~\bibnamefont {von
  Klitzing}}, \bibinfo {author} {\bibfnamefont {G.}~\bibnamefont {Dorda}},\
  and\ \bibinfo {author} {\bibfnamefont {M.}~\bibnamefont {Pepper}},\
  }\bibfield  {title} {\bibinfo {title} {{New method for high-accuracy
  determination of the fine-structure constant based on quantized Hall
  resistance}},\ }\href {https://doi.org/10.1103/PhysRevLett.45.494} {\bibfield
   {journal} {\bibinfo  {journal} {Phys. Rev. Lett.}\ }\textbf {\bibinfo
  {volume} {45}},\ \bibinfo {pages} {494} (\bibinfo {year} {1980})}\BibitemShut
  {NoStop}%
\bibitem [{\citenamefont {Tsui}\ and\ \citenamefont
  {Gossard}(1981)}]{Tsui1981}%
  \BibitemOpen
  \bibfield  {author} {\bibinfo {author} {\bibfnamefont {D.~C.}\ \bibnamefont
  {Tsui}}\ and\ \bibinfo {author} {\bibfnamefont {A.~C.}\ \bibnamefont
  {Gossard}},\ }\bibfield  {title} {\bibinfo {title} {{Resistance standard
  using qunatization of the Hall resistance of GaAs-Al$_x$Ga$_{1-x}$As
  heterostructures}},\ }\href {https://doi.org/10.1063/1.92408} {\bibfield
  {journal} {\bibinfo  {journal} {Appl. Phys. Lett.}\ }\textbf {\bibinfo
  {volume} {38}},\ \bibinfo {pages} {550} (\bibinfo {year} {1981})}\BibitemShut
  {NoStop}%
\bibitem [{\citenamefont {Wigner}(1951)}]{Wigner1951}%
  \BibitemOpen
  \bibfield  {author} {\bibinfo {author} {\bibfnamefont {E.~P.}\ \bibnamefont
  {Wigner}},\ }\bibfield  {title} {\bibinfo {title} {On a class of analytic
  functions from the quantum theory of collisions},\ }\href
  {https://doi.org/https://doi.org/10.2307/1969342} {\bibfield  {journal}
  {\bibinfo  {journal} {Ann. Math.}\ }\textbf {\bibinfo {volume} {53}},\
  \bibinfo {pages} {36} (\bibinfo {year} {1951})}\BibitemShut {NoStop}%
\bibitem [{\citenamefont {Dyson}(1962{\natexlab{a}})}]{Dyson1962a}%
  \BibitemOpen
  \bibfield  {author} {\bibinfo {author} {\bibfnamefont {F.~J.}\ \bibnamefont
  {Dyson}},\ }\bibfield  {title} {\bibinfo {title} {Statistical theory of the
  energy levels of complex systems. {I}},\ }\href
  {https://doi.org/10.1063/1.1703773} {\bibfield  {journal} {\bibinfo
  {journal} {J. Math. Phys.}\ }\textbf {\bibinfo {volume} {3}},\ \bibinfo
  {pages} {140} (\bibinfo {year} {1962}{\natexlab{a}})}\BibitemShut {NoStop}%
\bibitem [{\citenamefont {Dyson}(1962{\natexlab{b}})}]{Dyson1962b}%
  \BibitemOpen
  \bibfield  {author} {\bibinfo {author} {\bibfnamefont {F.~J.}\ \bibnamefont
  {Dyson}},\ }\bibfield  {title} {\bibinfo {title} {The threefold way.
  algebraic structure of symmetry groups and ensembles in quantum mechanics},\
  }\href {https://doi.org/10.1063/1.1703863} {\bibfield  {journal} {\bibinfo
  {journal} {J. Math. Phys.}\ }\textbf {\bibinfo {volume} {3}},\ \bibinfo
  {pages} {2299} (\bibinfo {year} {1962}{\natexlab{b}})}\BibitemShut {NoStop}%
\bibitem [{\citenamefont {Zirnbauer}(1996)}]{Zirnbauer1996}%
  \BibitemOpen
  \bibfield  {author} {\bibinfo {author} {\bibfnamefont {M.~R.}\ \bibnamefont
  {Zirnbauer}},\ }\bibfield  {title} {\bibinfo {title} {Riemannian symmetric
  superspaces and their origin in random-matrix theory},\ }\href
  {https://doi.org/10.1063/1.531675} {\bibfield  {journal} {\bibinfo  {journal}
  {J. Math. Phys.}\ }\textbf {\bibinfo {volume} {37}},\ \bibinfo {pages} {4986}
  (\bibinfo {year} {1996})}\BibitemShut {NoStop}%
\bibitem [{\citenamefont {Altland}\ and\ \citenamefont
  {Zirnbauer}(1997)}]{Zirnbauer1997}%
  \BibitemOpen
  \bibfield  {author} {\bibinfo {author} {\bibfnamefont {A.}~\bibnamefont
  {Altland}}\ and\ \bibinfo {author} {\bibfnamefont {M.~R.}\ \bibnamefont
  {Zirnbauer}},\ }\bibfield  {title} {\bibinfo {title} {Nonstandard symmetry
  classes in mesoscopic normal-superconducting hybrid structures},\ }\href
  {https://doi.org/10.1103/PhysRevB.55.1142} {\bibfield  {journal} {\bibinfo
  {journal} {Phys. Rev. B}\ }\textbf {\bibinfo {volume} {55}},\ \bibinfo
  {pages} {1142} (\bibinfo {year} {1997})}\BibitemShut {NoStop}%
\bibitem [{\citenamefont {Heinzner}\ \emph {et~al.}(2005)\citenamefont
  {Heinzner}, \citenamefont {Huckleberry},\ and\ \citenamefont
  {Zirnbauer}}]{Zirnbauer2005}%
  \BibitemOpen
  \bibfield  {author} {\bibinfo {author} {\bibfnamefont {P.}~\bibnamefont
  {Heinzner}}, \bibinfo {author} {\bibfnamefont {A.}~\bibnamefont
  {Huckleberry}},\ and\ \bibinfo {author} {\bibfnamefont {M.~R.}\ \bibnamefont
  {Zirnbauer}},\ }\bibfield  {title} {\bibinfo {title} {Symmetry classes of
  disordered fermions},\ }\href {https://doi.org/10.1007/s00220-005-1330-9}
  {\bibfield  {journal} {\bibinfo  {journal} {Commun. Math. Phys.}\ }\textbf
  {\bibinfo {volume} {257}},\ \bibinfo {pages} {725} (\bibinfo {year}
  {2005})}\BibitemShut {NoStop}%
\bibitem [{\citenamefont {Schnyder}\ \emph {et~al.}(2008)\citenamefont
  {Schnyder}, \citenamefont {Ryu}, \citenamefont {Furusaki},\ and\
  \citenamefont {Ludwig}}]{Schnyder2008}%
  \BibitemOpen
  \bibfield  {author} {\bibinfo {author} {\bibfnamefont {A.~P.}\ \bibnamefont
  {Schnyder}}, \bibinfo {author} {\bibfnamefont {S.}~\bibnamefont {Ryu}},
  \bibinfo {author} {\bibfnamefont {A.}~\bibnamefont {Furusaki}},\ and\
  \bibinfo {author} {\bibfnamefont {A.~W.~W.}\ \bibnamefont {Ludwig}},\
  }\bibfield  {title} {\bibinfo {title} {Classification of topological
  insulators and superconductors in three spatial dimensions},\ }\href
  {https://doi.org/10.1103/PhysRevB.78.195125} {\bibfield  {journal} {\bibinfo
  {journal} {Phys. Rev. B}\ }\textbf {\bibinfo {volume} {78}},\ \bibinfo
  {pages} {195125} (\bibinfo {year} {2008})}\BibitemShut {NoStop}%
\bibitem [{\citenamefont {Schnyder}\ \emph {et~al.}(2009)\citenamefont
  {Schnyder}, \citenamefont {Ryu}, \citenamefont {Furusaki},\ and\
  \citenamefont {Ludwig}}]{Schnyder2009}%
  \BibitemOpen
  \bibfield  {author} {\bibinfo {author} {\bibfnamefont {A.~P.}\ \bibnamefont
  {Schnyder}}, \bibinfo {author} {\bibfnamefont {S.}~\bibnamefont {Ryu}},
  \bibinfo {author} {\bibfnamefont {A.}~\bibnamefont {Furusaki}},\ and\
  \bibinfo {author} {\bibfnamefont {A.~W.~W.}\ \bibnamefont {Ludwig}},\
  }\bibfield  {title} {\bibinfo {title} {Classification of topological
  insulators and superconductors},\ }\href
  {https://doi.org/10.1103/PhysRevB.78.195125} {\bibfield  {journal} {\bibinfo
  {journal} {AIP Conf. Proc.}\ }\textbf {\bibinfo {volume} {1134}},\ \bibinfo
  {pages} {10} (\bibinfo {year} {2009})}\BibitemShut {NoStop}%
\bibitem [{\citenamefont {Kitaev}(2009)}]{Kitaev2009}%
  \BibitemOpen
  \bibfield  {author} {\bibinfo {author} {\bibfnamefont {A.~Y.}\ \bibnamefont
  {Kitaev}},\ }\bibfield  {title} {\bibinfo {title} {Periodic table for
  topological insulators and superconductors},\ }\href
  {https://doi.org/10.1063/1.3149495} {\bibfield  {journal} {\bibinfo
  {journal} {AIP Conf. Proc.}\ }\textbf {\bibinfo {volume} {1134}},\ \bibinfo
  {pages} {22} (\bibinfo {year} {2009})}\BibitemShut {NoStop}%
\bibitem [{\citenamefont {Levine}\ \emph {et~al.}(1983)\citenamefont {Levine},
  \citenamefont {Libby},\ and\ \citenamefont {Pruisken}}]{Levine1983}%
  \BibitemOpen
  \bibfield  {author} {\bibinfo {author} {\bibfnamefont {H.}~\bibnamefont
  {Levine}}, \bibinfo {author} {\bibfnamefont {S.~B.}\ \bibnamefont {Libby}},\
  and\ \bibinfo {author} {\bibfnamefont {A.~M.~M.}\ \bibnamefont {Pruisken}},\
  }\bibfield  {title} {\bibinfo {title} {Electron delocalization by a magnetic
  field in two dimensions},\ }\href
  {https://doi.org/10.1103/PhysRevLett.51.1915} {\bibfield  {journal} {\bibinfo
   {journal} {Phys. Rev. Lett.}\ }\textbf {\bibinfo {volume} {51}},\ \bibinfo
  {pages} {1915} (\bibinfo {year} {1983})}\BibitemShut {NoStop}%
\bibitem [{\citenamefont {Khmel’nitskii}(1983)}]{Khmelnitskii1983}%
  \BibitemOpen
  \bibfield  {author} {\bibinfo {author} {\bibfnamefont {D.}~\bibnamefont
  {Khmel’nitskii}},\ }\bibfield  {title} {\bibinfo {title} {Quantization of
  {Hall} conductivity},\ }\href
  {http://jetpletters.ru/ps/0/article_22668.shtml} {\bibfield  {journal}
  {\bibinfo  {journal} {JETP Lett.}\ }\textbf {\bibinfo {volume} {38}},\
  \bibinfo {pages} {552} (\bibinfo {year} {1983})}\BibitemShut {NoStop}%
\bibitem [{\citenamefont {Pruisken}(1984)}]{pruisken1984localization}%
  \BibitemOpen
  \bibfield  {author} {\bibinfo {author} {\bibfnamefont {A.~M.~M.}\
  \bibnamefont {Pruisken}},\ }\bibfield  {title} {\bibinfo {title} {On
  localization in the theory of the quantized {Hall} effect: {A}
  two-dimensional realization of the $\theta$-vacuum},\ }\href
  {https://doi.org/10.1016/0550-3213(84)90101-9} {\bibfield  {journal}
  {\bibinfo  {journal} {Nucl. Phys. B}\ }\textbf {\bibinfo {volume} {235}},\
  \bibinfo {pages} {277} (\bibinfo {year} {1984})}\BibitemShut {NoStop}%
\bibitem [{\citenamefont {Pruisken}(1985)}]{Pruisken1985}%
  \BibitemOpen
  \bibfield  {author} {\bibinfo {author} {\bibfnamefont {A.~M.~M.}\
  \bibnamefont {Pruisken}},\ }\bibfield  {title} {\bibinfo {title} {Dilute
  instanton gas as the precursor of the integer quantum {Hall} effect},\ }\href
  {https://doi.org/10.1103/PhysRevB.32.2636} {\bibfield  {journal} {\bibinfo
  {journal} {Phys. Rev. B}\ }\textbf {\bibinfo {volume} {32}},\ \bibinfo
  {pages} {2636} (\bibinfo {year} {1985})}\BibitemShut {NoStop}%
\bibitem [{\citenamefont
  {Pruisken}(1987{\natexlab{a}})}]{pruisken1987quasiparticles}%
  \BibitemOpen
  \bibfield  {author} {\bibinfo {author} {\bibfnamefont {A.~M.~M.}\
  \bibnamefont {Pruisken}},\ }\bibfield  {title} {\bibinfo {title}
  {Quasiparticles in the theory of the integral quantum {Hall} effect {(I)}},\
  }\href {https://doi.org/10.1016/0550-3213(87)90363-4} {\bibfield  {journal}
  {\bibinfo  {journal} {Nucl. Phys. B}\ }\textbf {\bibinfo {volume} {285}},\
  \bibinfo {pages} {719} (\bibinfo {year} {1987}{\natexlab{a}})}\BibitemShut
  {NoStop}%
\bibitem [{\citenamefont
  {Pruisken}(1987{\natexlab{b}})}]{pruisken1987quasiparticlesB}%
  \BibitemOpen
  \bibfield  {author} {\bibinfo {author} {\bibfnamefont {A.~M.~M.}\
  \bibnamefont {Pruisken}},\ }\bibfield  {title} {\bibinfo {title}
  {Quasiparticles in the theory of the integral quantum {Hall} effect {(II)}.
  {Renormalization of the Hall} conductance or instanton angle theta},\ }\href
  {https://doi.org/10.1016/0550-3213(87)90178-7} {\bibfield  {journal}
  {\bibinfo  {journal} {Nucl. Phys. B}\ }\textbf {\bibinfo {volume} {290}},\
  \bibinfo {pages} {61} (\bibinfo {year} {1987}{\natexlab{b}})}\BibitemShut
  {NoStop}%
\bibitem [{\citenamefont {Pruisken}\ and\ \citenamefont
  {Baranov}(1995)}]{pruisken1995cracking}%
  \BibitemOpen
  \bibfield  {author} {\bibinfo {author} {\bibfnamefont {A.~M.~M.}\
  \bibnamefont {Pruisken}}\ and\ \bibinfo {author} {\bibfnamefont {M.~A.}\
  \bibnamefont {Baranov}},\ }\bibfield  {title} {\bibinfo {title} {Cracking
  {Coulomb} interactions in the quantum {Hall} regime},\ }\href
  {https://doi.org/10.1209/0295-5075/31/9/007} {\bibfield  {journal} {\bibinfo
  {journal} {Europhysics Lett.}\ }\textbf {\bibinfo {volume} {31}},\ \bibinfo
  {pages} {543} (\bibinfo {year} {1995})}\BibitemShut {NoStop}%
\bibitem [{\citenamefont {Pruisken}\ and\ \citenamefont
  {Burmistrov}(2007{\natexlab{a}})}]{Pruisken2007}%
  \BibitemOpen
  \bibfield  {author} {\bibinfo {author} {\bibfnamefont {A.~M.~M.}\
  \bibnamefont {Pruisken}}\ and\ \bibinfo {author} {\bibfnamefont {I.~S.}\
  \bibnamefont {Burmistrov}},\ }\bibfield  {title} {\bibinfo {title} {$\theta$
  renormalization, electron--electron interactions and super universality in
  the quantum hall regime},\ }\href
  {https://doi.org/https://doi.org/10.1016/j.aop.2006.11.007} {\bibfield
  {journal} {\bibinfo  {journal} {Ann. Phys. (N.Y.)}\ }\textbf {\bibinfo
  {volume} {322}},\ \bibinfo {pages} {1265} (\bibinfo {year}
  {2007}{\natexlab{a}})}\BibitemShut {NoStop}%
\bibitem [{\citenamefont {Zirnbauer}()}]{Zirnbauer1999}%
  \BibitemOpen
  \bibfield  {author} {\bibinfo {author} {\bibfnamefont {M.~R.}\ \bibnamefont
  {Zirnbauer}},\ }\href@noop {} {\bibinfo {title} {Conformal field theory of
  the integer quantum {Hall} plateau transition}},\ \bibinfo {howpublished}
  {\url{ https://doi.org/10.48550/arXiv.hep-th/9905054}}\BibitemShut {NoStop}%
\bibitem [{\citenamefont {Kettemann}\ and\ \citenamefont
  {Tsvelik}(1999)}]{Kettemann1999}%
  \BibitemOpen
  \bibfield  {author} {\bibinfo {author} {\bibfnamefont {S.}~\bibnamefont
  {Kettemann}}\ and\ \bibinfo {author} {\bibfnamefont {A.~M.}\ \bibnamefont
  {Tsvelik}},\ }\bibfield  {title} {\bibinfo {title} {Information about the
  integer quantum {Hall} transition extracted from the autocorrelation function
  of spectral determinants},\ }\href
  {https://doi.org/10.1103/PhysRevLett.82.3689} {\bibfield  {journal} {\bibinfo
   {journal} {Phys. Rev. Lett.}\ }\textbf {\bibinfo {volume} {82}},\ \bibinfo
  {pages} {3689} (\bibinfo {year} {1999})}\BibitemShut {NoStop}%
\bibitem [{\citenamefont {Bhaseen}\ \emph {et~al.}(2000)\citenamefont
  {Bhaseen}, \citenamefont {Kogan}, \citenamefont {Soloviev}, \citenamefont
  {Taniguchi},\ and\ \citenamefont {Tsvelik}}]{Bhaseen2000}%
  \BibitemOpen
  \bibfield  {author} {\bibinfo {author} {\bibfnamefont {M.~J.}\ \bibnamefont
  {Bhaseen}}, \bibinfo {author} {\bibfnamefont {I.~I.}\ \bibnamefont {Kogan}},
  \bibinfo {author} {\bibfnamefont {O.~A.}\ \bibnamefont {Soloviev}}, \bibinfo
  {author} {\bibfnamefont {N.}~\bibnamefont {Taniguchi}},\ and\ \bibinfo
  {author} {\bibfnamefont {A.~M.}\ \bibnamefont {Tsvelik}},\ }\bibfield
  {title} {\bibinfo {title} {Towards a field theory of the plateau transitions
  in the integer quantum {Hall} effect},\ }\href
  {https://doi.org/https://doi.org/10.1016/S0550-3213(00)00276-5} {\bibfield
  {journal} {\bibinfo  {journal} {Nucl. Phys. B}\ }\textbf {\bibinfo {volume}
  {580}},\ \bibinfo {pages} {688} (\bibinfo {year} {2000})}\BibitemShut
  {NoStop}%
\bibitem [{\citenamefont
  {Tsvelik}(2001)}]{tsvelik2001wavefunctionsstatisticsquantum}%
  \BibitemOpen
  \bibfield  {author} {\bibinfo {author} {\bibfnamefont {A.~M.}\ \bibnamefont
  {Tsvelik}},\ }\href@noop {} {\bibinfo {title} {Wave functions statistics at
  quantum {Hall} critical point}},\ \bibinfo {howpublished}
  {\url{https://arxiv.org/abs/cond-mat/0112008}} (\bibinfo {year}
  {2001})\BibitemShut {NoStop}%
\bibitem [{\citenamefont {Tsvelik}(2007)}]{Tsvelik2007}%
  \BibitemOpen
  \bibfield  {author} {\bibinfo {author} {\bibfnamefont {A.~M.}\ \bibnamefont
  {Tsvelik}},\ }\bibfield  {title} {\bibinfo {title} {Evidence for the
  {PSL(2|2) Wess-Zumino-Novikov-Witten} model as a model for the plateau
  transition in the quantum {Hall} effect: {Evaluation} of numerical
  simulations},\ }\href {https://doi.org/10.1103/PhysRevB.75.184201} {\bibfield
   {journal} {\bibinfo  {journal} {Phys. Rev. B}\ }\textbf {\bibinfo {volume}
  {75}},\ \bibinfo {pages} {184201} (\bibinfo {year} {2007})}\BibitemShut
  {NoStop}%
\bibitem [{\citenamefont {Zirnbauer}(2019)}]{Zirnbauer2019}%
  \BibitemOpen
  \bibfield  {author} {\bibinfo {author} {\bibfnamefont {M.~R.}\ \bibnamefont
  {Zirnbauer}},\ }\bibfield  {title} {\bibinfo {title} {The integer quantum
  {Hall} plateau transition is a current algebra after all},\ }\href
  {https://doi.org/https://doi.org/10.1016/j.nuclphysb.2019.02.017} {\bibfield
  {journal} {\bibinfo  {journal} {Nucl. Phys. B}\ }\textbf {\bibinfo {volume}
  {941}},\ \bibinfo {pages} {458} (\bibinfo {year} {2019})}\BibitemShut
  {NoStop}%
\bibitem [{\citenamefont {Wegner}(1980)}]{Wegner1980}%
  \BibitemOpen
  \bibfield  {author} {\bibinfo {author} {\bibfnamefont {F.}~\bibnamefont
  {Wegner}},\ }\bibfield  {title} {\bibinfo {title} {Inverse participation
  ratio in $2+\epsilon$ dimensions},\ }\href
  {http://link.springer.com/article/10.1007/BF01325284} {\bibfield  {journal}
  {\bibinfo  {journal} {Z. Phys.B}\ }\textbf {\bibinfo {volume} {36}},\
  \bibinfo {pages} {209} (\bibinfo {year} {1980})}\BibitemShut {NoStop}%
\bibitem [{\citenamefont {Castellani}\ and\ \citenamefont
  {Peliti}(1986)}]{Castellani1986}%
  \BibitemOpen
  \bibfield  {author} {\bibinfo {author} {\bibfnamefont {C.}~\bibnamefont
  {Castellani}}\ and\ \bibinfo {author} {\bibfnamefont {L.}~\bibnamefont
  {Peliti}},\ }\bibfield  {title} {\bibinfo {title} {Multifractal wavefunction
  at the localisation threshold},\ }\href
  {http://stacks.iop.org/0305-4470/19/i=8/a=004} {\bibfield  {journal}
  {\bibinfo  {journal} {J. Phys. A}\ }\textbf {\bibinfo {volume} {19}},\
  \bibinfo {pages} {L429} (\bibinfo {year} {1986})}\BibitemShut {NoStop}%
\bibitem [{\citenamefont {Lerner}(1988)}]{Lerner1988}%
  \BibitemOpen
  \bibfield  {author} {\bibinfo {author} {\bibfnamefont {I.~V.}\ \bibnamefont
  {Lerner}},\ }\bibfield  {title} {\bibinfo {title} {Distribution functions of
  current density and local density of states in disordered quantum
  conductors},\ }\href
  {http://www.sciencedirect.com/science/article/pii/0375960188910274}
  {\bibfield  {journal} {\bibinfo  {journal} {Phys. Lett. A}\ }\textbf
  {\bibinfo {volume} {133}},\ \bibinfo {pages} {253} (\bibinfo {year}
  {1988})}\BibitemShut {NoStop}%
\bibitem [{\citenamefont {H{$\mathrm{\ddot{o}}$}f}\ and\ \citenamefont
  {Wegner}(1986)}]{Wegner1986}%
  \BibitemOpen
  \bibfield  {author} {\bibinfo {author} {\bibfnamefont {D.}~\bibnamefont
  {H{$\mathrm{\ddot{o}}$}f}}\ and\ \bibinfo {author} {\bibfnamefont
  {F.}~\bibnamefont {Wegner}},\ }\bibfield  {title} {\bibinfo {title}
  {Calculation of anomalous dimensions for the nonlinear sigma model},\ }\href
  {http://www.sciencedirect.com/science/article/pii/0550321386905754}
  {\bibfield  {journal} {\bibinfo  {journal} {Nucl. Phys. B}\ }\textbf
  {\bibinfo {volume} {275}},\ \bibinfo {pages} {561} (\bibinfo {year}
  {1986})}\BibitemShut {NoStop}%
\bibitem [{\citenamefont {Gruzberg}\ \emph {et~al.}(2013)\citenamefont
  {Gruzberg}, \citenamefont {Mirlin},\ and\ \citenamefont
  {Zirnbauer}}]{Gruzberg2013}%
  \BibitemOpen
  \bibfield  {author} {\bibinfo {author} {\bibfnamefont {I.~A.}\ \bibnamefont
  {Gruzberg}}, \bibinfo {author} {\bibfnamefont {A.~D.}\ \bibnamefont
  {Mirlin}},\ and\ \bibinfo {author} {\bibfnamefont {M.~R.}\ \bibnamefont
  {Zirnbauer}},\ }\bibfield  {title} {\bibinfo {title} {Classification and
  symmetry properties of scaling dimensions at {Anderson} transitions},\ }\href
  {https://doi.org/10.1103/PhysRevB.87.125144} {\bibfield  {journal} {\bibinfo
  {journal} {Phys. Rev. B}\ }\textbf {\bibinfo {volume} {87}},\ \bibinfo
  {pages} {125144} (\bibinfo {year} {2013})}\BibitemShut {NoStop}%
\bibitem [{\citenamefont {Karcher}\ \emph
  {et~al.}(2022{\natexlab{a}})\citenamefont {Karcher}, \citenamefont
  {Gruzberg},\ and\ \citenamefont {Mirlin}}]{Karcher2022}%
  \BibitemOpen
  \bibfield  {author} {\bibinfo {author} {\bibfnamefont {J.~F.}\ \bibnamefont
  {Karcher}}, \bibinfo {author} {\bibfnamefont {I.~A.}\ \bibnamefont
  {Gruzberg}},\ and\ \bibinfo {author} {\bibfnamefont {A.~D.}\ \bibnamefont
  {Mirlin}},\ }\bibfield  {title} {\bibinfo {title} {Generalized
  multifractality at the spin quantum {Hall} transition: {Percolation} mapping
  and pure-scaling observables},\ }\href
  {https://doi.org/10.1103/PhysRevB.105.184205} {\bibfield  {journal} {\bibinfo
   {journal} {Phys. Rev. B}\ }\textbf {\bibinfo {volume} {105}},\ \bibinfo
  {pages} {184205} (\bibinfo {year} {2022}{\natexlab{a}})}\BibitemShut
  {NoStop}%
\bibitem [{\citenamefont {Karcher}\ \emph
  {et~al.}(2022{\natexlab{b}})\citenamefont {Karcher}, \citenamefont
  {Gruzberg},\ and\ \citenamefont {Mirlin}}]{Karcher2022b}%
  \BibitemOpen
  \bibfield  {author} {\bibinfo {author} {\bibfnamefont {J.~F.}\ \bibnamefont
  {Karcher}}, \bibinfo {author} {\bibfnamefont {I.~A.}\ \bibnamefont
  {Gruzberg}},\ and\ \bibinfo {author} {\bibfnamefont {A.~D.}\ \bibnamefont
  {Mirlin}},\ }\bibfield  {title} {\bibinfo {title} {Generalized
  multifractality at metal-insulator transitions and in metallic phases of
  two-dimensional disordered systems},\ }\href
  {https://doi.org/10.1103/PhysRevB.106.104202} {\bibfield  {journal} {\bibinfo
   {journal} {Phys. Rev. B}\ }\textbf {\bibinfo {volume} {106}},\ \bibinfo
  {pages} {104202} (\bibinfo {year} {2022}{\natexlab{b}})}\BibitemShut
  {NoStop}%
\bibitem [{\citenamefont {Karcher}\ \emph
  {et~al.}(2023{\natexlab{a}})\citenamefont {Karcher}, \citenamefont
  {Gruzberg},\ and\ \citenamefont {Mirlin}}]{Karcher2023}%
  \BibitemOpen
  \bibfield  {author} {\bibinfo {author} {\bibfnamefont {J.~F.}\ \bibnamefont
  {Karcher}}, \bibinfo {author} {\bibfnamefont {I.~A.}\ \bibnamefont
  {Gruzberg}},\ and\ \bibinfo {author} {\bibfnamefont {A.~D.}\ \bibnamefont
  {Mirlin}},\ }\bibfield  {title} {\bibinfo {title} {Generalized
  multifractality in two-dimensional disordered systems of chiral symmetry
  classes},\ }\href {https://doi.org/10.1103/PhysRevB.107.104202} {\bibfield
  {journal} {\bibinfo  {journal} {Phys. Rev. B}\ }\textbf {\bibinfo {volume}
  {107}},\ \bibinfo {pages} {104202} (\bibinfo {year}
  {2023}{\natexlab{a}})}\BibitemShut {NoStop}%
\bibitem [{\citenamefont {Mirlin}(2000)}]{Mirlin2000}%
  \BibitemOpen
  \bibfield  {author} {\bibinfo {author} {\bibfnamefont {A.~D.}\ \bibnamefont
  {Mirlin}},\ }\bibfield  {title} {\bibinfo {title} {Statistics of energy
  levels and eigenfunctions in disordered systems},\ }\href
  {http://www.sciencedirect.com/science/article/pii/S0370157399000915}
  {\bibfield  {journal} {\bibinfo  {journal} {Phys. Rep.}\ }\textbf {\bibinfo
  {volume} {326}},\ \bibinfo {pages} {259} (\bibinfo {year}
  {2000})}\BibitemShut {NoStop}%
\bibitem [{\citenamefont {Evers}\ and\ \citenamefont
  {Mirlin}(2008)}]{EversMirlin}%
  \BibitemOpen
  \bibfield  {author} {\bibinfo {author} {\bibfnamefont {F.}~\bibnamefont
  {Evers}}\ and\ \bibinfo {author} {\bibfnamefont {A.~D.}\ \bibnamefont
  {Mirlin}},\ }\bibfield  {title} {\bibinfo {title} {Anderson transitions},\
  }\href {http://link.aps.org/doi/10.1103/RevModPhys.80.1355} {\bibfield
  {journal} {\bibinfo  {journal} {Rev. Mod. Phys.}\ }\textbf {\bibinfo {volume}
  {80}},\ \bibinfo {pages} {1355} (\bibinfo {year} {2008})}\BibitemShut
  {NoStop}%
\bibitem [{\citenamefont {Bondesan}\ \emph {et~al.}(2017)\citenamefont
  {Bondesan}, \citenamefont {Wieczorek},\ and\ \citenamefont
  {Zirnbauer}}]{Bondesan2017}%
  \BibitemOpen
  \bibfield  {author} {\bibinfo {author} {\bibfnamefont {R.}~\bibnamefont
  {Bondesan}}, \bibinfo {author} {\bibfnamefont {D.}~\bibnamefont
  {Wieczorek}},\ and\ \bibinfo {author} {\bibfnamefont {M.}~\bibnamefont
  {Zirnbauer}},\ }\bibfield  {title} {\bibinfo {title} {Gaussian free fields at
  the integer quantum {Hall} plateau transition},\ }\href
  {https://doi.org/https://doi.org/10.1016/j.nuclphysb.2017.02.011} {\bibfield
  {journal} {\bibinfo  {journal} {Nucl. Phys. B}\ }\textbf {\bibinfo {volume}
  {918}},\ \bibinfo {pages} {52} (\bibinfo {year} {2017})}\BibitemShut
  {NoStop}%
\bibitem [{\citenamefont {Karcher}\ \emph {et~al.}(2021)\citenamefont
  {Karcher}, \citenamefont {Charles}, \citenamefont {Gruzberg},\ and\
  \citenamefont {Mirlin}}]{Karcher2021}%
  \BibitemOpen
  \bibfield  {author} {\bibinfo {author} {\bibfnamefont {J.~F.}\ \bibnamefont
  {Karcher}}, \bibinfo {author} {\bibfnamefont {N.}~\bibnamefont {Charles}},
  \bibinfo {author} {\bibfnamefont {I.~A.}\ \bibnamefont {Gruzberg}},\ and\
  \bibinfo {author} {\bibfnamefont {A.~D.}\ \bibnamefont {Mirlin}},\ }\bibfield
   {title} {\bibinfo {title} {Generalized multifractality at spin quantum
  {Hall} transition},\ }\href
  {https://doi.org/https://doi.org/10.1016/j.aop.2021.168584} {\bibfield
  {journal} {\bibinfo  {journal} {Ann. Phys. (N.Y.)}\ }\textbf {\bibinfo
  {volume} {435}},\ \bibinfo {pages} {168584} (\bibinfo {year} {2021})},\
  \bibinfo {note} {special issue on Philip W. Anderson}\BibitemShut {NoStop}%
\bibitem [{\citenamefont {Padayasi}\ and\ \citenamefont
  {Gruzberg}(2023)}]{Padayasi2023}%
  \BibitemOpen
  \bibfield  {author} {\bibinfo {author} {\bibfnamefont {J.}~\bibnamefont
  {Padayasi}}\ and\ \bibinfo {author} {\bibfnamefont {I.}~\bibnamefont
  {Gruzberg}},\ }\bibfield  {title} {\bibinfo {title} {Conformal invariance and
  multifractality at anderson transitions in arbitrary dimensions},\ }\href
  {https://doi.org/10.1103/PhysRevLett.131.266401} {\bibfield  {journal}
  {\bibinfo  {journal} {Phys. Rev. Lett.}\ }\textbf {\bibinfo {volume} {131}},\
  \bibinfo {pages} {266401} (\bibinfo {year} {2023})}\BibitemShut {NoStop}%
\bibitem [{\citenamefont {Obuse}\ \emph {et~al.}(2008)\citenamefont {Obuse},
  \citenamefont {Subramaniam}, \citenamefont {Furusaki}, \citenamefont
  {Gruzberg},\ and\ \citenamefont {Ludwig}}]{Obuse2008}%
  \BibitemOpen
  \bibfield  {author} {\bibinfo {author} {\bibfnamefont {H.}~\bibnamefont
  {Obuse}}, \bibinfo {author} {\bibfnamefont {A.~R.}\ \bibnamefont
  {Subramaniam}}, \bibinfo {author} {\bibfnamefont {A.}~\bibnamefont
  {Furusaki}}, \bibinfo {author} {\bibfnamefont {I.~A.}\ \bibnamefont
  {Gruzberg}},\ and\ \bibinfo {author} {\bibfnamefont {A.~W.~W.}\ \bibnamefont
  {Ludwig}},\ }\bibfield  {title} {\bibinfo {title} {Boundary multifractality
  at the integer quantum {Hall} plateau transition: {Implications} for the
  critical theory},\ }\href {https://doi.org/10.1103/PhysRevLett.101.116802}
  {\bibfield  {journal} {\bibinfo  {journal} {Phys. Rev. Lett.}\ }\textbf
  {\bibinfo {volume} {101}},\ \bibinfo {pages} {116802} (\bibinfo {year}
  {2008})}\BibitemShut {NoStop}%
\bibitem [{\citenamefont {Gruzberg}\ \emph {et~al.}(2017)\citenamefont
  {Gruzberg}, \citenamefont {Kl\"umper}, \citenamefont {Nuding},\ and\
  \citenamefont {Sedrakyan}}]{Sedrakyan2017}%
  \BibitemOpen
  \bibfield  {author} {\bibinfo {author} {\bibfnamefont {I.~A.}\ \bibnamefont
  {Gruzberg}}, \bibinfo {author} {\bibfnamefont {A.}~\bibnamefont {Kl\"umper}},
  \bibinfo {author} {\bibfnamefont {W.}~\bibnamefont {Nuding}},\ and\ \bibinfo
  {author} {\bibfnamefont {A.}~\bibnamefont {Sedrakyan}},\ }\bibfield  {title}
  {\bibinfo {title} {Geometrically disordered network models, quenched quantum
  gravity, and critical behavior at quantum {Hall} plateau transitions},\
  }\href {https://doi.org/10.1103/PhysRevB.95.125414} {\bibfield  {journal}
  {\bibinfo  {journal} {Phys. Rev. B}\ }\textbf {\bibinfo {volume} {95}},\
  \bibinfo {pages} {125414} (\bibinfo {year} {2017})}\BibitemShut {NoStop}%
\bibitem [{\citenamefont {Kl\"umper}\ \emph {et~al.}(2019)\citenamefont
  {Kl\"umper}, \citenamefont {Nuding},\ and\ \citenamefont
  {Sedrakyan}}]{Sedrakyan2019}%
  \BibitemOpen
  \bibfield  {author} {\bibinfo {author} {\bibfnamefont {A.}~\bibnamefont
  {Kl\"umper}}, \bibinfo {author} {\bibfnamefont {W.}~\bibnamefont {Nuding}},\
  and\ \bibinfo {author} {\bibfnamefont {A.}~\bibnamefont {Sedrakyan}},\
  }\bibfield  {title} {\bibinfo {title} {Random network models with variable
  disorder of geometry},\ }\href {https://doi.org/10.1103/PhysRevB.100.140201}
  {\bibfield  {journal} {\bibinfo  {journal} {Phys. Rev. B}\ }\textbf {\bibinfo
  {volume} {100}},\ \bibinfo {pages} {140201(R)} (\bibinfo {year}
  {2019})}\BibitemShut {NoStop}%
\bibitem [{\citenamefont {Conti}\ \emph {et~al.}(2021)\citenamefont {Conti},
  \citenamefont {Topchyan}, \citenamefont {R.Tateo},\ and\ \citenamefont
  {Sedrakyan}}]{Sedrakyan2021}%
  \BibitemOpen
  \bibfield  {author} {\bibinfo {author} {\bibfnamefont {R.}~\bibnamefont
  {Conti}}, \bibinfo {author} {\bibfnamefont {H.}~\bibnamefont {Topchyan}},
  \bibinfo {author} {\bibnamefont {R.Tateo}},\ and\ \bibinfo {author}
  {\bibfnamefont {A.}~\bibnamefont {Sedrakyan}},\ }\bibfield  {title} {\bibinfo
  {title} {{Geometry of random potentials: Induction of two-dimensional gravity
  in quantum Hall plateau transitions}},\ }\href
  {https://doi.org/10.1103/PhysRevB.103.L041302} {\bibfield  {journal}
  {\bibinfo  {journal} {Phys. Rev. B}\ }\textbf {\bibinfo {volume} {103}},\
  \bibinfo {pages} {L041302} (\bibinfo {year} {2021})}\BibitemShut {NoStop}%
\bibitem [{\citenamefont {Dresselhaus}\ \emph {et~al.}(2022)\citenamefont
  {Dresselhaus}, \citenamefont {Sbierski},\ and\ \citenamefont
  {Gruzberg}}]{Dresselhaus2022}%
  \BibitemOpen
  \bibfield  {author} {\bibinfo {author} {\bibfnamefont {E.~J.}\ \bibnamefont
  {Dresselhaus}}, \bibinfo {author} {\bibfnamefont {B.}~\bibnamefont
  {Sbierski}},\ and\ \bibinfo {author} {\bibfnamefont {I.~A.}\ \bibnamefont
  {Gruzberg}},\ }\bibfield  {title} {\bibinfo {title} {Scaling collapse of
  longitudinal conductance near the integer quantum {Hall} transition},\ }\href
  {https://doi.org/10.1103/PhysRevLett.129.026801} {\bibfield  {journal}
  {\bibinfo  {journal} {Phys. Rev. Lett.}\ }\textbf {\bibinfo {volume} {129}},\
  \bibinfo {pages} {026801} (\bibinfo {year} {2022})}\BibitemShut {NoStop}%
\bibitem [{\citenamefont {Topchyan}\ \emph {et~al.}(2024)\citenamefont
  {Topchyan}, \citenamefont {Gruzberg}, \citenamefont {Nuding}, \citenamefont
  {Kl\"umper},\ and\ \citenamefont {Sedrakyan}}]{Topchyan2024}%
  \BibitemOpen
  \bibfield  {author} {\bibinfo {author} {\bibfnamefont {H.}~\bibnamefont
  {Topchyan}}, \bibinfo {author} {\bibfnamefont {I.}~\bibnamefont {Gruzberg}},
  \bibinfo {author} {\bibfnamefont {W.}~\bibnamefont {Nuding}}, \bibinfo
  {author} {\bibfnamefont {A.}~\bibnamefont {Kl\"umper}},\ and\ \bibinfo
  {author} {\bibfnamefont {A.}~\bibnamefont {Sedrakyan}},\ }\bibfield  {title}
  {\bibinfo {title} {The integer quantum {Hall} transition: an {S-matrix}
  approach to random networks},\ }\bibfield  {journal} {\bibinfo  {journal}
  {arXiv:2407.04132}\ }\href {https://doi.org/10.48550/arXiv.2407.04132}
  {10.48550/arXiv.2407.04132} (\bibinfo {year} {2024})\BibitemShut {NoStop}%
\bibitem [{\citenamefont {Volovik}(1997)}]{Volovik1997}%
  \BibitemOpen
  \bibfield  {author} {\bibinfo {author} {\bibfnamefont {G.}~\bibnamefont
  {Volovik}},\ }\bibfield  {title} {\bibinfo {title} {On edge states in
  superconductors with time inversion symmetry breaking},\ }\href@noop {}
  {\bibfield  {journal} {\bibinfo  {journal} {JETP Lett.}\ }\textbf {\bibinfo
  {volume} {66}},\ \bibinfo {pages} {522} (\bibinfo {year} {1997})}\BibitemShut
  {NoStop}%
\bibitem [{\citenamefont {Kagalovsky}\ \emph {et~al.}(1999)\citenamefont
  {Kagalovsky}, \citenamefont {Horovitz}, \citenamefont {Avishai},\ and\
  \citenamefont {Chalker}}]{Kagolovsky1999}%
  \BibitemOpen
  \bibfield  {author} {\bibinfo {author} {\bibfnamefont {V.}~\bibnamefont
  {Kagalovsky}}, \bibinfo {author} {\bibfnamefont {B.}~\bibnamefont
  {Horovitz}}, \bibinfo {author} {\bibfnamefont {Y.}~\bibnamefont {Avishai}},\
  and\ \bibinfo {author} {\bibfnamefont {J.~T.}\ \bibnamefont {Chalker}},\
  }\bibfield  {title} {\bibinfo {title} {Quantum {Hall} plateau transitions in
  disordered superconductors},\ }\href
  {https://doi.org/10.1103/PhysRevLett.82.3516} {\bibfield  {journal} {\bibinfo
   {journal} {Phys. Rev. Lett.}\ }\textbf {\bibinfo {volume} {82}},\ \bibinfo
  {pages} {3516} (\bibinfo {year} {1999})}\BibitemShut {NoStop}%
\bibitem [{\citenamefont {Senthil}\ \emph {et~al.}(1999)\citenamefont
  {Senthil}, \citenamefont {Marston},\ and\ \citenamefont
  {Fisher}}]{Senthil1999}%
  \BibitemOpen
  \bibfield  {author} {\bibinfo {author} {\bibfnamefont {T.}~\bibnamefont
  {Senthil}}, \bibinfo {author} {\bibfnamefont {J.~B.}\ \bibnamefont
  {Marston}},\ and\ \bibinfo {author} {\bibfnamefont {M.~P.~A.}\ \bibnamefont
  {Fisher}},\ }\bibfield  {title} {\bibinfo {title} {Spin quantum {Hall} effect
  in unconventional superconductors},\ }\href
  {https://doi.org/10.1103/PhysRevB.60.4245} {\bibfield  {journal} {\bibinfo
  {journal} {Phys. Rev. B}\ }\textbf {\bibinfo {volume} {60}},\ \bibinfo
  {pages} {4245} (\bibinfo {year} {1999})}\BibitemShut {NoStop}%
\bibitem [{\citenamefont {Volovik}\ and\ \citenamefont
  {Yakovenko}(1989)}]{GEVolovik_1989}%
  \BibitemOpen
  \bibfield  {author} {\bibinfo {author} {\bibfnamefont {G.~E.}\ \bibnamefont
  {Volovik}}\ and\ \bibinfo {author} {\bibfnamefont {V.~M.}\ \bibnamefont
  {Yakovenko}},\ }\bibfield  {title} {\bibinfo {title} {Fractional charge, spin
  and statistics of solitons in superfluid {$^3$He} film},\ }\href
  {https://doi.org/10.1088/0953-8984/1/31/025} {\bibfield  {journal} {\bibinfo
  {journal} {J. Phys.: Condens. Matter}\ }\textbf {\bibinfo {volume} {1}},\
  \bibinfo {pages} {5263} (\bibinfo {year} {1989})}\BibitemShut {NoStop}%
\bibitem [{\citenamefont {Volovik}\ \emph {et~al.}(1989)\citenamefont
  {Volovik}, \citenamefont {Solov'ev},\ and\ \citenamefont
  {Yakovenko}}]{GEVolovik_1989_2}%
  \BibitemOpen
  \bibfield  {author} {\bibinfo {author} {\bibfnamefont {G.~E.}\ \bibnamefont
  {Volovik}}, \bibinfo {author} {\bibfnamefont {A.}~\bibnamefont {Solov'ev}},\
  and\ \bibinfo {author} {\bibfnamefont {V.~M.}\ \bibnamefont {Yakovenko}},\
  }\bibfield  {title} {\bibinfo {title} {Spin and statistics of soliton in a
  superfluid {$^3$He-A} film},\ }\href
  {http://jetpletters.ru/ps/0/article_16831.shtml} {\bibfield  {journal}
  {\bibinfo  {journal} {Pis'ma Zh. Eksp. Teor. Fiz.}\ }\textbf {\bibinfo
  {volume} {49}},\ \bibinfo {pages} {55} (\bibinfo {year} {1989})}\BibitemShut
  {NoStop}%
\bibitem [{\citenamefont {Gruzberg}\ \emph {et~al.}(1999)\citenamefont
  {Gruzberg}, \citenamefont {Ludwig},\ and\ \citenamefont
  {Read}}]{Gruzberg1999}%
  \BibitemOpen
  \bibfield  {author} {\bibinfo {author} {\bibfnamefont {I.~A.}\ \bibnamefont
  {Gruzberg}}, \bibinfo {author} {\bibfnamefont {A.~W.~W.}\ \bibnamefont
  {Ludwig}},\ and\ \bibinfo {author} {\bibfnamefont {N.}~\bibnamefont {Read}},\
  }\bibfield  {title} {\bibinfo {title} {Exact exponents for the spin quantum
  {Hall} transition},\ }\href {https://doi.org/10.1103/PhysRevLett.82.4524}
  {\bibfield  {journal} {\bibinfo  {journal} {Phys. Rev. Lett.}\ }\textbf
  {\bibinfo {volume} {82}},\ \bibinfo {pages} {4524} (\bibinfo {year}
  {1999})}\BibitemShut {NoStop}%
\bibitem [{\citenamefont {Beamond}\ \emph {et~al.}(2002)\citenamefont
  {Beamond}, \citenamefont {Cardy},\ and\ \citenamefont
  {Chalker}}]{Beamond2002}%
  \BibitemOpen
  \bibfield  {author} {\bibinfo {author} {\bibfnamefont {E.~J.}\ \bibnamefont
  {Beamond}}, \bibinfo {author} {\bibfnamefont {J.}~\bibnamefont {Cardy}},\
  and\ \bibinfo {author} {\bibfnamefont {J.~T.}\ \bibnamefont {Chalker}},\
  }\bibfield  {title} {\bibinfo {title} {Quantum and classical localization,
  the spin quantum {Hall} effect, and generalizations},\ }\href
  {https://doi.org/10.1103/PhysRevB.65.214301} {\bibfield  {journal} {\bibinfo
  {journal} {Phys. Rev. B}\ }\textbf {\bibinfo {volume} {65}},\ \bibinfo
  {pages} {214301} (\bibinfo {year} {2002})}\BibitemShut {NoStop}%
\bibitem [{\citenamefont {Mirlin}\ \emph {et~al.}(2003)\citenamefont {Mirlin},
  \citenamefont {Evers},\ and\ \citenamefont {Mildenberger}}]{Mirlin2003}%
  \BibitemOpen
  \bibfield  {author} {\bibinfo {author} {\bibfnamefont {A.~D.}\ \bibnamefont
  {Mirlin}}, \bibinfo {author} {\bibfnamefont {F.}~\bibnamefont {Evers}},\ and\
  \bibinfo {author} {\bibfnamefont {A.}~\bibnamefont {Mildenberger}},\
  }\bibfield  {title} {\bibinfo {title} {Wavefunction statistics and
  multifractality at the spin quantum {Hall} transition},\ }\href
  {https://doi.org/10.1088/0305-4470/36/12/323} {\bibfield  {journal} {\bibinfo
   {journal} {J. Phys. A: Math. and Gen.}\ }\textbf {\bibinfo {volume} {36}},\
  \bibinfo {pages} {3255} (\bibinfo {year} {2003})}\BibitemShut {NoStop}%
\bibitem [{\citenamefont {Evers}\ \emph {et~al.}(2003)\citenamefont {Evers},
  \citenamefont {Mildenberger},\ and\ \citenamefont {Mirlin}}]{Evers2003}%
  \BibitemOpen
  \bibfield  {author} {\bibinfo {author} {\bibfnamefont {F.}~\bibnamefont
  {Evers}}, \bibinfo {author} {\bibfnamefont {A.}~\bibnamefont
  {Mildenberger}},\ and\ \bibinfo {author} {\bibfnamefont {A.~D.}\ \bibnamefont
  {Mirlin}},\ }\bibfield  {title} {\bibinfo {title} {Multifractality at the
  spin quantum {Hall} transition},\ }\href
  {https://doi.org/10.1103/PhysRevB.67.041303} {\bibfield  {journal} {\bibinfo
  {journal} {Phys. Rev. B}\ }\textbf {\bibinfo {volume} {67}},\ \bibinfo
  {pages} {041303} (\bibinfo {year} {2003})}\BibitemShut {NoStop}%
\bibitem [{\citenamefont {Subramaniam}\ \emph {et~al.}(2008)\citenamefont
  {Subramaniam}, \citenamefont {Gruzberg},\ and\ \citenamefont
  {Ludwig}}]{Subramaniam2008}%
  \BibitemOpen
  \bibfield  {author} {\bibinfo {author} {\bibfnamefont {A.~R.}\ \bibnamefont
  {Subramaniam}}, \bibinfo {author} {\bibfnamefont {I.~A.}\ \bibnamefont
  {Gruzberg}},\ and\ \bibinfo {author} {\bibfnamefont {A.~W.~W.}\ \bibnamefont
  {Ludwig}},\ }\bibfield  {title} {\bibinfo {title} {Boundary criticality and
  multifractality at the two-dimensional spin quantum {Hall} transition},\
  }\href {https://doi.org/10.1103/PhysRevB.78.245105} {\bibfield  {journal}
  {\bibinfo  {journal} {Phys. Rev. B}\ }\textbf {\bibinfo {volume} {78}},\
  \bibinfo {pages} {245105} (\bibinfo {year} {2008})}\BibitemShut {NoStop}%
\bibitem [{\citenamefont {Puschmann}\ \emph {et~al.}(2021)\citenamefont
  {Puschmann}, \citenamefont {Hernang\'omez-P\'erez}, \citenamefont {Lang},
  \citenamefont {Bera},\ and\ \citenamefont {Evers}}]{Puschmann2021}%
  \BibitemOpen
  \bibfield  {author} {\bibinfo {author} {\bibfnamefont {M.}~\bibnamefont
  {Puschmann}}, \bibinfo {author} {\bibfnamefont {D.}~\bibnamefont
  {Hernang\'omez-P\'erez}}, \bibinfo {author} {\bibfnamefont {B.}~\bibnamefont
  {Lang}}, \bibinfo {author} {\bibfnamefont {S.}~\bibnamefont {Bera}},\ and\
  \bibinfo {author} {\bibfnamefont {F.}~\bibnamefont {Evers}},\ }\bibfield
  {title} {\bibinfo {title} {Quartic multifractality and finite-size
  corrections at the spin quantum {Hall} transition},\ }\href
  {https://doi.org/10.1103/PhysRevB.103.235167} {\bibfield  {journal} {\bibinfo
   {journal} {Phys. Rev. B}\ }\textbf {\bibinfo {volume} {103}},\ \bibinfo
  {pages} {235167} (\bibinfo {year} {2021})}\BibitemShut {NoStop}%
\bibitem [{\citenamefont {Karcher}\ \emph
  {et~al.}(2023{\natexlab{b}})\citenamefont {Karcher}, \citenamefont
  {Gruzberg},\ and\ \citenamefont {Mirlin}}]{Karcher2023a}%
  \BibitemOpen
  \bibfield  {author} {\bibinfo {author} {\bibfnamefont {J.~F.}\ \bibnamefont
  {Karcher}}, \bibinfo {author} {\bibfnamefont {I.~A.}\ \bibnamefont
  {Gruzberg}},\ and\ \bibinfo {author} {\bibfnamefont {A.~D.}\ \bibnamefont
  {Mirlin}},\ }\bibfield  {title} {\bibinfo {title} {Metal-insulator transition
  in a two-dimensional system of chiral unitary class},\ }\href
  {https://doi.org/10.1103/PhysRevB.107.L020201} {\bibfield  {journal}
  {\bibinfo  {journal} {Phys. Rev. B}\ }\textbf {\bibinfo {volume} {107}},\
  \bibinfo {pages} {L020201} (\bibinfo {year}
  {2023}{\natexlab{b}})}\BibitemShut {NoStop}%
\bibitem [{\citenamefont
  {Zirnbauer}(2024)}]{zirnbauer2024infraredlimito3nonlinear}%
  \BibitemOpen
  \bibfield  {author} {\bibinfo {author} {\bibfnamefont {M.~R.}\ \bibnamefont
  {Zirnbauer}},\ }\href@noop {} {\bibinfo {title} {On the infrared limit of the
  {O(3) nonlinear $\sigma$-model at $\theta = \pi$}}},\ \bibinfo {howpublished}
  {\url{https://arxiv.org/abs/2408.12215}} (\bibinfo {year} {2024})\BibitemShut
  {NoStop}%
\bibitem [{\citenamefont {Pauli}\ and\ \citenamefont {Villars}(1949)}]{PV1949}%
  \BibitemOpen
  \bibfield  {author} {\bibinfo {author} {\bibfnamefont {W.}~\bibnamefont
  {Pauli}}\ and\ \bibinfo {author} {\bibfnamefont {F.}~\bibnamefont
  {Villars}},\ }\bibfield  {title} {\bibinfo {title} {On the invariant
  regularization in relativistic quantum theory},\ }\href
  {https://doi.org/10.1103/RevModPhys.21.434} {\bibfield  {journal} {\bibinfo
  {journal} {Rev. Mod. Phys.}\ }\textbf {\bibinfo {volume} {21}},\ \bibinfo
  {pages} {434} (\bibinfo {year} {1949})}\BibitemShut {NoStop}%
\bibitem [{\citenamefont {Jeng}\ \emph
  {et~al.}(2001{\natexlab{a}})\citenamefont {Jeng}, \citenamefont {Ludwig},
  \citenamefont {Senthil},\ and\ \citenamefont {Chamon}}]{Jeng2001a}%
  \BibitemOpen
  \bibfield  {author} {\bibinfo {author} {\bibfnamefont {M.}~\bibnamefont
  {Jeng}}, \bibinfo {author} {\bibfnamefont {A.~W.~W.}\ \bibnamefont {Ludwig}},
  \bibinfo {author} {\bibfnamefont {T.}~\bibnamefont {Senthil}},\ and\ \bibinfo
  {author} {\bibfnamefont {C.}~\bibnamefont {Chamon}},\ }\bibfield  {title}
  {\bibinfo {title} {Interaction effects on quasiparticle localization in dirty
  superconductors},\ }\href@noop {} {\bibfield  {journal} {\bibinfo  {journal}
  {Bull. Am. Phys. Soc.}\ }\textbf {\bibinfo {volume} {46}},\ \bibinfo {pages}
  {231} (\bibinfo {year} {2001}{\natexlab{a}})}\BibitemShut {NoStop}%
\bibitem [{\citenamefont {Jeng}\ \emph
  {et~al.}(2001{\natexlab{b}})\citenamefont {Jeng}, \citenamefont {Ludwig},
  \citenamefont {Senthil},\ and\ \citenamefont {Chamon}}]{Jeng2001}%
  \BibitemOpen
  \bibfield  {author} {\bibinfo {author} {\bibfnamefont {M.}~\bibnamefont
  {Jeng}}, \bibinfo {author} {\bibfnamefont {A.~W.~W.}\ \bibnamefont {Ludwig}},
  \bibinfo {author} {\bibfnamefont {T.}~\bibnamefont {Senthil}},\ and\ \bibinfo
  {author} {\bibfnamefont {C.}~\bibnamefont {Chamon}},\ }\href@noop {}
  {\bibinfo {title} {Interaction effects on quasiparticle localization in dirty
  superconductors}},\ \bibinfo {howpublished} {\url{
  https://doi.org/10.48550/arXiv.cond-mat/0112044}} (\bibinfo {year}
  {2001}{\natexlab{b}})\BibitemShut {NoStop}%
\bibitem [{\citenamefont {Dell'Anna}(2006)}]{DellAnna2006}%
  \BibitemOpen
  \bibfield  {author} {\bibinfo {author} {\bibfnamefont {L.}~\bibnamefont
  {Dell'Anna}},\ }\bibfield  {title} {\bibinfo {title} {Disordered d-wave
  superconductors with interactions},\ }\href
  {https://doi.org/https://doi.org/10.1016/j.nuclphysb.2006.09.024} {\bibfield
  {journal} {\bibinfo  {journal} {Nucl. Phys. B}\ }\textbf {\bibinfo {volume}
  {758}},\ \bibinfo {pages} {255} (\bibinfo {year} {2006})}\BibitemShut
  {NoStop}%
\bibitem [{\citenamefont {Liao}\ \emph {et~al.}(2017)\citenamefont {Liao},
  \citenamefont {Levchenko},\ and\ \citenamefont {Foster}}]{Liao2017}%
  \BibitemOpen
  \bibfield  {author} {\bibinfo {author} {\bibfnamefont {Y.}~\bibnamefont
  {Liao}}, \bibinfo {author} {\bibfnamefont {A.}~\bibnamefont {Levchenko}},\
  and\ \bibinfo {author} {\bibfnamefont {M.~S.}\ \bibnamefont {Foster}},\
  }\bibfield  {title} {\bibinfo {title} {Response theory of the ergodic
  many-body delocalized phase: {Keldysh Finkel'stein} sigma models and the
  10-fold way},\ }\href
  {https://doi.org/https://doi.org/10.1016/j.aop.2017.08.020} {\bibfield
  {journal} {\bibinfo  {journal} {Ann. Phys. (N.Y.)}\ }\textbf {\bibinfo
  {volume} {386}},\ \bibinfo {pages} {97} (\bibinfo {year} {2017})}\BibitemShut
  {NoStop}%
\bibitem [{\citenamefont {Babkin}\ and\ \citenamefont
  {Burmistrov}(2022)}]{Babkin2022}%
  \BibitemOpen
  \bibfield  {author} {\bibinfo {author} {\bibfnamefont {S.~S.}\ \bibnamefont
  {Babkin}}\ and\ \bibinfo {author} {\bibfnamefont {I.~S.}\ \bibnamefont
  {Burmistrov}},\ }\bibfield  {title} {\bibinfo {title} {Generalized
  multifractality in the spin quantum hall symmetry class with interaction},\
  }\href {https://doi.org/10.1103/PhysRevB.106.125424} {\bibfield  {journal}
  {\bibinfo  {journal} {Phys. Rev. B}\ }\textbf {\bibinfo {volume} {106}},\
  \bibinfo {pages} {125424} (\bibinfo {year} {2022})}\BibitemShut {NoStop}%
\bibitem [{\citenamefont {Pruisken}\ and\ \citenamefont
  {Burmistrov}(2005{\natexlab{a}})}]{Pruisken2005}%
  \BibitemOpen
  \bibfield  {author} {\bibinfo {author} {\bibfnamefont {A.}~\bibnamefont
  {Pruisken}}\ and\ \bibinfo {author} {\bibfnamefont {I.}~\bibnamefont
  {Burmistrov}},\ }\bibfield  {title} {\bibinfo {title} {The instanton vacuum
  of generalized {CP$^{N-1}$} models},\ }\href
  {https://doi.org/https://doi.org/10.1016/j.aop.2004.08.009} {\bibfield
  {journal} {\bibinfo  {journal} {Ann. Phys. (N.Y.)}\ }\textbf {\bibinfo
  {volume} {316}},\ \bibinfo {pages} {285} (\bibinfo {year}
  {2005}{\natexlab{a}})}\BibitemShut {NoStop}%
\bibitem [{\citenamefont {’t Hooft}(1976)}]{tHooft1976}%
  \BibitemOpen
  \bibfield  {author} {\bibinfo {author} {\bibfnamefont {G.}~\bibnamefont {’t
  Hooft}},\ }\bibfield  {title} {\bibinfo {title} {Computation of the quantum
  effects due to a four-dimensional pseudoparticle},\ }\href
  {https://doi.org/10.1103/PhysRevD.14.3432} {\bibfield  {journal} {\bibinfo
  {journal} {Phys. Rev. D}\ }\textbf {\bibinfo {volume} {14}},\ \bibinfo
  {pages} {3432} (\bibinfo {year} {1976})}\BibitemShut {NoStop}%
\bibitem [{\citenamefont {Pruisken}\ and\ \citenamefont
  {Burmistrov}(2007{\natexlab{b}})}]{pruisken2007theta}%
  \BibitemOpen
  \bibfield  {author} {\bibinfo {author} {\bibfnamefont {A.~M.~M.}\
  \bibnamefont {Pruisken}}\ and\ \bibinfo {author} {\bibfnamefont {I.~S.}\
  \bibnamefont {Burmistrov}},\ }\bibfield  {title} {\bibinfo {title} {$\theta$
  renormalization, electron--electron interactions and super universality in
  the quantum {Hall} regime},\ }\href
  {https://doi.org/10.1016/j.aop.2006.11.007} {\bibfield  {journal} {\bibinfo
  {journal} {Ann. Phys. (N.Y.)}\ }\textbf {\bibinfo {volume} {322}},\ \bibinfo
  {pages} {1265} (\bibinfo {year} {2007}{\natexlab{b}})}\BibitemShut {NoStop}%
\bibitem [{\citenamefont {Pruisken}\ and\ \citenamefont
  {Burmistrov}(2005{\natexlab{b}})}]{pruisken2005instanton}%
  \BibitemOpen
  \bibfield  {author} {\bibinfo {author} {\bibfnamefont {A.~M.~M.}\
  \bibnamefont {Pruisken}}\ and\ \bibinfo {author} {\bibfnamefont {I.~S.}\
  \bibnamefont {Burmistrov}},\ }\bibfield  {title} {\bibinfo {title} {The
  instanton vacuum of generalized $\text{CP}^{N-1}$ models},\ }\href
  {https://doi.org/10.1016/j.aop.2004.08.009} {\bibfield  {journal} {\bibinfo
  {journal} {Ann. Phys. (N.Y.)}\ }\textbf {\bibinfo {volume} {316}},\ \bibinfo
  {pages} {285} (\bibinfo {year} {2005}{\natexlab{b}})}\BibitemShut {NoStop}%
\bibitem [{\citenamefont {Marinov}(1980)}]{marinov1980invariant}%
  \BibitemOpen
  \bibfield  {author} {\bibinfo {author} {\bibfnamefont {M.}~\bibnamefont
  {Marinov}},\ }\bibfield  {title} {\bibinfo {title} {Invariant volumes of
  compact groups},\ }\href {https://doi.org/10.1088/0305-4470/13/11/009}
  {\bibfield  {journal} {\bibinfo  {journal} {Journal of Physics A:
  Mathematical and General}\ }\textbf {\bibinfo {volume} {13}},\ \bibinfo
  {pages} {3357} (\bibinfo {year} {1980})}\BibitemShut {NoStop}%
\bibitem [{\citenamefont {Boya}\ \emph {et~al.}(2003)\citenamefont {Boya},
  \citenamefont {Sudarshan},\ and\ \citenamefont {Tilma}}]{boya2003volumes}%
  \BibitemOpen
  \bibfield  {author} {\bibinfo {author} {\bibfnamefont {L.~J.}\ \bibnamefont
  {Boya}}, \bibinfo {author} {\bibfnamefont {E.}~\bibnamefont {Sudarshan}},\
  and\ \bibinfo {author} {\bibfnamefont {T.}~\bibnamefont {Tilma}},\ }\bibfield
   {title} {\bibinfo {title} {Volumes of compact manifolds},\ }\href
  {https://doi.org/10.1016/S0034-4877(03)80038-1} {\bibfield  {journal}
  {\bibinfo  {journal} {Reports on Mathematical Physics}\ }\textbf {\bibinfo
  {volume} {52}},\ \bibinfo {pages} {401} (\bibinfo {year} {2003})}\BibitemShut
  {NoStop}%
\bibitem [{\citenamefont {Morris}\ \emph
  {et~al.}(1985{\natexlab{a}})\citenamefont {Morris}, \citenamefont {Ross},\
  and\ \citenamefont {Sachrajda}}]{Morris1985a}%
  \BibitemOpen
  \bibfield  {author} {\bibinfo {author} {\bibfnamefont {T.~R.}\ \bibnamefont
  {Morris}}, \bibinfo {author} {\bibfnamefont {D.~A.}\ \bibnamefont {Ross}},\
  and\ \bibinfo {author} {\bibfnamefont {C.~T.}\ \bibnamefont {Sachrajda}},\
  }\bibfield  {title} {\bibinfo {title} {Higher-order quantum corrections in
  the presence of an instanton background field},\ }\href
  {https://doi.org/10.1016/0550-3213(85)90131-2} {\bibfield  {journal}
  {\bibinfo  {journal} {Nuclear Physics B}\ }\textbf {\bibinfo {volume}
  {255}},\ \bibinfo {pages} {115} (\bibinfo {year}
  {1985}{\natexlab{a}})}\BibitemShut {NoStop}%
\bibitem [{\citenamefont {Morris}\ \emph
  {et~al.}(1985{\natexlab{b}})\citenamefont {Morris}, \citenamefont {Ross},\
  and\ \citenamefont {Sachrajda}}]{Morris1985b}%
  \BibitemOpen
  \bibfield  {author} {\bibinfo {author} {\bibfnamefont {T.~R.}\ \bibnamefont
  {Morris}}, \bibinfo {author} {\bibfnamefont {D.~A.}\ \bibnamefont {Ross}},\
  and\ \bibinfo {author} {\bibfnamefont {C.~T.}\ \bibnamefont {Sachrajda}},\
  }\bibfield  {title} {\bibinfo {title} {Instanton calculus and the
  $\beta$-function in supersymmetric yang-mills theories},\ }\href
  {https://doi.org/10.1016/0370-2693(85)90960-8} {\bibfield  {journal}
  {\bibinfo  {journal} {Phys. Lett. B}\ }\textbf {\bibinfo {volume} {158}},\
  \bibinfo {pages} {223} (\bibinfo {year} {1985}{\natexlab{b}})}\BibitemShut
  {NoStop}%
\bibitem [{\citenamefont {Morris}\ \emph
  {et~al.}(1986{\natexlab{a}})\citenamefont {Morris}, \citenamefont {Ross},\
  and\ \citenamefont {Sachrajda}}]{Morris1986a}%
  \BibitemOpen
  \bibfield  {author} {\bibinfo {author} {\bibfnamefont {T.~R.}\ \bibnamefont
  {Morris}}, \bibinfo {author} {\bibfnamefont {D.~A.}\ \bibnamefont {Ross}},\
  and\ \bibinfo {author} {\bibfnamefont {C.~T.}\ \bibnamefont {Sachrajda}},\
  }\bibfield  {title} {\bibinfo {title} {{Instantons and the renormalisation
  group in supersymmetric Yang-Mills theories}},\ }\href
  {https://doi.org/10.1016/0550-3213(86)90476-1} {\bibfield  {journal}
  {\bibinfo  {journal} {Nucl. Phys. B}\ }\textbf {\bibinfo {volume} {264}},\
  \bibinfo {pages} {111} (\bibinfo {year} {1986}{\natexlab{a}})}\BibitemShut
  {NoStop}%
\bibitem [{\citenamefont {Morris}\ \emph
  {et~al.}(1986{\natexlab{b}})\citenamefont {Morris}, \citenamefont {Ross},\
  and\ \citenamefont {Sachrajda}}]{Morris1986b}%
  \BibitemOpen
  \bibfield  {author} {\bibinfo {author} {\bibfnamefont {T.~R.}\ \bibnamefont
  {Morris}}, \bibinfo {author} {\bibfnamefont {D.~A.}\ \bibnamefont {Ross}},\
  and\ \bibinfo {author} {\bibfnamefont {C.~T.}\ \bibnamefont {Sachrajda}},\
  }\bibfield  {title} {\bibinfo {title} {Instantons, the beta-function and
  renormalisation scheme dependence},\ }\href
  {https://doi.org/10.1016/0370-2693(86)90212-1} {\bibfield  {journal}
  {\bibinfo  {journal} {Phys. Lett. B}\ }\textbf {\bibinfo {volume} {172}},\
  \bibinfo {pages} {40} (\bibinfo {year} {1986}{\natexlab{b}})}\BibitemShut
  {NoStop}%
\bibitem [{\citenamefont {Mello}(1990)}]{mello1990averages}%
  \BibitemOpen
  \bibfield  {author} {\bibinfo {author} {\bibfnamefont {P.}~\bibnamefont
  {Mello}},\ }\bibfield  {title} {\bibinfo {title} {Averages on the unitary
  group and applications to the problem of disordered conductors},\ }\href
  {https://doi.org/10.1088/0305-4470/23/18/013} {\bibfield  {journal} {\bibinfo
   {journal} {Journal of Physics A: Mathematical and General}\ }\textbf
  {\bibinfo {volume} {23}},\ \bibinfo {pages} {4061} (\bibinfo {year}
  {1990})}\BibitemShut {NoStop}%
\bibitem [{\citenamefont {Hikami}(1981)}]{Hikami1981}%
  \BibitemOpen
  \bibfield  {author} {\bibinfo {author} {\bibfnamefont {S.}~\bibnamefont
  {Hikami}},\ }\bibfield  {title} {\bibinfo {title} {Three-loop
  $\beta$-functions of non-linear $\sigma$-models on symmetric spaces},\ }\href
  {https://doi.org/10.1016/0370-2693(81)90989-8} {\bibfield  {journal}
  {\bibinfo  {journal} {Phys. Lett. B}\ }\textbf {\bibinfo {volume} {98}},\
  \bibinfo {pages} {208} (\bibinfo {year} {1981})}\BibitemShut {NoStop}%
\bibitem [{\citenamefont {Evers}(1997)}]{Evers1997}%
  \BibitemOpen
  \bibfield  {author} {\bibinfo {author} {\bibfnamefont {F.}~\bibnamefont
  {Evers}},\ }\bibfield  {title} {\bibinfo {title} {Relaxation on critical
  percolation clusters, self-avoiding random walks, and the quantum {Hall}
  effect},\ }\href {https://doi.org/10.1103/PhysRevE.55.2321} {\bibfield
  {journal} {\bibinfo  {journal} {Phys. Rev. E}\ }\textbf {\bibinfo {volume}
  {55}},\ \bibinfo {pages} {2321} (\bibinfo {year} {1997})}\BibitemShut
  {NoStop}%
\bibitem [{\citenamefont {Cardy}(2000)}]{Cardy2000}%
  \BibitemOpen
  \bibfield  {author} {\bibinfo {author} {\bibfnamefont {J.}~\bibnamefont
  {Cardy}},\ }\bibfield  {title} {\bibinfo {title} {Linking numbers for
  self-avoiding loops and percolation: Application to the spin quantum hall
  transition},\ }\href {https://doi.org/10.1103/PhysRevLett.84.3507} {\bibfield
   {journal} {\bibinfo  {journal} {Phys. Rev. Lett.}\ }\textbf {\bibinfo
  {volume} {84}},\ \bibinfo {pages} {3507} (\bibinfo {year}
  {2000})}\BibitemShut {NoStop}%
\bibitem [{\citenamefont {Babkin}\ \emph {et~al.}(2023)\citenamefont {Babkin},
  \citenamefont {Karcher}, \citenamefont {Burmistrov},\ and\ \citenamefont
  {Mirlin}}]{Babkin2023}%
  \BibitemOpen
  \bibfield  {author} {\bibinfo {author} {\bibfnamefont {S.~S.}\ \bibnamefont
  {Babkin}}, \bibinfo {author} {\bibfnamefont {J.~F.}\ \bibnamefont {Karcher}},
  \bibinfo {author} {\bibfnamefont {I.~S.}\ \bibnamefont {Burmistrov}},\ and\
  \bibinfo {author} {\bibfnamefont {A.~D.}\ \bibnamefont {Mirlin}},\ }\bibfield
   {title} {\bibinfo {title} {Generalized surface multifractality in
  two-dimensional disordered systems},\ }\href
  {https://doi.org/10.1103/PhysRevB.108.104205} {\bibfield  {journal} {\bibinfo
   {journal} {Phys. Rev. B}\ }\textbf {\bibinfo {volume} {108}},\ \bibinfo
  {pages} {104205} (\bibinfo {year} {2023})}\BibitemShut {NoStop}%
\bibitem [{\citenamefont {Slevin}\ and\ \citenamefont
  {Ohtsuki}(2014)}]{Slevin2014}%
  \BibitemOpen
  \bibfield  {author} {\bibinfo {author} {\bibfnamefont {K.}~\bibnamefont
  {Slevin}}\ and\ \bibinfo {author} {\bibfnamefont {T.}~\bibnamefont
  {Ohtsuki}},\ }\bibfield  {title} {\bibinfo {title} {Critical exponent for the
  {Anderson} transition in the three-dimensional orthogonal universality
  class},\ }\href {http://stacks.iop.org/1367-2630/16/i=1/a=015012} {\bibfield
  {journal} {\bibinfo  {journal} {New J. Phys.}\ }\textbf {\bibinfo {volume}
  {16}},\ \bibinfo {pages} {015012} (\bibinfo {year} {2014})}\BibitemShut
  {NoStop}%
\bibitem [{\citenamefont {Subramaniam}\ \emph {et~al.}(2006)\citenamefont
  {Subramaniam}, \citenamefont {Gruzberg}, \citenamefont {Ludwig},
  \citenamefont {Evers}, \citenamefont {Mildenberger},\ and\ \citenamefont
  {Mirlin}}]{Subramaniam2006}%
  \BibitemOpen
  \bibfield  {author} {\bibinfo {author} {\bibfnamefont {A.~R.}\ \bibnamefont
  {Subramaniam}}, \bibinfo {author} {\bibfnamefont {I.~A.}\ \bibnamefont
  {Gruzberg}}, \bibinfo {author} {\bibfnamefont {A.~W.~W.}\ \bibnamefont
  {Ludwig}}, \bibinfo {author} {\bibfnamefont {F.}~\bibnamefont {Evers}},
  \bibinfo {author} {\bibfnamefont {A.}~\bibnamefont {Mildenberger}},\ and\
  \bibinfo {author} {\bibfnamefont {A.~D.}\ \bibnamefont {Mirlin}},\ }\bibfield
   {title} {\bibinfo {title} {Surface criticality and multifractality at
  localization transitions},\ }\href
  {https://doi.org/10.1103/PhysRevLett.96.126802} {\bibfield  {journal}
  {\bibinfo  {journal} {Phys. Rev. Lett.}\ }\textbf {\bibinfo {volume} {96}},\
  \bibinfo {pages} {126802} (\bibinfo {year} {2006})}\BibitemShut {NoStop}%
\bibitem [{\citenamefont {Mildenberger}\ \emph {et~al.}(2007)\citenamefont
  {Mildenberger}, \citenamefont {Subramaniam}, \citenamefont {Narayanan},
  \citenamefont {Evers}, \citenamefont {Gruzberg},\ and\ \citenamefont
  {Mirlin}}]{mildenberger2007}%
  \BibitemOpen
  \bibfield  {author} {\bibinfo {author} {\bibfnamefont {A.}~\bibnamefont
  {Mildenberger}}, \bibinfo {author} {\bibfnamefont {A.~R.}\ \bibnamefont
  {Subramaniam}}, \bibinfo {author} {\bibfnamefont {R.}~\bibnamefont
  {Narayanan}}, \bibinfo {author} {\bibfnamefont {F.}~\bibnamefont {Evers}},
  \bibinfo {author} {\bibfnamefont {I.~A.}\ \bibnamefont {Gruzberg}},\ and\
  \bibinfo {author} {\bibfnamefont {A.~D.}\ \bibnamefont {Mirlin}},\ }\bibfield
   {title} {\bibinfo {title} {Boundary multifractality in critical
  one-dimensional systems with long-range hopping},\ }\href
  {https://doi.org/10.1103/PhysRevB.75.094204} {\bibfield  {journal} {\bibinfo
  {journal} {Phys. Rev. B}\ }\textbf {\bibinfo {volume} {75}},\ \bibinfo
  {pages} {094204} (\bibinfo {year} {2007})}\BibitemShut {NoStop}%
\bibitem [{\citenamefont {Senthil}\ and\ \citenamefont
  {Fisher}(2000)}]{Senthil2000}%
  \BibitemOpen
  \bibfield  {author} {\bibinfo {author} {\bibfnamefont {T.}~\bibnamefont
  {Senthil}}\ and\ \bibinfo {author} {\bibfnamefont {M.~P.~A.}\ \bibnamefont
  {Fisher}},\ }\bibfield  {title} {\bibinfo {title} {Quasiparticle localization
  in superconductors with spin-orbit scattering},\ }\href
  {https://doi.org/10.1103/PhysRevB.61.9690} {\bibfield  {journal} {\bibinfo
  {journal} {Phys. Rev. B}\ }\textbf {\bibinfo {volume} {61}},\ \bibinfo
  {pages} {9690} (\bibinfo {year} {2000})}\BibitemShut {NoStop}%
\bibitem [{\citenamefont {Bocquet}\ \emph {et~al.}(2000)\citenamefont
  {Bocquet}, \citenamefont {Serban},\ and\ \citenamefont
  {Zirnbauer}}]{Bocquet2000}%
  \BibitemOpen
  \bibfield  {author} {\bibinfo {author} {\bibfnamefont {M.}~\bibnamefont
  {Bocquet}}, \bibinfo {author} {\bibfnamefont {D.}~\bibnamefont {Serban}},\
  and\ \bibinfo {author} {\bibfnamefont {M.~R.}\ \bibnamefont {Zirnbauer}},\
  }\bibfield  {title} {\bibinfo {title} {Disordered 2d quasiparticles in class
  {D: D}irac fermions with random mass, and dirty superconductors},\ }\href
  {https://doi.org/10.1016/S0550-3213(00)00208-X} {\bibfield  {journal}
  {\bibinfo  {journal} {Nucl. Phys. B}\ }\textbf {\bibinfo {volume} {578}},\
  \bibinfo {pages} {628} (\bibinfo {year} {2000})}\BibitemShut {NoStop}%
\bibitem [{\citenamefont {Chalker}\ \emph {et~al.}(2001)\citenamefont
  {Chalker}, \citenamefont {Read}, \citenamefont {Kagalovsky}, \citenamefont
  {Horovitz}, \citenamefont {Avishai},\ and\ \citenamefont
  {Ludwig}}]{Chalker2001}%
  \BibitemOpen
  \bibfield  {author} {\bibinfo {author} {\bibfnamefont {J.~T.}\ \bibnamefont
  {Chalker}}, \bibinfo {author} {\bibfnamefont {N.}~\bibnamefont {Read}},
  \bibinfo {author} {\bibfnamefont {V.}~\bibnamefont {Kagalovsky}}, \bibinfo
  {author} {\bibfnamefont {B.}~\bibnamefont {Horovitz}}, \bibinfo {author}
  {\bibfnamefont {Y.}~\bibnamefont {Avishai}},\ and\ \bibinfo {author}
  {\bibfnamefont {A.~W.~W.}\ \bibnamefont {Ludwig}},\ }\bibfield  {title}
  {\bibinfo {title} {Thermal metal in network models of a disordered
  two-dimensional superconductor},\ }\href
  {https://doi.org/10.1103/PhysRevB.65.012506} {\bibfield  {journal} {\bibinfo
  {journal} {Phys. Rev. B}\ }\textbf {\bibinfo {volume} {65}},\ \bibinfo
  {pages} {012506} (\bibinfo {year} {2001})}\BibitemShut {NoStop}%
\bibitem [{\citenamefont {Read}\ and\ \citenamefont {Ludwig}(2001)}]{Read2001}%
  \BibitemOpen
  \bibfield  {author} {\bibinfo {author} {\bibfnamefont {N.}~\bibnamefont
  {Read}}\ and\ \bibinfo {author} {\bibfnamefont {A.~W.~W.}\ \bibnamefont
  {Ludwig}},\ }\bibfield  {title} {\bibinfo {title} {Absence of a metallic
  phase in random-bond ising models in two dimensions: Applications to
  disordered superconductors and paired quantum hall states},\ }\href
  {https://doi.org/10.1103/PhysRevB.63.024404} {\bibfield  {journal} {\bibinfo
  {journal} {Phys. Rev. B}\ }\textbf {\bibinfo {volume} {63}},\ \bibinfo
  {pages} {024404} (\bibinfo {year} {2001})}\BibitemShut {NoStop}%
\bibitem [{\citenamefont {Gruzberg}\ \emph {et~al.}(2005)\citenamefont
  {Gruzberg}, \citenamefont {Read},\ and\ \citenamefont
  {Vishveshwara}}]{Gruzberg2005}%
  \BibitemOpen
  \bibfield  {author} {\bibinfo {author} {\bibfnamefont {I.~A.}\ \bibnamefont
  {Gruzberg}}, \bibinfo {author} {\bibfnamefont {N.}~\bibnamefont {Read}},\
  and\ \bibinfo {author} {\bibfnamefont {S.}~\bibnamefont {Vishveshwara}},\
  }\bibfield  {title} {\bibinfo {title} {Localization in disordered
  superconducting wires with broken spin-rotation symmetry},\ }\href
  {https://doi.org/10.1103/PhysRevB.71.245124} {\bibfield  {journal} {\bibinfo
  {journal} {Phys. Rev. B}\ }\textbf {\bibinfo {volume} {71}},\ \bibinfo
  {pages} {245124} (\bibinfo {year} {2005})}\BibitemShut {NoStop}%
\end{thebibliography}%
	
\end{document}